\documentclass[useAMS,usenatbib]{mn2e}

\usepackage{color}
\usepackage{colortbl}
\usepackage{multirow}
\usepackage{dcolumn}
\usepackage{amsmath}
\usepackage{amssymb}
\usepackage{graphicx}
\usepackage{epsfig}
\usepackage{changebar}
\usepackage{natbib}
\usepackage{multirow}
\usepackage{subfigure}
\usepackage{txfonts}
\usepackage{rotating}

\usepackage{longtable,lscape}

\newcolumntype{d}[1]{D{.}{\cdot}{#1}}

\newcolumntype{.}{D{.}{.}{-1}}

\newcommand{\lsun}{L$_\odot$}
\newcommand{\msun}{M$_\odot$}
\newcommand{\rsun}{R$_\odot$}

\newcommand{\lbol}{\emph{L}$_{\rm{bol}}$}
\newcommand{\mclump}{\emph{M}$_{\rm{clump}}$}
\newcommand{\mco}{\emph{M}$_{\rm{^{13}CO}}$}

\newcommand{\reff}{\emph{R}$_{\rm{eff}}$}

\newcommand{\vlsr}{V$_{\rm{LSR}}$}

\newcommand{\mum}{$\umu$m}

\newcommand{\kms}{km\,s$^{-1}$}

\newcommand{\hi}{H~{\sc i}}
\newcommand{\hii}{H~{\sc ii}}
\newcommand{\uchii}{UC\,H~{\sc ii}}

\newcommand{\poi}{Poisson}

\newcommand{\sex}{\texttt{SExtractor}}

\newcommand{\submm}{submillimetre}

\newcommand{\KS}{Kolmogorov-Smirnov}

\title[ATLASGAL --- properties of massive star forming clumps]{ATLASGAL --- towards a complete sample of massive star forming clumps\thanks{The full version of Tables\,1, 3, 6 and 7, and Fig.\,3 are only available in electronic form at the CDS via anonymous ftp to cdsarc.u-strasbg.fr (130.79.125.5) or via http://cdsweb.u-strasbg.fr/cgi-bin/qcat?J/MNRAS/.}}
\author[J. S. Urquhart et al.]{J.\,S.\,Urquhart$^{1}$\thanks{E-mail:
jurquhart@mpifr-bonn.mpg.de (MPIfR)}, T.\,J.\,T.\,Moore$^{2}$, T.\,Csengeri$^{1}$, F.\,Wyrowski$^{1}$, F.\,Schuller$^{3}$, M.\,G.\,Hoare${^4}$, \newauthor S.\,L.\,Lumsden${^4}$, J.\,C.\,Mottram$^{5}$, M.\,A.\,Thompson$^{6}$, K.\,M.\,Menten$^{1}$, C.\,M.\,Walmsley$^{7, 8}$, \newauthor L.\,Bronfman$^{9}$, S.\,Pfalzner$^{1}$, C.\,K\"onig$^{1}$ and M.\,Wienen$^{1}$\\
$^{1}$ Max-Planck-Institut f\"ur Radioastronomie, Auf dem H\"ugel
  69, D-53121 Bonn, Germany \\
$^{2}$Astrophysics Research Institute, Liverpool John Moores University, Liverpool Science Park, 146 Brownlow Hill, Liverpool, L3\,5RF, UK\\ 
$^{3}$European Southern Observatory, Alonso de Cordova 3107, Vitacura, Santiago, Chile\\
$^{4}$School of Physics and Astrophysics, University of Leeds, Leeds, LS2\,9JT, UK \\
$^{5}$ Leiden Observatory, Leiden University, NL-2300 RA Leiden, the Netherlands \\
$^{6}$ Centre for Astrophysics Research, Science and Technology Research Institute, University of Hertfordshire, College Lane, Hatfield, AL10 9AB, UK \\
$^{7}$Osservatorio Astrofisico di Arcetri, Largo E. Fermi, 5, 50125 Firenze, Italy\\
$^{8}$Dublin Institute for Advanced Studies, Burlington Road 10, Dublin 4, Ireland\\
$^{9}$Departamento de Astronom\'{i}a, Universidad de Chile, Casilla 36-D, Santiago, Chile\\
}

\begin{document}

\date{Accepted ??. Received ??; in original form ??}

\pagerange{\pageref{firstpage}--\pageref{lastpage}} \pubyear{2014}

\maketitle

\label{firstpage}

\begin{abstract}

By matching infrared-selected, massive young stellar objects (MYSOs) and compact HII regions in the RMS survey to massive clumps found in the submillimetre ATLASGAL survey, we have identified $\sim$1000 embedded young massive stars between $280\degr$$<$\,$\ell$\,$<$\,$350\degr$ and  $10\degr$$<$\,$\ell$\,$<$\,$60\degr$ with $|\,b\,|<1.5\degr$. Combined with an existing sample of radio-selected methanol masers and compact HII regions, the result is a catalogue of $\sim$1700 massive stars embedded within $\sim$1300 clumps located across the inner Galaxy, containing three observationally distinct subsamples, methanol-maser, MYSO and HII-region associations, covering the most important tracers of massive star formation, thought to represent key stages of evolution. We find that massive star formation is strongly correlated with the regions of highest column density in spherical, centrally condensed clumps. We find no significant differences between the three samples in clump structure or the relative location of the embedded stars, which suggests that the structure of a clump is set before the onset of star formation, and changes little as the embedded object evolves towards the main sequence. There is a strong linear correlation between clump mass and bolometric luminosity, with the most massive stars forming in the most massive clumps. We find that the MYSO and HII-region subsamples are likely to cover a similar range of evolutionary stages and that the majority are near the end of their main accretion phase. We find few infrared-bright MYSOs associated with the most massive clumps, probably due to very short pre-main sequence lifetimes in the most luminous sources.

\end{abstract}

\section{Introduction}
\label{sect:intro}

Massive stars ($>$ 8\,\msun\ and 10$^3$\,\lsun) play an important role in many astrophysical processes, from the formation of the first solid material in the early Universe (\citealt{dunne2003}) to their substantial influence upon the evolution of their host galaxies and future generations of star formation (\citealt{kennicutt2005}). These stars have a profound impact on their local environment through powerful outflows, strong stellar winds, optical/far-UV radiation and supernovae, shaping the interstellar medium (ISM) and regulating star formation. Our understanding of how galaxies evolve over cosmological timescales depends critically on various assumptions that are made about massive star formation, such as how the feedback introduced by massive stars affects the star-formation rate, or the universality of the initial mass function (IMF). However, the underlying physical processes involved in the formation of massive stars are still poorly defined and this presents a major hurdle to furthering understanding of the massive star formation itself and how galaxies evolve (e.g., \citealt{silk2012} and \citealt{scannapieco2012}). 

Outside the Milky Way, we are generally restricted to studies of global star-formation properties integrated over entire star-forming complexes or even groups of complexes, which renders it impossible to study either feedback processes or the environmental dependence of star formation (\citealt{kennicutt2012}). Within the Milky Way we are able to probe massive star-forming regions in far greater detail, which provides the best opportunity of understanding the processes involved in massive star formation. Furthermore, the Milky Way consists of a large range of environments including the Galactic centre with its extreme UV-radiation and cosmic-ray fluxes, intense star-forming regions found in the spiral arms (e.g., W43, W49 \& W51; often described as ``mini-starbursts'') and the reduced metallicity of the outer Galaxy (\citealt{kruijssen2013}). Many of these environments have extragalactic analogues and an understanding of how they affect star formation in the Milky Way will provide crucial insights into star formation in the era of galaxy formation. 

Despite their importance, our understanding of the initial conditions required and processes involved in the formation and early evolution of massive stars is still rather poor in comparison to low-mass stars. Massive stars are rare and the regions where they form are significantly more distant than low-mass star forming regions. Massive stars form almost exclusively in clusters (\citealt{de-wit2004}), making it hard to distinguish between the properties of the cluster and individual members. Furthermore, the Kelvin-Helmholtz contraction time is shorter than the accretion time for massive stars. Consequently, massive protostars reach the main sequence while still deeply embedded in their natal environment and so all of the earliest stages can only be probed at far-infrared and (sub)millimetre wavelengths.

It is still not yet clear how massive stars obtain their final mass as much of this needs to be acquired after they have reached the main sequence when radiation pressure would be expected to halt accretion. this problem can be overcome through accretion via a disc \citep[e.g.][]{kuiper2011}, however, the mechanism by which this happens is still an open question. Currently, there are two competing accretion models by which an embedded protostar can acquire the mass required to evolve into a massive star; these are the ``monolithic accretion'' and ``competitive accretion'' models (e.g., \citealt{mckee2002,mckee2003} and \citealt{bonnell1997,bonnell2001}, respectively).
 
Our ability to make significant progress in this field has been dramatically enhanced in recent years with the completion of a large number of Galactic-plane surveys that cover the whole wavelength range from the near-infrared to the radio (e.g., the UKIDSS GPS, \citealt{lucas2008};  GLIMPSE, \citealt{benjamin2003_ori}; MSX, \citealt{price2001}; MIPSGAL, \citealt{carey2009}; \textit{Herschel} Hi-GAL, \citealt{molinari2010}); APEX Telescope Large Area Survey of the Galaxy (ATLASGAL), \citealt{schuller2009}; Bolocam Galactic Plane Survey (BGPS), \citealt{aguirre2011}; methanol multibeam (MMB) survey,  \citealt{caswell2010b}; and the Co-ordinated Radio and Infrared Survey for High-Mass Star Formation (CORNISH), \citealt{purcell2013} and  \citealt{hoare2012}). These surveys probe the large spatial volumes required to address the major problem in studying Galactic massive star formation --- its intrinsic rarity. At last we can identify and study statistically significant samples over the full range of evolutionary stages, from the pre-stellar through to the post-compact HII region stage when the star emerges from its natal environment. 

\subsection{Evolutionary sequence}

In spite of the observational challenges associated with massive star formation,  an outline of an evolutionary sequence has emerged over the past decade from a combination of observations and numerical modeling (e.g., \citealt{zinnecker2007}). This sequence can be roughly divided into four distinct stages: 

\begin{enumerate}

\item Pre-stellar: consisting of a massive, dense clump that is gravitationally bound but shows no evidence of being associated with any embedded point source at infrared wavelengths or other tracers.

\item Clump-core collapse: the clumps collapse and fragment into cores that form a protostar surrounded by an accretion disc. As the protostar gains mass it begins to warm its natal envelope, exciting class II methanol-maser emission through radiative pumping in the disc (\citealt{sobolev2007}). 

\item Protostellar swelling and contraction: Numerical simulations with high accretion rates have been performed by \citet{hosokawa2009} and \citet{hosokawa2010}. These models reveal that the high accretion rates lead to a dramatic swelling up of the protostar from a few solar radii to $\sim$100\,\rsun\ as its stellar mass increases from 5 to 10\,\msun; the increased volume of the protostar results in lower internal densities and temperatures, which delays the onset of core hydrogen burning (see also \citealt{hoare2007a}). Above 10\,\msun\ the Kelvin-Helmholtz timescale is lower than the accretion timescale and as a consequence the protostar begins to contract and the internal temperature gradually increases. When the temperature exceeds 10$^7$\,K core hydrogen burning begins and the star joins the main sequence with a mass of $\sim$30\,\msun. This implies there are no stars more massive than this at ZAMS onset and accretion must continue post-ZAMS for larger stars. Observationally, protostars that have not yet begun to form a \hii\ region  will have luminosities between 10$^{3}$ and 10$^5$\,\lsun; we will refer to these objects as massive young stellar objects (MYSOs; \citealt{wynn-williams1982}). The RMS survey has identified $\sim$600 MYSOs located throughout the Galaxy (\citealt{lumsden2013,mottram2011a}). The most luminous of these correspond to the range predicted by the Hosokawa et al. models (i.e., $\leq$10$^{5}$\,\lsun).

\item  Formation of the \hii\ region: once core hydrogen burning has begun and the star has joined the main sequence it will begin to emit a sufficient amount of ionizing radiation and a \hii\ region will form around it. This will begin as a hyper-compact \hii\ region but will rapidly expand, transitioning through the ultra-compact (UC) and compact \hii\ region phases before breaking out of its natal clump and evolving into the more classical \hii\ regions that are observable at optical wavelengths. The thermal bremsstrahlung emission from the ionized gas within the \hii\ regions can be readily detected at radio wavelengths (e.g., \citealt{wood1989} and \citealt{kurtz1994}). Radio continuum observations can therefore be used to distinguish between MYSOs and \hii\ regions (\citealt{hoare2007a,hoare2007}).

\end{enumerate}

In order to test this evolutionary sequence we need to compile a large sample of massive star-forming regions with a sufficient number of sources that the statistical properties of each stage can be determined. 
 
\subsection{Identifying an evolutionary sample of massive star forming clumps}

All of the earliest stages of massive star formation take place while the forming stars are deeply embedded in dense clumps (\citealt{garay2004}), where they may remain hidden, even at mid-infrared wavelengths \citep[e.g.,][]{parsons2009}. However, the thermal emission from dust at submm wavelengths is optically thin and is therefore an excellent tracer of column density and total mass of a clump (\citealt{schuller2009}). The ATLASGAL survey is a blind 870-\mum\ survey covering 420 square degrees of the inner Galactic plane. Its unbiased coverage and fairly uniform sensitivity provides an ideal resource for identifying a large and representative sample of massive, star-forming structures (\citealt{contreras2013}, Urquhart et al. 2014b). Furthermore, ATLASGAL is complete to all dense clumps (Log($N$(H$_2$)) $> 22$\,cm$^{-2}$) with masses ($M_{\rm{clump}}$) $>$ 1000\,\msun\ located within the inner Galactic disc at heliocentric distances up to $\sim$20\,kpc (\citealt{urquhart2013a}; hereafter Paper\,I) and provides a uniform set of measurements that can be used to test the proposed evolutionary sequence.

In two previous studies we have used the ATLASGAL Compact Source Catalogue (CSC; \citealt{contreras2013}) as a starting point and cross-matched the positions of dense clumps with those of methanol masers identified by the MMB survey (Paper\,I) and compact \hii\ regions identified by  the CORNISH VLA 5\,GHz continuum survey (\citealt{urquhart2013b}; hereafter Paper\,II). These two studies identified a large sample of $\sim$800 massive star-forming clumps. This was  used to derive a minimum surface density threshold for massive star formation and provided some of the strongest evidence that methanol masers trace one of the earliest stages in the formation of massive stars. Furthermore, we used the Galactic distribution of the methanol masers to estimate the star-formation efficiency (SFE) across the inner Galaxy and found it to be approximately three times lower in the Galactic Centre than in the disc (cf. \citealt{longmore2012b}). 

In this paper, the final in this series, we investigate the properties of the ATLASGAL clumps associated with MYSOs and \hii\ regions identified by the Red MSX Source (RMS) survey (\citealt{hoare2005,urquhart2008,lumsden2013}). Inclusion of this sample of young massive stars will produce the largest and most comprehensive sample of massive star-forming clumps to date and cover the whole accretion process from the earliest, most deeply embedded stage to the creation and expansion of an ionized nebula that disrupts the natal environment, halting accretion and setting the final mass of the star.

The structure of this paper is as follows: in Sect.\,2 we give a brief summary of the individual surveys used and describe the matching procedure and reliability checks performed. In Sect.\,3 we discuss the correlation of RMS-associated clumps with the samples of methanol masers and \hii\ regions presented in Papers\,I and II and look at the Galactic distribution of the whole sample. In this section we also distinguish between clumps associated with two or more of the different tracers from those associated with a single tracer (i.e., methanol-maser, mid-infrared bright and compact radio emission) in order to identify relatively clean samples with which to look for evolutionary trends that might be present in these data. In Sect.\,4 we derive properties of the associated clumps and compare the statistical properties of the observationally defined subsamples. In Sect.\,5 we investigate the empirical relationships between clump mass, bolometric luminosity and radius with respect to the different subsamples and discuss their implications. We present a summary of our main findings and give an overview of this series of papers in Sect.\,6. 

\section{Survey descriptions and matching statistics}

\begin{figure}
\begin{center}
\includegraphics[width=0.48\textwidth, trim= 0 0 0 0]{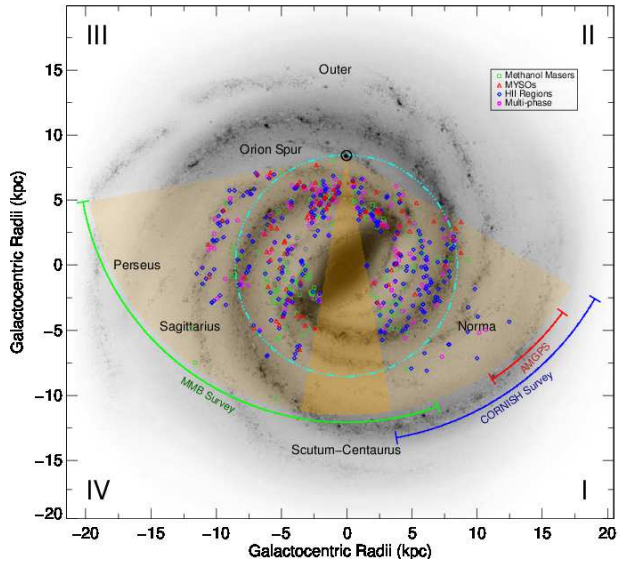}

\caption{\label{fig:gal_distribution_nodata} Schematic of the Galactic disc as viewed from the Northern Galactic Pole showing the positions of the Galactic bar and the spiral arms. The orange wedges show the coverage of the ATLASGAL survey assuming a mass sensitivity of 1,000\,\msun\ at a distance of 20\,kpc. The brighter wedge shows the inner most 10\degr\ in longitude not included in the RMS survey. The spiral arms are labeled in black and Galactic quadrants are given by the roman numerals in the corners of the image. The Galactic longitude covered by the AMGPS, MMB and CORNISH surveys are indicated by red, blue and green curves, respectively. The Sun is located at the apex of the wedges and is indicated by the $\odot$ symbol.  The dot-dashed cyan circle shows the Solar circle and the inner most thick green circle shows the location of the 5\,kpc molecular ring where most of the molecular gas in the Galaxy resides. Background image courtesy of NASA/JPL-Caltech/R. Hurt (SSC/Caltech).} 

\end{center}
\end{figure}

\subsection{ATLASGAL}

ATLASGAL is one of the first systematic surveys of the inner Galactic plane in the \submm\ band. The survey was carried out with the Large APEX Bolometer Camera (LABOCA; \citealt{siringo2009}), an array of 295 bolometers observing at 870\,$\umu$m (345\,GHz). The APEX telescope has a full-width half-maximum (FWHM) beam size of 19.2\arcsec\ at this frequency and has a positional accuracy of $\sim$2\arcsec. The initial survey covered a Galactic longitude region of $300\degr < \ell < 60\degr$ and $|b| < 1.5\degr$ but this was later extended to include $280\degr < \ell < 300\degr$ and $-2\degr < b < 1\degr$ to take account of the warp present in this part of the Galactic plane (\citealt{schuller2009}). This extension increasing the overall coverage of the survey to 420\,sq.\,degrees and includes the Sagittarius spiral arm tangent and the Carina star-forming complex. 

Using the source extraction algorithm \sex\ (\citealt{bertin1996}) a total of 10000 dust clumps have been identified in the ATLASGAL emission maps. An initial catalogue of the inner 150\,sq. degrees of the Galactic plane and a detailed description of the method and consistency checks are presented in  \citet[][i.e., 330\degr\ $ <\ell <$ 21\degr]{contreras2013} while a description of the rest of the catalogue is presented in a subsequent research note (Urquhart et al., 2014). The whole catalogue is 99\,per\,cent complete at $\sim$6$\sigma$, which corresponds to a flux sensitivity of 0.3-0.5\,Jy\,beam$^{-1}$ (see Fig.\,1 of \citealt{csengeri2014} for survey sensitivity as a function of Galactic longitude). The effective radius of the clumps range from 20-100\arcsec, with a mean of and median value of $\sim$30\arcsec, and so the majority of sources are extended with respect to the APEX beam. 

We present a schematic of the Galactic plane in Fig.\,\ref{fig:gal_distribution_nodata} showing the ATLASGAL coverage of the inner Galactic disc. This survey covers $\sim$2/3 of the surface area of the Galactic disc (assuming the molecular disc has a Galactocentric radius, $R_{\rm{GC}}$, $<15$\,kpc) and 97\,per\,cent of the surface area located inside the Solar circle and all of the 5\,kpc molecular ring where the vast majority of the molecular gas in the Galaxy is located (i.e., 4\,kpc $< R_{\rm{GC}} < 6$\,kpc; \citealt{stecker1975}). The full ATLASGAL compact source catalogue (CSC) therefore provides a comprehensive census of dense dust clumps located in the inner Galaxy  and is complete to all potential massive star-forming compact clumps with masses greater than 1000\,\msun\ out to a heliocentric distance of 20\,kpc  (see Sect.\,4.5 and Paper\,I for details).

\begin{figure*}
\begin{center}

\includegraphics[width=0.49\textwidth, trim= 0 0 0 0]{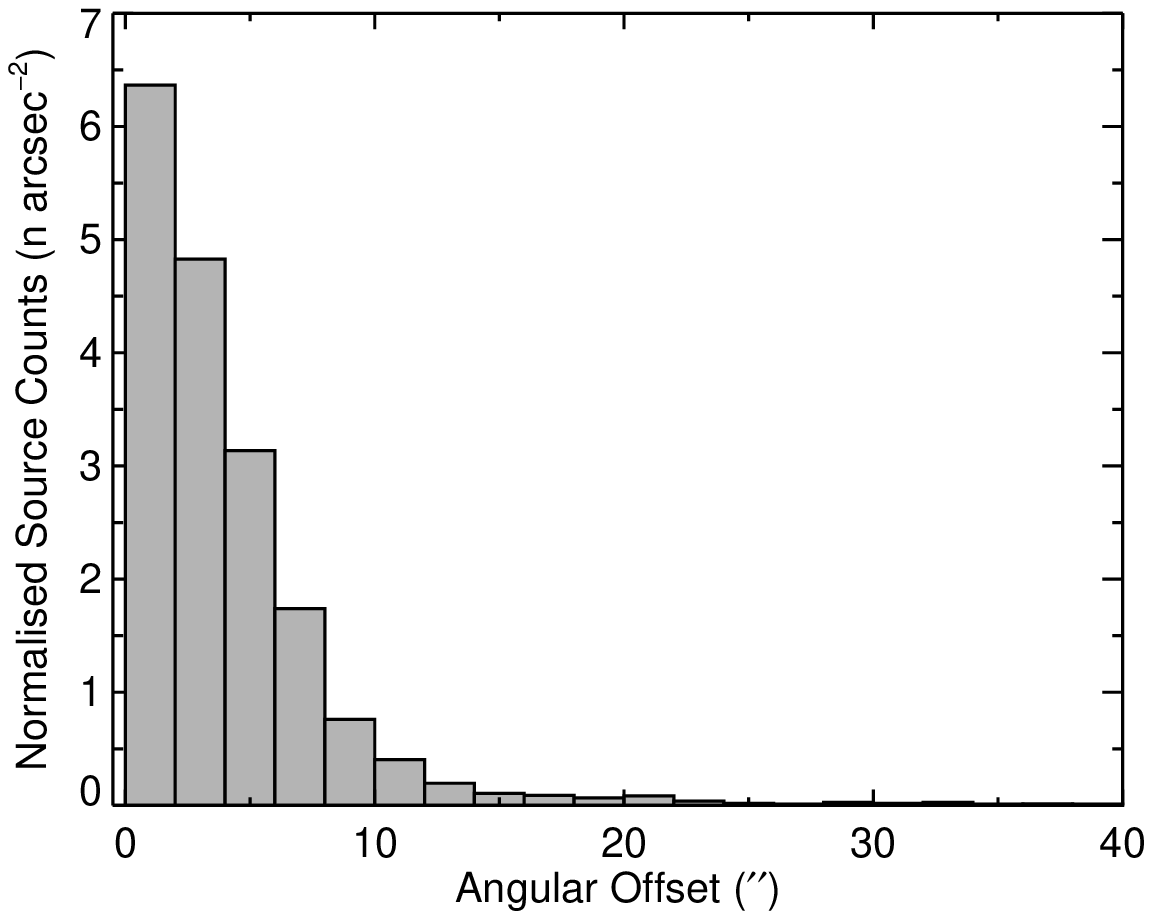}
\includegraphics[width=0.49\textwidth, trim= 0 0 0 0]{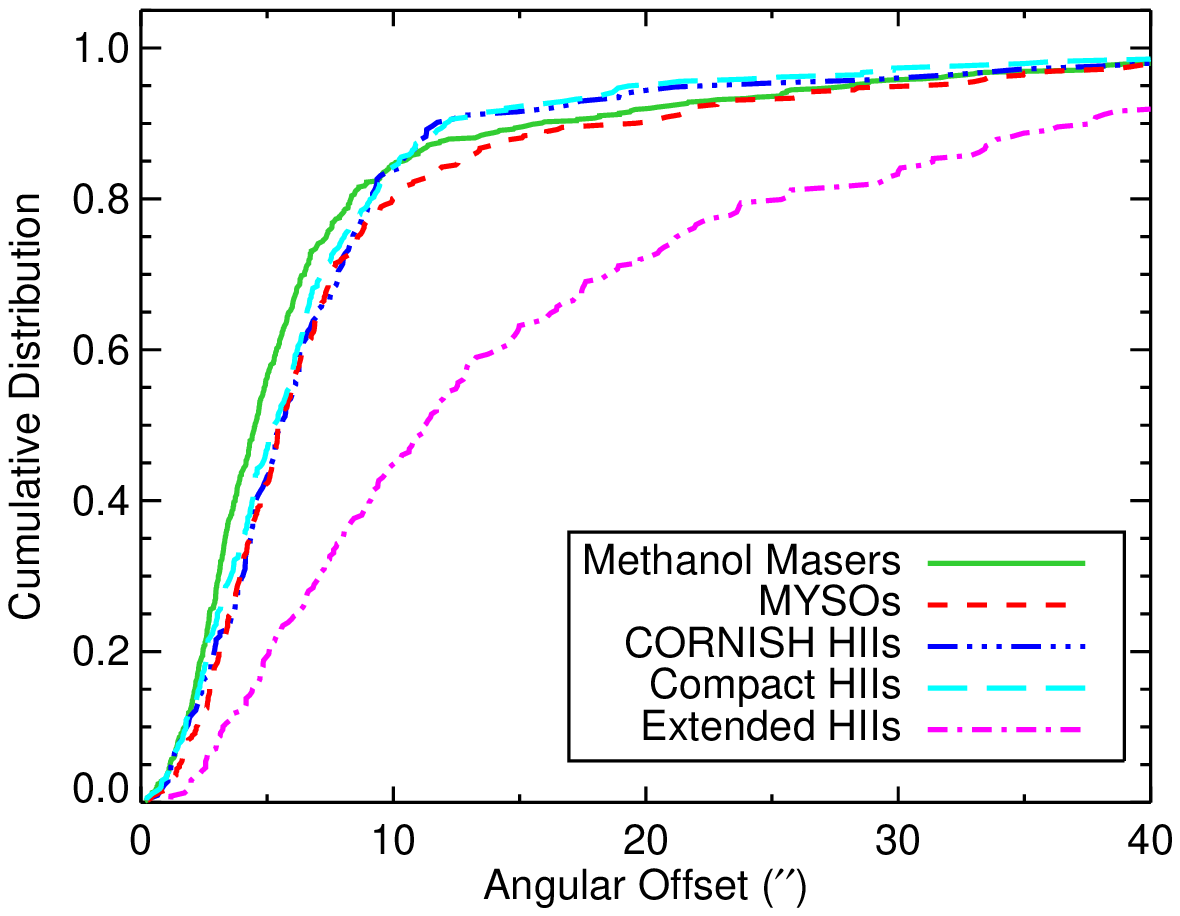}

\caption{\label{fig:angular_offset}   Left panel: Normalised source counts for the ATLASGAL-RMS matches are shown as a function of offset between their position and that of the peak of the \submm\ emission of their associated ATLASGAL source. We have truncated the $x$-axis of this plot at 40\arcsec\ as there are only 34 matches that have a larger angular separation  ($\lesssim$80\arcsec) and the surface density effectively falls to zero. The bin size is 2\arcsec. Right panel: Cumulative distribution of angular offset of the methanol masers, MYSOs and \hii\ regions drawn from the MMB, AMGPS, CORNISH and RMS surveys.}

\end{center}
\end{figure*}

\subsection{RMS Survey}
\label{sect:rms_description}

The RMS survey is a Galaxy-wide sample of massive young stellar candidates selected by comparing near- and mid-infrared colours to those of previously identified MYSOs and \hii\ regions (\citealt{lumsden2002}). Although near-infrared images were used to help aid in the colour selection a detection in this wavelength range was not required since many will be deeply embedded and do not have a near-infrared counterpart. This initially, mid-infrared-selected sample has been followed-up by a comprehensive multi-wavelength programme of observations. These observations were carefully chosen to distinguish between genuine embedded MYSOs and \hii\ regions from other dust-enshrouded contaminating sources (e.g., evolved stars, planetary nebula and nearby low-mass YSOs). These included radio continuum observations (\citealt{urquhart_radio_south, urquhart_radio_north}), a mixture of targeted mid-infrared imaging (e.g., \citealt{mottram2007}) and archival images obtained by the \textit{Spitzer} GLIMPSE legacy project (\citealt{benjamin2005_org}), near-infrared spectroscopy (e.g., \citealt{clarke2006,cooper2013}) and molecular-line observations to obtain kinematic distances (e.g., \citealt{urquhart_13co_south,urquhart_13co_north,urquhart2011b}). 

A detailed description of these follow-up observations and discussion of the properties of the final catalogue are presented in \citet{lumsden2013}. The RMS survey has identified approximately $\sim$1700 MYSOs and \hii\ regions and represents the largest and best-characterised sample compiled to date. Distances are available for $\sim$90\,per\,cent of the catalogue; these are a compilation of maser parallax (e.g., \citealt{reid2009}), spectrophotometric (e.g., \citealt{Moises2011}) and kinematic distances taken from the literature (e.g., \citealt{roman2009,green2011b}) and the RMS project's own analysis (see \citealt{urquhart2012} and \citealt{urquhart2014} for more details).  

\setlength{\tabcolsep}{6pt}

\begin{table*}

\begin{center}\caption{ATLASGAL-RMS associations and the properties of the embedded sources.}
\label{tbl:atlas_rms_matches}
\begin{minipage}{\linewidth}
\small
\begin{tabular}{ll.l.c}
\hline \hline
\multicolumn{1}{c}{ATLASGAL name}&  \multicolumn{1}{c}{RMS name}&\multicolumn{1}{c}{Offset}&	\multicolumn{1}{c}{RMS}  &\multicolumn{1}{c}{Log[$L_{\rm{bol}}$]}  &\multicolumn{1}{c}{Compact radio}  \\

\multicolumn{1}{c}{}&  \multicolumn{1}{c}{}&\multicolumn{1}{c}{(\arcsec)}&	\multicolumn{1}{c}{classification}  &\multicolumn{1}{c}{(\lsun)}  &\multicolumn{1}{c}{emission}  \\

\hline
AGAL010.323$-$00.161	&	G010.3208$-$00.1570B	&	3.5	&	YSO	&	4.4	&	no	\\
AGAL010.439+00.009	&	G010.4413+00.0101	&	7.6	&	\hii\ region	&	4.4	&	no	\\
AGAL010.472+00.027	&	G010.4718+00.0256	&	1.1	&	\hii\ region	&	5.5	&	yes	\\
AGAL010.624$-$00.384	&	G010.6235$-$00.3834	&	2.6	&	\hii\ region	&	5.6	&	yes	\\
AGAL010.884+00.122	&	G010.8856+00.1221	&	3.5	&	YSO	&	3.8	&	no	\\
AGAL010.957+00.022	&	G010.9592+00.0217	&	3.2	&	\hii\ region	&	5.1	&	yes	\\
AGAL011.034+00.061	&	G011.0340+00.0629	&	6.0	&	\hii\ region	&	5.0	&	yes	\\
AGAL011.171$-$00.064	&	G011.1723$-$00.0656	&	4.9	&	\hii\ region	&	4.4	&	yes	\\
AGAL011.497$-$01.485	&	G011.5001$-$01.4857	&	7.0	&	YSO	&	3.3	&	no	\\
AGAL011.756$-$00.149	&	G011.7588$-$00.1502	&	7.3	&	\hii\ region	&	4.1	&	no	\\
\hline\\
\end{tabular}\\

Notes: Only a small portion of the data is provided here, the full table is available in electronic form at the CDS via anonymous ftp to cdsarc.u-strasbg.fr (130.79.125.5) or via http://cdsweb.u-strasbg.fr/cgi-bin/qcat?J/MNRAS/.
\
\end{minipage}

\end{center}
\end{table*}
\setlength{\tabcolsep}{6pt}

\subsection{ATLASGAL-RMS associations}
\label{sect:atlas_rms_assoc}

Of the $\sim$1700 MYSOs and \hii\ regions identified by the RMS survey 1232 are located in the region covered by ATLASGAL; this corresponds to approximately 80\,per\,cent of the RMS catalogue. Following the same matching procedure developed for the CORNISH comparison we have used the ATLASGAL emission masks to map these embedded RMS sources to their host dust emission clumps (see Sect.\,3.1 of Paper\,II for more details). The emission masks are produced by the extraction algorithm and consist of an image the same dimensions as the original emission map but where the pixel values have been changed to integer values that relate each pixel to the catalogue identification number. Matching the source position with its associated pixel therefore provides an unambiguous match to the ATLASGAL source, even in cases where the clump morphology is irregular.  We have inspected the mid-infrared images, dust emission and positions of the RMS sources for all detections and identified a small number of sources (22) where the match did not appear to be reliable. These tended to have the largest angular separation (between $\sim$40-250\arcsec; see Fig.\,\ref{fig:angular_offset} for typical angular offsets) between the dust and the RMS source and were found to be more extended \hii\ regions that have broken out of their natal environment. Any properties derived for their associated clump are unlikely to provide any reliable information about the relationship between the host clump and its associated star formation and so these have been removed from the sample.

After the exclusion of this relatively small number of less reliable matches we are left with a sample of 1056 MYSO and \hii\ regions for which distances and luminosities have been determined. We further filter this sample to exclude sources with luminosities lower than 1000\,\lsun\  (see \citealt{mottram2011a} and Sect.\,\ref{sect:bol_lum} for details on the luminosity determination); this is to ensure that the sample contains only embedded sources that are going to become massive stars. This reduces the sample to 964, which corresponds to $\sim$90\,per\,cent of the RMS catalogue located within the ATLASGAL region and includes approximately two-thirds of the whole RMS sample of these objects for which a distance is available. These RMS sources are associated with 768 ATLASGAL clumps, with 151 clumps harbouring two or more embedded massive stars. We failed to find a match for 154 RMS sources but the majority of these were found to be more extended \hii\ regions that have already disrupted their natal environment to the extent that the dust emission falls below the ATLASGAL sensitivity or lower luminosity YSOs. Of the other 92 RMS source matched to an ATLASGAL clump (i.e., 1056$-$964), distances are not available for 75 matches and 17 have bolometric luminosities below the threshold (i.e., $<$1000\,\lsun). See Table\,\ref{tbl:atlas_rms_matches_breakdown} for a summary of these ATLASGAL-RMS matches. 

\setlength{\tabcolsep}{6pt}

\begin{table}

\begin{center}\caption{Summary of the ATLASGAL-RMS associations.}
\label{tbl:atlas_rms_matches_breakdown}
\begin{minipage}{\linewidth}
\small
\begin{tabular}{lr}
\hline \hline
\multicolumn{1}{c}{ATLASGAL-RMS matches}&  \multicolumn{1}{c}{Number} \\

\hline
Total &	1056	\\
Luminosity $>$ 1000\,\lsun &	964	\\
Luminosity $<$ 1000\,\lsun &	17	\\
No distance available&	75	\\
\hline\\
\end{tabular}\\

\end{minipage}

\end{center}
\end{table}
\setlength{\tabcolsep}{6pt}
 
The associated RMS sample consists of 361 MYSOs, 25 MYSO/\hii\ regions and 578 \hii\ regions. The MYSO/\hii\ regions classification is used when there is insufficient data to distinguish between these two stages. We checked the matches for a possible systematic offset between the RMS and ATLASGAL matches and although there is evidence of a subtle effect this is small compared to the APEX pointing uncertainty and so can be neglected. 

The RMS sample of \hii\ regions consists of a mixture of compact and UC \hii\ regions that are associated with compact radio emission and more extended \hii\ regions that are radio-quiet but can be classified from their mid-infrared morphology.\footnote{The radio emission from these extended \hii\ regions is either filtered out by the interferometric radio continuum observations (i.e., \citealt{urquhart_radio_south, urquhart_radio_north}) or their surface brightness, which decreases as they expand, has fallen below the sensitivity of these observations (see discussion in Paper\,II for more details).} It is therefore useful to further distinguish between the compact and extended \hii\ regions depending on their association or absence of radio emission; in this way we classify 339 \hii\ regions as compact (associated with radio emission) and 239 \hii\ regions as extended. In Table\,\ref{tbl:atlas_rms_matches} we present a complete list of ATLASGAL-RMS matches and the angular offset between the embedded object and the peak of the dust emission, the RMS source classification, bolometric luminosity and a flag to indicate the RMS sources that are associated with compact radio emission.

In the left panel of Fig.\,\ref{fig:angular_offset}, we plot the surface density of sources as a function of the angular offset of the tracer position from the peak of the host-clump submillimetre emission, for all matches with separations less than 40\arcsec. The distribution is clearly peaked towards smaller offsets, revealing a strong correlation between the positions of the embedded MYSOs and \hii\ regions and the submillimetre emission peak. Approximately, 70\,per\,cent of the matches have angular separations less than 10\arcsec\ and over 85\,per\,cent have separations less than 20\arcsec. Only 34 matches have angular separations between $\sim$40 and 80\,arc\,seconds. The majority of these are classified as radio quiet \hii\ regions (20), which suggests they are more evolved and may be close to breaking out of their natal clump.  Of the remaining fourteen large-separation matches, five are compact \hii\ regions and nine are MYSOs; these appear to be geniune associations and their location on the edges of these clumps may be the result of an interaction between the clump and its environment (e.g., \citealt{thompson2012}). 

In Papers\,I and II we found similar strong positional correlations between the peak of the dust emission and both the methanol masers and radio-continuum traced compact \hii\ regions. The right panel of Fig.\,\ref{fig:angular_offset} shows the cumulative distribution of the angular separations of the RMS sources, together with those of the previous two source samples. It is clear that there is no significant difference in the distribution of separations between the three samples which are all strongly correlated with the position of the peak dust emission. The young embedded massive stars pin-pointed by these tracers are preferentially found towards the centre of their host clumps, where the highest column and volume densities are likely to be found. Furthermore, finding little difference between the different samples, which are expected to cover all of the important embedded stages, we suggest that the relative position of the forming massive star within its host clump does not change significantly over the course of its evolution. 

All three distributions are, however, significantly different to that of the radio-quiet (extended) \hii\ regions.$^1$ This is consistent with our hypothesis that the radio-quiet \hii\ regions are more evolved than the more compact radio-loud \hii\ regions. As the \hii\ region expands it is likely to have a disruptive influence on the surrounding molecular gas and dust, resulting in the peak of the dust emission being pushed away from the centre of the \hii\ region (see \citealt{thompson2006} for detailed discussion of these processes). If we exclude these extended \hii\ regions then we find that $\sim$85\,per\,cent of the embedded objects are located within 10\arcsec\ and that $\sim$95\,per\,cent of sources are located within 20\arcsec\ of the sub-millimetre peak.

\subsection{Methanol-maser associations between $\ell$\,=\,20\degr-60\degr}
\label{sect:methanol_extra}

In Paper\,I we presented a detailed comparison between methanol masers identified by the MMB survey and the properties of their host clumps. This study was restricted to the MMB survey alone as it provides an large sample of sources and unbiased coverage of a large region of the Galaxy and uniform sensitivity, make it ideal for conducting a detailed statistical study. However, while the the MMB catalogue covers the all of the southern Galactic plane ($|b| < 2$) it only covers the innermost 20\degr\ of the northern Galactic plane (i.e., $\ell < 20\degr$) and in order to examine the larger-scale Galactic distribution we need to expand our sample of methanol masers and their host clumps to include the whole of the northern Galactic plane covered by ATLASGAL.

Arecibo Methanol Maser Galactic Plane Survey (AMGPS; \citealt{pandian2007}) is a blind survey covering approximately 20 square degrees of the northern Galactic plane ($\ell=35.2$\degr-53.7\degr\ and $|b|<0.41$\degr). The majority of the detected masers have been followed up with MERLIN to obtain accurate positions (\citealt{pandian2011}). We have used these updated positions to determine their association with ATLASGAL sources and, where a match has been identified, to calculate the angular offset between the peak of the submillimetre emission and the maser. We have matched 59 of the 86 methanol masers reported by \citet{pandian2007}; these are associated with 57 ATLASGAL clumps with three masers found to be associated with a single clump.  

For the regions not covered by the MMB or AMGPS surveys we have used the catalogue of methanol masers compiled by \citet{pestalozzi2005}. We note that this catalogue has been compiled from a mixture of interferometric and single-dish data and, as a consequence, the positional accuracy is not as consistent as the MMB and AMGPS surveys. As methanol masers are thought to be intimately associated with the star-formation process one would expect them to be coincident with an embedded thermal source and, although the extinction will be high at mid-infrared and shorter wavelengths, these objects should be detected in the far infrared. We have therefore cross-matched these with \textit{Herschel} 70 and 160\,\mum\ HiGAL images (\citealt{molinari2010}) and removed any matches where the methanol maser is not coincident with a far-infrared point source. This insures that only methanol masers with accurate positions are included in the analysis that follows. In total 53 methanol masers from the \citet{pestalozzi2005} catalogue are matched with 53 ATLASGAL clumps. 

\subsubsection{Combined methanol maser sample}

In total, we have matched an additional 112 methanol masers from these two catalogues that are associated with 110 ATLASGAL clumps in the $\ell=20$\degr-60\degr\ region not covered by the MMB survey. This brings the total number of methanol-maser-associated clumps to 740 and represents the largest sample of methanol masers and their host clumps. However, this sample includes 146 methanol masers located within 10\degr\ of the Galactic Centre (i.e., $| \ell |$ $<$10\degr), which is not covered by the RMS or CORNISH surveys and so these are not included in the analyse presented in this paper. This reduces the number of methanol maser associations considered here to 594.

\subsection{Survey sensitivities and coverage}
\label{sect:sensitivity}

We have previously mentioned that the ATLASGAL is complete to all compact clumps with masses in excess of 1000\,\msun\ to a heliocentric distance of 20\,kpc across the inner Galactic disc. This is shown by the orange shaded region in Fig.\,1. The only significant feature beyond 20\,kpc in this longitude range is the farside segment of the Scutum-Centaurus arm, however, this is known to be rising out of the plane (\citealt{dame2011}) in this direction and is unlikely to be covered in the ATLASGAL survey. 

\citet{lumsden2013} estimated that the RMS survey is $\sim$90\,per\,cent complete for all MYSOs and compact \hii\ regions with luminosities over a few 10$^4$\,\lsun\ to also to a distance of 20\,kpc; this is roughly equivalent to a zero age main sequence (ZAMS) star of spectral type of B0 or earlier. The CORNISH survey has a 5$\sigma$ sensitivity of 2\,mJy\,beam$^{-1}$, which is sufficient to detected an unresolved, optically thin, \hii\ region powered by a ZAMS star with a spectral type of B0.5 or earlier to the same distances as the RMS and ATLASGAL surveys (\citealt{hoare2012,kurtz1994}). Finally, the MMB survey has a flux sensitivity of 0.17\,Jy\,beam$^{-1}$, which corresponds to a luminosity of $\sim$1000\,Jy\,kpc$^2$ at 20\,kpc (see Fig.\,12 of Paper\,I). The AMGPS has a similar flux sensitivity as the MMB and is therefore likely to have a similar luminosity sensitivity.

With the exception of some of the outer Galactic, warped disc, all the constituent surveys are therefore effectively complete, above the relevant limits and within the observed longitude ranges, out to the edge of the Galaxy.  However, to avoid introducing a distance bias when comparing the properties of the clumps associated with the different tracers we will use a distance-limited sample that we define as having heliocentric distances between 3 and 5\,kpc; this range has been chosen as it includes the highest source densities of all three tracers and covers a large range of clump properties (masses, column densities and sizes) and therefore facilitates statistical robust comparisons between different subsamples (see Sect.\,4.1 for more details).

Although the surveys have comparable completeness the overlap in Galactic longitude and latitude between them is not homogeneous (see Fig.\,1 for longitude range covered by each survey) and so care needs to be taken to avoid introducing a selection bias. We have a uniform coverage of our sample of MYSOs and \hii\ region over the whole ATLASGAL region where $|\ell| > 10$\degr, however, we only have unbiased coverage for the methanol masers in the regions covered by the MMB and AMGPS (i.e., $280\degr < \ell < 20\degr$ and $35\degr < \ell < 54\degr$). It is this more limited range in longitude that will be used in the analysis to come when comparing the relative number of each type of association.

\section{Massive star forming clumps}
\label{sect:correlation_mmb_cornish}

Combining the associations discussed in the previous section with those presented in Papers\,I and II, we have identified $\sim$1700 embedded sources that are associated with 1256 ATLASGAL clumps. Of these clumps, 1131 are located within the $280\degr<\ell < 350\degr$  and $10\degr<\ell < 60\degr$ region where the overlap between the different surveys is reasonably uniform, which corresponds to approximately 15\,per\,cent of the 7255 of ATLASGAL clumps within this region {see Fig.\,\ref{fig:gal_distribution_nodata} for survey coverage and source distribution}. The embedded sources located in the overlapping survey region break down as follows: 594 methanol masers, 361 MYSOs, 611 \hii\ regions and 25 YSO/\hii\ regions, where the distinction between the MYSOs and \hii\ regions comes from the RMS and CORNISH classifications. This represents an increase in the number of methanol masers discussed in Paper\,I by $\sim$112 and a threefold increase in the number of \hii\ regions discussed by Paper\,II  as well as expanding the study to include the MYSO stage. This sample of massive star forming (MSF) clumps is the largest of its kind and incorporates a large fraction of the whole Galactic population of these objects (see Sect.\,2.1 for discussion of survey coverage) and also includes many of the sources used in previous studies of massive star formation presented in the literature (e.g., radio continuum studies of \uchii\ regions --- \citet{kurtz1994} and \citet{wood1989}; mid-infrared bright samples of high-mass protostellar objects (HMPO) --- \citet{sridharan2002}; IRAS colour selected samples --- \citet{molinari1996}  and \citet{bronfman1996}; and methanol masers --- \citet{walsh1998}). ATLASGAL is therefore able to put all of these studies into a larger global framework. 

\begin{figure*}
\begin{center}

\includegraphics[width=0.33\textwidth, trim= 15 0 20 0,clip=true]{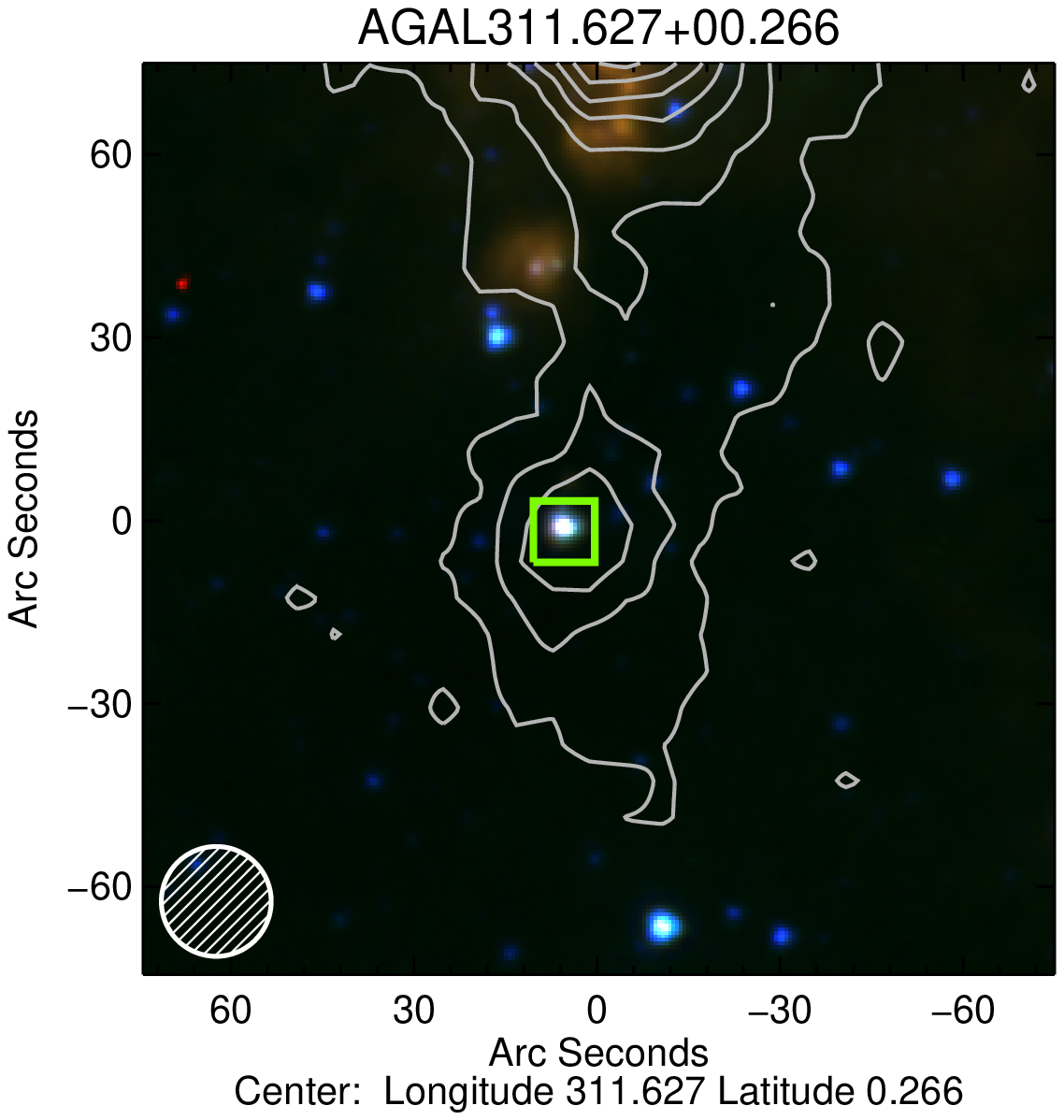}
\includegraphics[width=0.33\textwidth, trim= 15 0 20 0,clip=true]{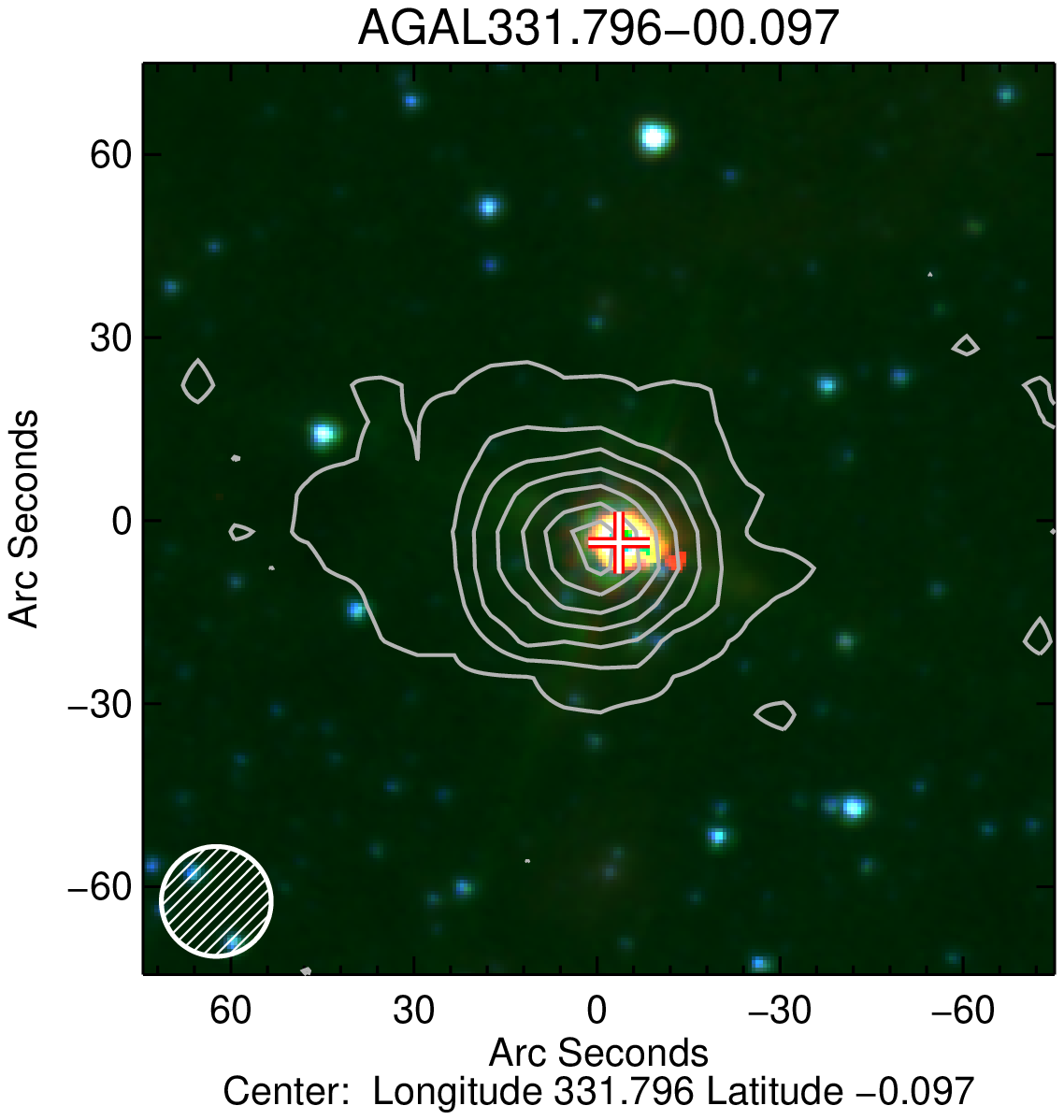}
\includegraphics[width=0.33\textwidth, trim= 15 0 20 0,clip=true]{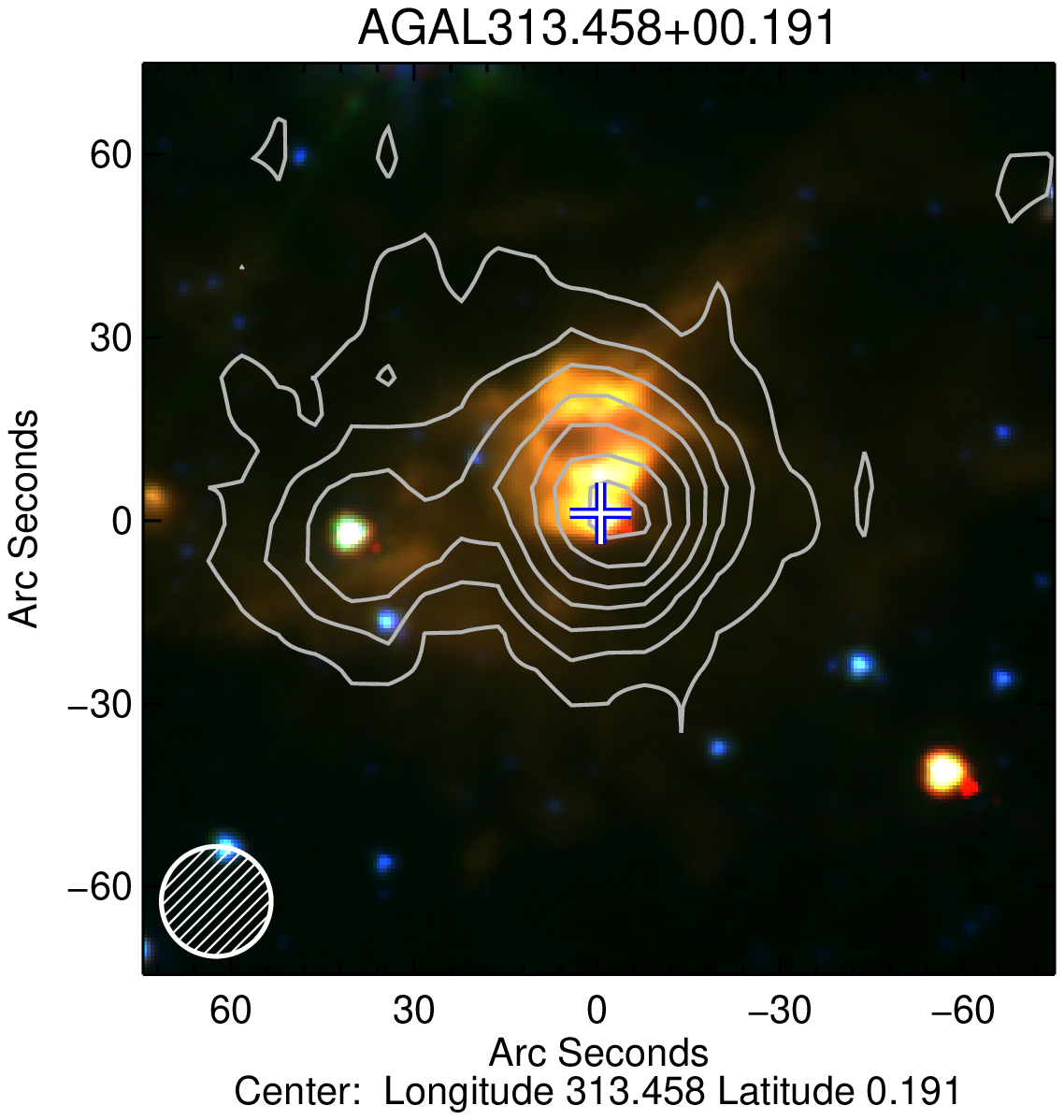}
\includegraphics[width=0.33\textwidth, trim= 15 0 20 0,clip=true]{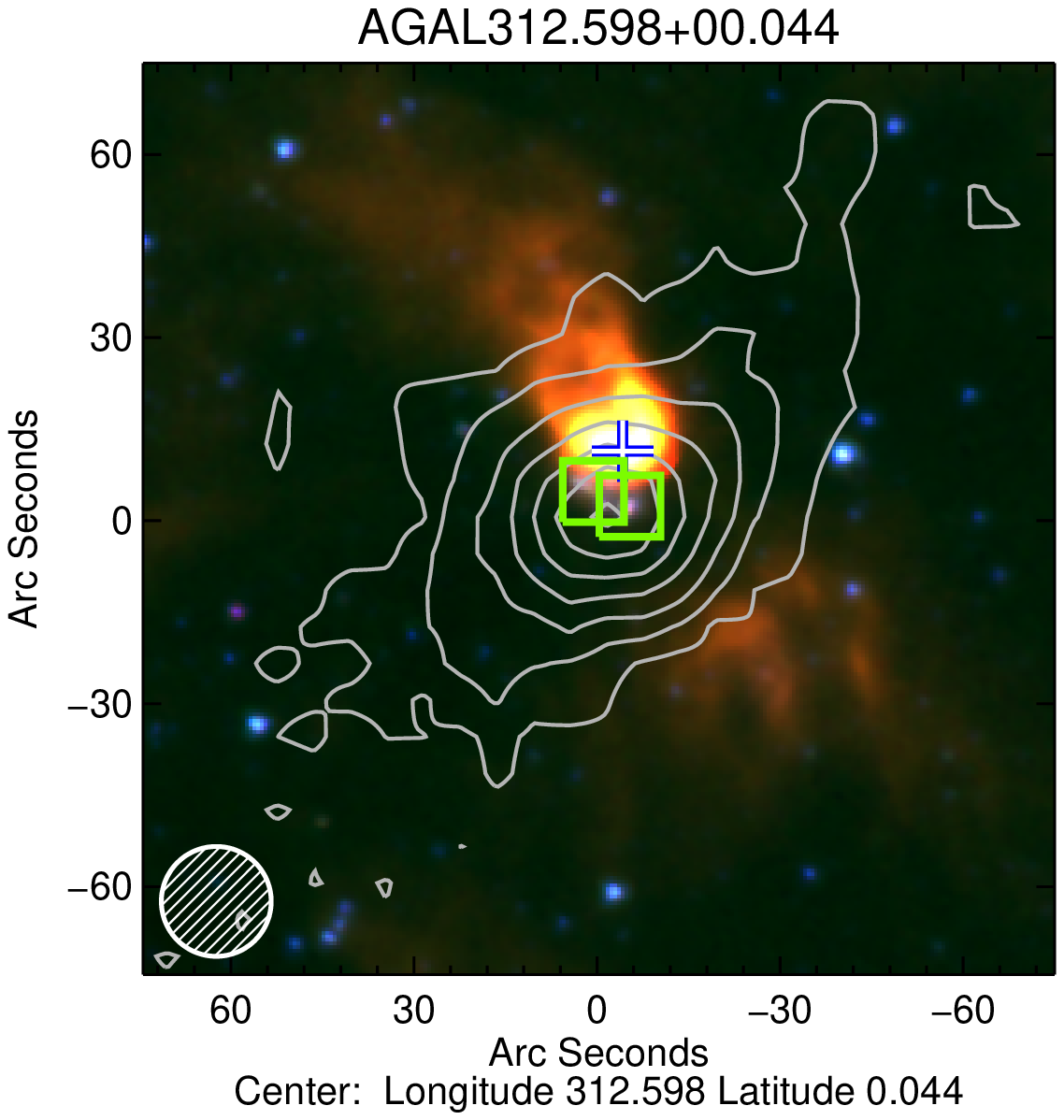}
\includegraphics[width=0.33\textwidth, trim= 15 0 20 0,clip=true]{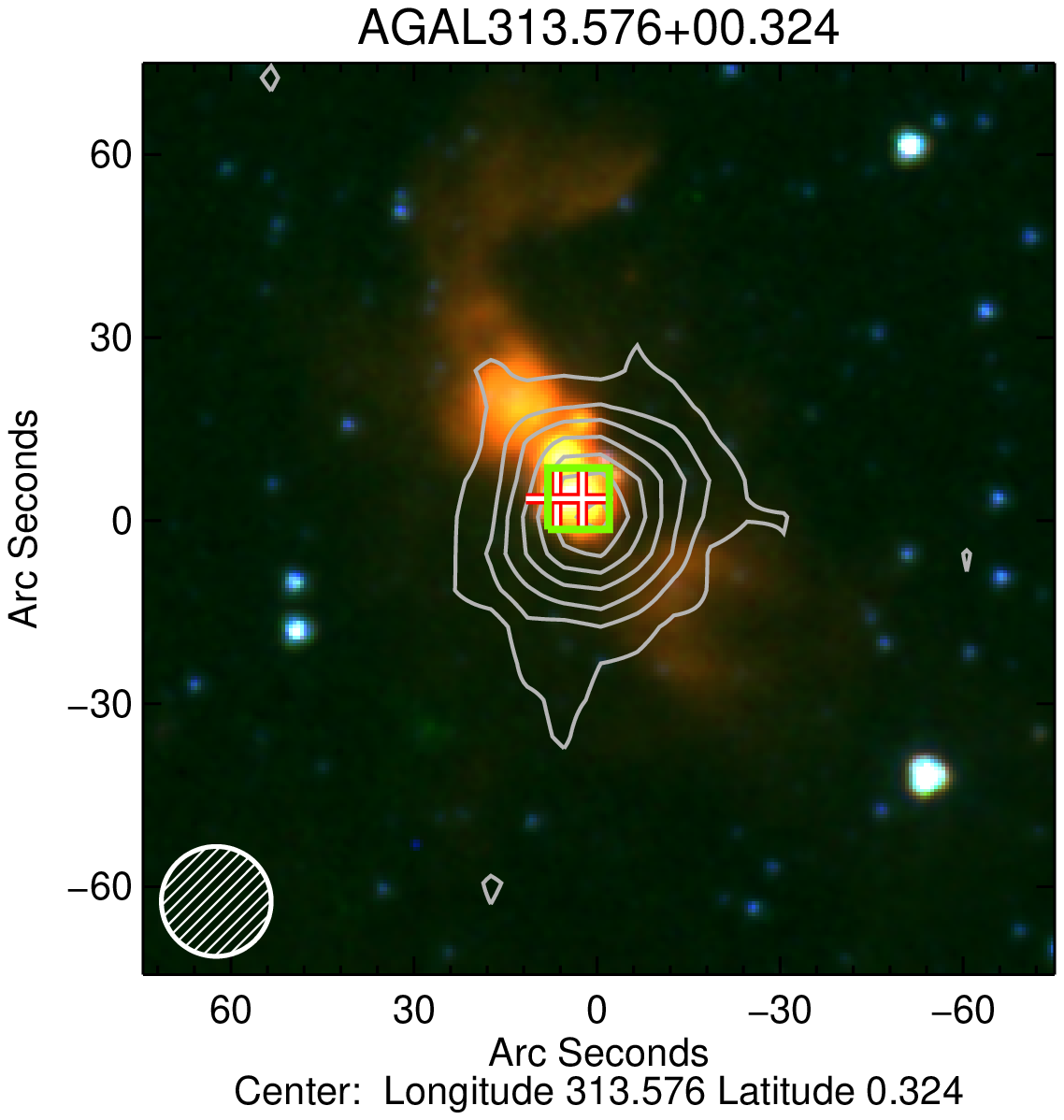}
\includegraphics[width=0.33\textwidth, trim= 15 0 20 0,clip=true]{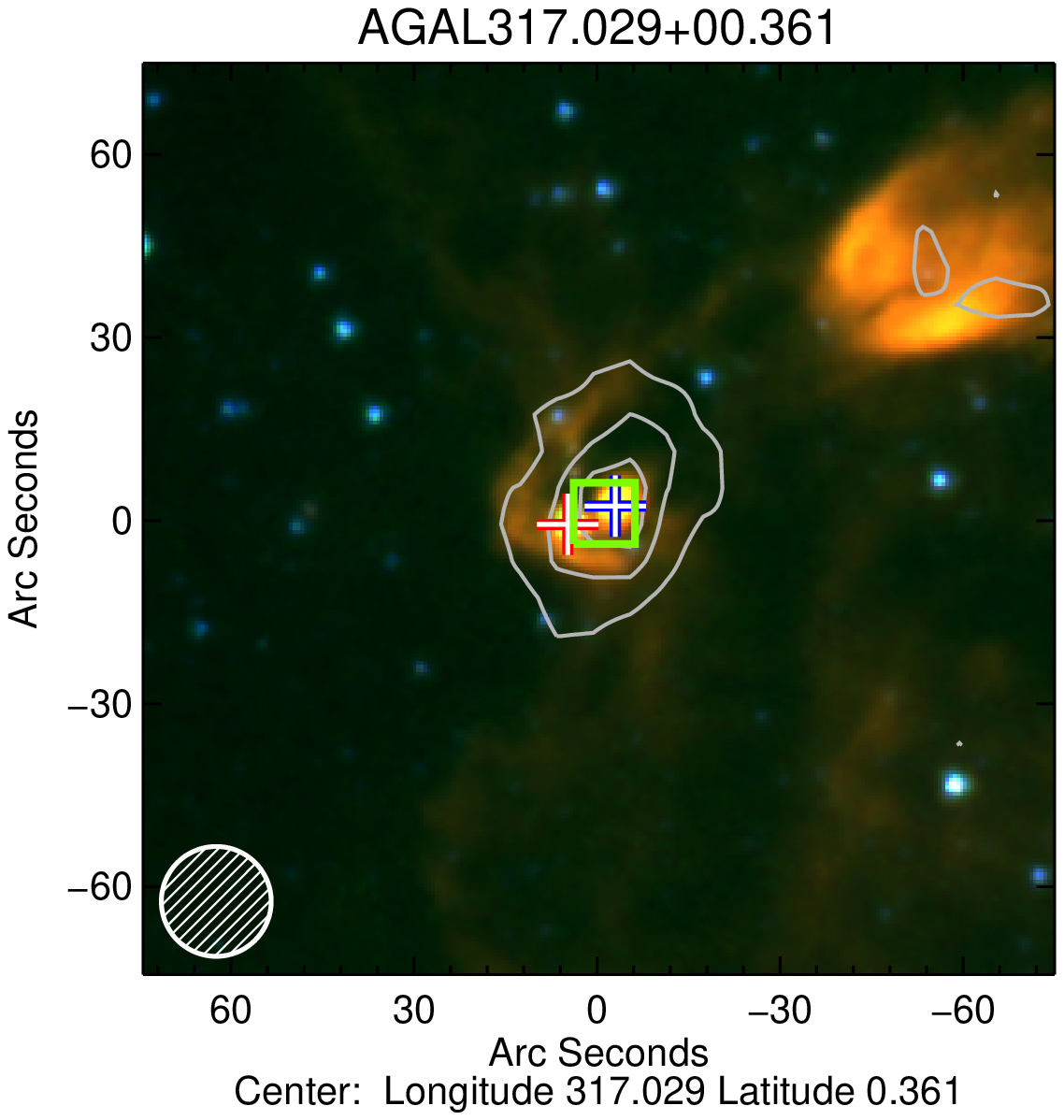}

\caption{\label{fig:irac_images}   Examples of the 870-$\umu$m dust emission and local mid-infrared environments of the matched sources. The upper panels show clumps associated with a single massive star-formation tracer, while the lower panels show examples of clumps where two or more tracers are found to be embedded (multi-phase associations). The positions of the methanol masers, MYSOs and \hii\ regions are indicated by green squares, and red and blue crosses, respectively. The mid-infrared images are composed of the GLIMPSE 4.5, 5.8 and 8.0\,\mum\ IRAC bands (coloured blue, green and red, respectively). The contours overlaid on all panels show the 870\,$\umu$m dust emission while the hatched circle in the lower-left corner of each image shows the size of the ATLASGAL beam. The contour levels start at 2$\sigma$ and increase in steps determined by a dynamic power law.} 

\end{center}
\end{figure*}

In Fig.\,\ref{fig:irac_images} we present three-colour, mid-infrared images that show  examples of the massive star-forming environments and position of the embedded object with respect to the dust distribution of the clump. Approximately a third of the matched ATLASGAL clumps are associated with two or more of the tracers and therefore are likely to host objects with a range of evolutionary stages; this is discussed in more detail in Sect.\,\ref{sect:multiplicity}. In the upper panels we show examples of clumps associated with a single observational tracer, while in the lower panels we show examples of clumps hosting two or more of the different tracers. We will refer to these as multi-phase clumps. Many of the most intense star-forming regions in the Galaxy fall into the multi-phase subsample (e.g., W43, W49 and G305). For the analysis that follows we will define four distinct subsamples: methanol maser, MYSO, \hii\ region and multi-phase associated clumps. The properties for the individual clumps are given in Table\,\ref{tbl:derived_clump_para} while the statistical properties of the four subsamples and combined sample of MSF clumps are given in Tables\,\ref{tbl:derived_para} and \ref{tbl:derived_para_all}, respectively.\footnote{The aspect ratio and compactness parameter are discussed in the Sect.\,3.2 while derivation and analysis of the rest of the parameters (i.e., distance, clump mass etc) is given in Sect.\,4.}  

\setlength{\tabcolsep}{1pt}

\begin{table*}

\begin{center}\caption{Derived clump parameters. The bolometric luminosity given is the sum of all RMS sources embedded in each clump.}
\label{tbl:derived_clump_para}
\begin{minipage}{\linewidth}
\small
\begin{tabular}{lc............}
\hline \hline
  \multicolumn{1}{c}{ATLASGAL}&  \multicolumn{1}{c}{Association}&\multicolumn{1}{c}{Aspect}  &\multicolumn{1}{c}{Peak flux}  &\multicolumn{1}{c}{Int. flux}  &\multicolumn{1}{c}{Compactness}  &\multicolumn{1}{c}{V$_{\rm{LSR}}$}   &	\multicolumn{1}{c}{Distance} &\multicolumn{1}{c}{R$_{\rm{GC}}$}&\multicolumn{1}{c}{Radius}&\multicolumn{1}{c}{Log[$N$(H$_2$)]} &	\multicolumn{1}{c}{Log[$M_{\rm{clump}}$]} &\multicolumn{1}{c}{Log[$L_{\rm{bol}}$]} &	\multicolumn{1}{c}{Log[$M_{\rm{vir}}$]} \\
  
    \multicolumn{1}{c}{name }&  \multicolumn{1}{c}{type } & \multicolumn{1}{c}{ratio}  &\multicolumn{1}{c}{(Jy beam$^{-1}$)}  &\multicolumn{1}{c}{(Jy)}  &\multicolumn{1}{c}{parameter}  &	\multicolumn{1}{c}{(km\,s$^{-1}$)}&	\multicolumn{1}{c}{(kpc)} &\multicolumn{1}{c}{(kpc)}&\multicolumn{1}{c}{(pc)}&\multicolumn{1}{c}{(cm$^{-2}$)} &	\multicolumn{1}{c}{(\msun)} &\multicolumn{1}{c}{(\lsun)} &	\multicolumn{1}{c}{(\msun)} \\
\hline
AGAL028.244+00.012	&	HII	&	2.1	&	1.04	&	12.11	&	11.6	&	106.5	&	8.5	&	4.2	&	5.06	&	22.42	&	3.7	&	4.42	&	3.07	\\
AGAL028.288$-$00.362	&	HII	&	1.4	&	2.74	&	25.73	&	9.4	&	48.8	&	11.6	&	5.8	&	7.72	&	22.84	&	4.3	&	5.53	&	3.67	\\
AGAL028.301$-$00.382	&	MM/YSO	&	1.8	&	1.56	&	17.05	&	10.9	&	85.5	&	9.8	&	4.6	&	5.68	&	22.59	&	4.0	&	4.03	&	3.64	\\
AGAL028.336+00.117	&	YSO	&	1.6	&	1.92	&	9.99	&	5.2	&	81.0	&	5.0	&	4.8	&	2.14	&	22.68	&	3.1	&	3.94	&	2.91	\\
AGAL028.608+00.019	&	HII	&	1.5	&	2.93	&	20.21	&	6.9	&	101.2	&	8.7	&	4.3	&	5.29	&	22.87	&	3.9	&	4.69	&	3.50	\\
AGAL028.649+00.027	&	HII	&	1.1	&	3.10	&	13.53	&	4.4	&	103.6	&	8.6	&	4.2	&	3.31	&	22.89	&	3.7	&	4.59	&	3.33	\\
AGAL028.687+00.177	&	HII	&	1.1	&	0.71	&	1.85	&	2.6	&	102.9	&	9.8	&	4.7	&	0.91	&	22.25	&	3.0	&	4.35	&	2.58	\\
AGAL028.816+00.366	&	HII	&	1.6	&	2.73	&	8.02	&	2.9	&	86.4	&	9.6	&	4.6	&	3.25	&	22.83	&	3.6	&	4.13	&	3.26	\\
AGAL028.861+00.066	&	MM/YSO	&	1.6	&	3.82	&	23.13	&	6.1	&	102.8	&	8.6	&	4.3	&	5.51	&	22.98	&	4.0	&	5.11	&	3.73	\\
AGAL029.436$-$00.174	&	YSO	&	1.6	&	0.91	&	3.22	&	3.5	&	85.6	&	5.2	&	4.7	&	1.28	&	22.36	&	2.7	&	3.77	&	2.52	\\
\hline\\
\end{tabular}\\
Notes: Only a small portion of the data is provided here, the full table is available in electronic form at the CDS via anonymous ftp to cdsarc.u-strasbg.fr (130.79.125.5) or via http://cdsweb.u-strasbg.fr/cgi-bin/qcat?J/MNRAS/.
\end{minipage}

\end{center}
\end{table*}
\setlength{\tabcolsep}{6pt}

\begin{table*}

\begin{center}\caption{Summary of derived parameters for the methanol maser, MYSO and \hii\ regions associated subsamples. The ``$<>$'' indicate that the given value is the mean of the logarithmic values.}
\label{tbl:derived_para}
\begin{minipage}{\linewidth}
\small
\begin{tabular}{lc......}
\hline \hline
  \multicolumn{1}{l}{Parameter}&  \multicolumn{1}{c}{Number}&	\multicolumn{1}{c}{Mean}  &	\multicolumn{1}{c}{Standard error} &\multicolumn{1}{c}{Standard deviation} &	\multicolumn{1}{c}{Median} & \multicolumn{1}{c}{Min}& \multicolumn{1}{c}{Max}\\
\hline

\multicolumn{8}{l}{Heliocentric distance (kpc)}\\
\hline

Methanol masers  &         269&7.61&0.25 & 4.16 & 7.70 & 0.02 & 19.62\\
MYSOs  &   210&5.57&0.23 & 3.29 & 4.51 & 1.12 & 15.28\\
\hii\ regions &          373&8.21&0.21 & 4.06 & 8.47 & 1.12 & 18.47\\
Multi-phase &          230&7.20&0.25 & 3.82 & 5.91 & 1.65 & 17.08\\

\hline
\multicolumn{8}{l}{Effective radius (pc)}\\
\hline

Methanol masers &  223&1.17&0.06 & 0.97 & 0.91 & 0.00 & 5.59\\
MYSOs &     164&0.79&0.04 & 0.54 & 0.64 & 0.15 & 2.83\\
\hii\ regions & 320&1.25&0.04 & 0.78 & 1.07 & 0.17 & 5.01\\   
Multi-phase  & 210&1.40&0.06 & 0.93 & 1.19 & 0.20 & 5.23\\

\hline
\multicolumn{8}{l}{Physical Offset (pc)}\\
\hline

Methanol masers &          546&0.27&0.02 & 0.41 & 0.14 & 0.00 & 3.55\\
MYSOs &          359&0.23&0.02 & 0.29 & 0.13 & 0.01 & 2.53\\
Compact \hii\ regions &          339&0.28&0.02 & 0.37 & 0.19 & 0.00 & 3.36\\
Extended \hii\ regions &          236&0.55&0.04 & 0.64 & 0.33 & 0.03 & 4.14\\

\hline
\multicolumn{8}{l}{NH$_3$ line width (\kms)}\\
\hline

Methanol masers &          105&2.38&0.08 & 0.79 & 2.20 & 1.30 & 5.00  \\
MYSOs  &          		105&1.87&0.06 & 0.64 & 1.80 & 0.60 & 4.80\\
\hii\ regions &          173&2.53&0.08 & 1.06 & 2.40 & 0.70 & 
11.00\\
 Multi-phase &          129&2.54&0.08 & 0.90 & 2.50 & 0.90 & 5.90\\

\hline
\multicolumn{8}{l}{Kinetic temperature (K)}\\
\hline

Methanol masers &           98&23.21&0.61 & 6.03 & 22.19 & 13.85 & 
42.42 \\
MYSOs &          95&20.71&0.46 & 4.53 & 20.07 & 13.85 & 
33.64 \\

\hii\ region &          151&23.70&0.40 & 4.93 & 22.51 & 
15.65 & 41.15\\
Multi-phase &          123&24.42&0.46 & 5.12 & 23.46 & 
14.52 & 42.08\\

\hline
\multicolumn{8}{l}{$<$Log[Clump mass (\msun)]$>$}\\
\hline

Methanol masers & 264&3.24&0.04 & 0.65 & 3.28 & 1.03 & 
4.99\\
MYSOs & 210&2.91&0.04 & 0.51 & 2.91 & 1.52 & 
4.21\\
\hii\ regions &  373&3.40&0.03 & 0.52 & 3.45 & 1.72 & 
4.77\\
Multi-phase  &230&3.52&0.04 & 0.56 & 3.50 & 1.93
 & 5.34\\ 
      
\hline
\multicolumn{8}{l}{$<$Log[H$_2$ Column density (cm$^{-2}$)]$>$ }\\
\hline

Methanol masers & 314&22.61&0.02 & 0.38 & 22.54 & 21.94
 & 23.81\\
YSOs &          210&22.56&0.02 & 0.33 & 22.51 & 22.04
 & 23.68\\
\hii\ regions & 375&22.66&0.02 & 0.39 & 22.62 & 21.92
 & 24.14\\
Multi-phase &  231&22.90&0.03 & 0.46 & 22.84
 & 21.96 & 24.30\\

\hline
\multicolumn{8}{l}{Virial ratio}\\
\hline

Methanol masers  &  88&0.72&0.11 & 1.02 & 0.53 & 0.04 & 9.29\\
YSOs &           91&0.77&0.05 & 0.47 & 0.67 & 0.18 & 2.49\\
\hii\ regions&      163&0.65&0.05 & 0.61 & 0.48 & 0.11 & 4.00\\
Multi-phase   &       124&0.58&0.04 & 0.40 & 0.51 & 0.06 & 3.02\\
              
\hline
\multicolumn{8}{l}{$<$Log[Bolometric luminosity (\lsun)]$>$}\\
\hline

Methanol masers &  199&3.70&0.07 & 0.96 & 3.84 & -2.40 & 5.28\\
YSOs 		&	361&3.91&0.03 & 0.53 & 3.87 & 3.01 & 5.66\\
\hii\ regions &	578&4.39&0.03 & 0.61 & 4.40 & 3.02 & 6.38\\

\hline
\multicolumn{8}{l}{\lbol-\mclump\ ratio (\lsun/\msun)}\\
\hline

 Methanol masers  			&	199&6.92&0.72 & 10.10 & 4.23 & 0.12 & 109.61\\
 YSOs &  						210&15.74&1.35 & 19.60 & 9.37 & 0.63 & 143.01\\
 \hii\ regions &   				343&16.45&1.07 & 19.80 & 10.99 & 0.71 & 184.54\\
Multi-phase   & 219&12.72&1.58 & 23.36 & 10.81 & 0.56 & 106.82\\
\hline

\end{tabular}\\
\end{minipage}

\end{center}
\end{table*}

\setlength{\tabcolsep}{6pt}

\begin{table*}

\begin{center}\caption{Summary of derived parameters for all massive star forming clumps.}
\label{tbl:derived_para_all}
\begin{minipage}{\linewidth}
\small
\begin{tabular}{lc......}
\hline \hline
  \multicolumn{1}{l}{Parameter}&  \multicolumn{1}{c}{Number}&	\multicolumn{1}{c}{Mean}  &	\multicolumn{1}{c}{Standard error} &\multicolumn{1}{c}{Standard deviation} &	\multicolumn{1}{c}{Median} & \multicolumn{1}{c}{Min}& \multicolumn{1}{c}{Max}\\
\hline
Angular offset ($^{\prime\prime}$) &         1695 &8.90&0.26 & 10.71 & 5.40 & 0.16 & 
82.58\\
Aspect Ratio &         1130&1.50&0.01 & 0.48 & 1.38 & 1.00 & 8.97\\
$Y$-factor &         1130&5.67&0.12 & 3.92 & 4.59 & 1.14 & 46.05\\
Distance (kpc) &         1082&7.33&0.12 & 4.00 & 6.02 & 0.02 & 19.62\\
Effective radius (pc) &          917&1.18&0.03 & 0.85 & 0.95 & 0.00 & 5.59\\
Physical Offset (pc) &          1480&0.31&0.01 & 0.44 & 0.18 & 0.00 & 4.14\\
Line width (\kms) &          512&2.37&0.04 & 0.93 & 2.20 & 0.60 & 11.00\\
Kinetic Temperature (K) &          467&23.18&0.25 & 5.30 & 22.26 & 13.85 & 42.42\\
$<$Log[Clump Mass (\msun)]$>$ &          1077&3.29&0.02 & 0.60 & 3.33 & 1.03 & 5.34\\
$<$Log[H$_2$ column density (cm$^{-2}$)]$>$ &         1130&22.67&0.01 & 0.41 & 22.61 & 
21.92 & 24.30\\
Viral Ratio &         466&0.67&0.03 & 0.64 & 0.53 & 0.04 & 9.29\\
$<$Log[Bol. lum (\lsun)]$>$ &         1163&4.11&0.02 & 0.72 & 4.11 & -2.40 & 6.38\\
\lbol-\mclump\ ratio (\lsun/\msun) &    971&14.11&0.57 & 17.68 & 8.91 & 0.12 & 184.54\\
\hline\\
\end{tabular}\\

\end{minipage}

\end{center}
\end{table*}

\subsection{Clump structure and morphology}
\label{sect:size}

The clump aspect ratios (i.e., $\theta_{\rm{maj}}/\theta_{\rm{min}}$) and the ratio of the integrated to peak flux densities (compactness parameter or $Y$-factor) are two observationally derived clump parameters that can be used to obtain some insight into the structure of this sample of massive star forming clumps.\footnote{The $Y$-factor is defined as the ratio of integrated to peak \submm\ fluxes and provides an estimate of how centrally condensed the emission is towards the centre of the clump.} Moreover, these are relatively distance-independent parameters, allowing us to compare the properties of the massive star-forming clumps with the quiescent ATLASGAL clump population.

\begin{figure*}
\begin{center}
\includegraphics[width=0.48\textwidth, trim= 0 0 0 0]{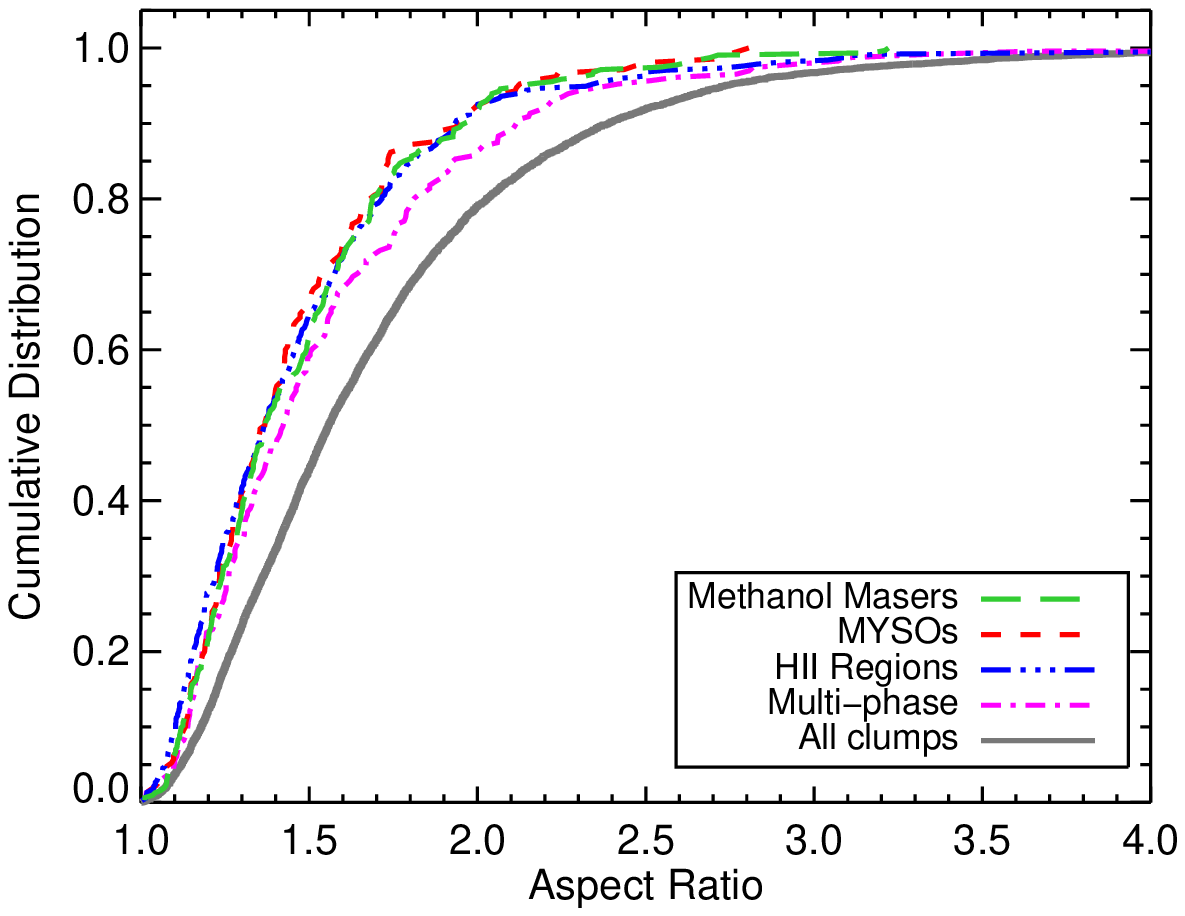}
\includegraphics[width=0.48\textwidth, trim= 0 0 0 0]{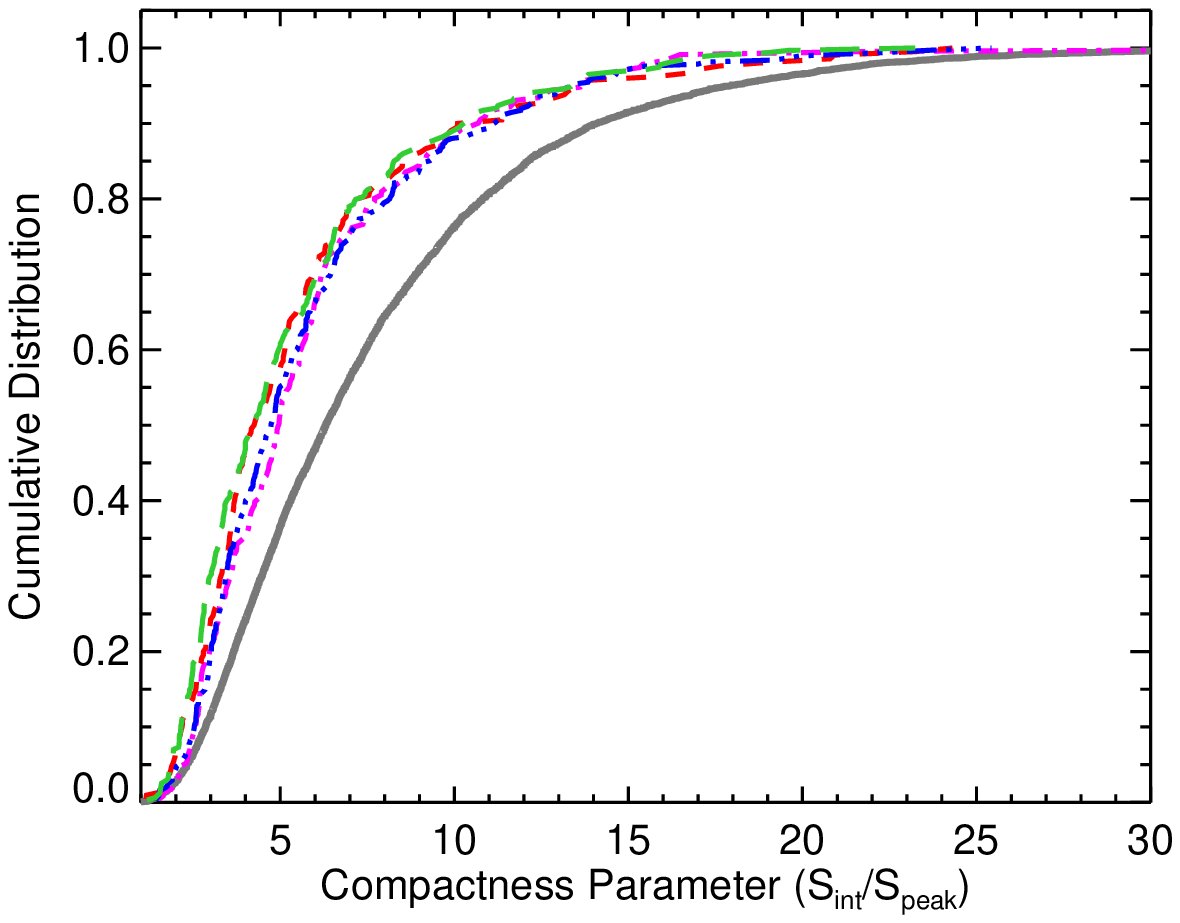}

\caption{\label{fig:clump_structure} The cumulative distributions of the aspect ratio and compactness parameter ($Y$-factor) are presented in the left and right panels, respectively. In these plots we show the distributions of the four subsamples of MSF clumps and the unassociated clumps separately (see legend left panel for colours and line styles for both plots).} 

\end{center}
\end{figure*}

Fig.\,\ref{fig:clump_structure} shows the cumulative distributions of the aspect ratios and compactness parameter ($Y$-factor) for the four subsamples and the unassociated clumps. The distributions of both parameters in the associated clumps are noticeably different to those of the general population of ATLASGAL clumps, with the former being significantly more centrally condensed and spherical (cf. \citealt{csengeri2014}). The mean and median values for the associated clumps are $1.50\pm0.01$ and 1.38 for the aspect ratio and $5.67\pm0.12$ and 4.59 for the $Y$-factor, compared to $1.70\pm0.01$ and 1.56 and $7.68\pm0.07$ and 6.31 obtained for the quiescent clumps in the ATLASGAL CSC, respectively.  Using a \KS\ (KS) test to compare these two samples, we find that the null hypothesis that these are drawn from the same parent population can be rejected for the aspect ratio and $Y$-factor with greater than 3$\sigma$ confidence ($r \ll 0.001$ for both parameters, where $r$ is the probability that the two samples are similar).  

A KS test confirms that there is no significant difference between any of the methanol maser, MYSO or \hii-region associated subsamples. The only difference between the four subsamples is found for the aspect ratio of the multi-phase clumps where a trend to slightly larger values is observed.  The star formation taking place within these clumps is more complicated and is likely to have had an affect on the clump structure. 
 
An important caveat is that the location of the peak emission and its intensity are likely to be affected by the presence of the embedded source itself as it warms its local environment {making heated clumps more strongly peaked}. Comparisons of the temperatures of star forming and quiescent clumps find average temperatures of $\sim$20 and 15\,K, respectively  (e.g., \citealt{giannetti2013,sanchez-monge2013}) and that the clump temperature is correlated with the luminosity of the embedded object with warmer clumps being associated with the more luminous and evolved sources  (\citealt{urquhart2011b}). 

The lack of any significant differences between the three subsamples suggests that the structure of the clump is not sensitive to the evolutionary stage of the embedded object. Furthermore, the quiescent clumps are themselves centrally condensed with aspect ratios $\sim$2, just not to the same extent as the MSF clumps. The presence of an embedded thermal source might lower the aspect ratio and $Y$-factor, but does not alter the fact that the massive stars are forming towards the centre of their host clumps. 

The similarities between the compactness and shape of the MSF clumps lead us to conclude that the structure of the clump \emph{envelope} does not evolve significantly during the early embedded stages in the massive star-formation process. This is consistent with the results of previous studies (e.g., HMPO; \citealt{williams2004} and \uchii\ regions; \citealt{thompson2006}), but for a much larger sample which strengthens their findings.  This result suggests that, in order for the quiescent clumps to form massive stars, they need either to be initially, or to become much more centrally condensed and spherical in structure. However, once the star formation begins, it proceeds more rapidly than the clump can adjust and therefore the morphological clump properties change little as it proceeds. This suggests a fast star-formation process as explored by \citet{csengeri2014}.

\subsection{Multiplicity}
\label{sect:multiplicity}

As mentioned above, we have found $\sim$300 massive star-forming clumps that are associated with two or more of the three tracers ($\sim$30\,per\,cent of the MSF clumps). As discussed in the next paragraph in approximately a third of cases the tracers are in fact associated with the same embedded sources. However, in the remaining two-thirds this multiplicity of tracers present in a clump is likely to indicate that multiple evolutionary stages may be present. This is not altogether surprising, given that most of the observational evidence supports the hypothesis that all massive stars form in clusters and there is no reason to expect all of the star formation taking place in a clump to be coeval. It is still an open question whether the low-mass stars form before, after or at the same time as their high-mass counterparts in a cluster. However, this may have serious implications when trying to attribute observed properties to a particular evolutionary stage. This problem may be particularly acute in cases where a particular sample has been selected using a single observational tracer (e.g., radio continuum or the presence of a methanol maser) as there may be multiple evolutionary stages present in the same clump that remain hidden. 

\begin{figure}
\begin{center}

\includegraphics[width=0.45\textwidth, trim= 0 100 0 100, clip=true, angle=270]{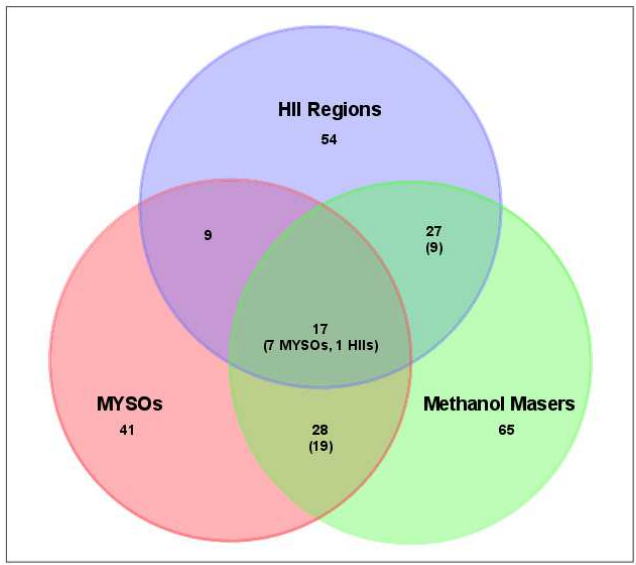}

\caption{\label{fig:ven_diagram} Venn diagram showing the distribution of the methanol masers, MYSOs and \hii\ regions identified with their associated ATLASGAL clumps with distances between 3 and 5\,kpc (i.e., distance-limited sample). For the multi-phase clumps that are associated with a methanol maser an additional number is given in parenthesis; this gives the number of clumps where the embedded MYSO or \hii\ region is found to be positionally coincident with the methanol maser (angular separation $<2\arcsec$) and likely to be tracing the same object.} 

\end{center}
\end{figure}

To investigate the issue of multiplicity further we present a Venn diagram in Fig.\,\ref{fig:ven_diagram} showing how the methanol masers, MYSOs, \hii\ regions are distributed amongst the associated ATLASGAL clumps. We use a distance-limited sample of all massive star-forming clumps located in the MMB and AMGPS survey regions (see Sect.\,\ref{sect:sensitivity}). This sample consists of 241 clumps and avoiding any potential bias due to the different completeness as a function of distance of the tracers used (see Sect.\,\ref{sect:distances} for more details). Furthermore, we exclude the clumps associated with the 25 RMS sources classified as MYSO/\hii\ as some ambiguity about their nature remains. We will focus the following statistical analysis on this distance-limited sample. Before starting this analysis we note that methanol masers have been found to be positionally coincident sometimes with MYSOs and \hii\ regions (e.g., \citealt{walsh1998, sridharan2002}). We therefore need to determine whether the detection of a methanol maser indicates the presence of a distinct embedded object or is likely to be physically associated with an embedded MYSO or \hii\ region.

\setlength{\tabcolsep}{6pt}

\begin{table*}

\begin{center}\caption{List of RMS-methanol maser associations with angular separations $<$2\arcsec.}
\label{tbl:rms_mmb_matches}
\begin{minipage}{\linewidth}
\small
\begin{tabular}{llll.}
\hline \hline
  \multicolumn{1}{c}{ATLASGAL}&  \multicolumn{1}{c}{RMS}&	\multicolumn{1}{c}{Methanol}  &\multicolumn{1}{c}{RMS}  &\multicolumn{1}{c}{Offset}  \\
  \multicolumn{1}{c}{name}&  \multicolumn{1}{c}{name}&	\multicolumn{1}{c}{name}  &\multicolumn{1}{c}{classification}  &\multicolumn{1}{c}{(\arcsec)}  \\
 
\hline
AGAL010.323$-$00.161	&	G010.3208$-$00.1570B	&	G010.323$-$00.160	&	YSO	&	0.35	\\
AGAL010.472+00.027	&	G010.4718+00.0256	&	G010.472+00.027	&	\hii\ region	&	0.40	\\
AGAL010.957+00.022	&	G010.9592+00.0217	&	G010.958+00.022	&	\hii\ region	&	0.58	\\
AGAL011.034+00.061	&	G011.0340+00.0629	&	G011.034+00.062	&	\hii\ region	&	0.98	\\
AGAL011.990$-$00.271	&	G011.9920$-$00.2731	&	G011.992$-$00.272	&	YSO	&	1.48	\\
AGAL012.024$-$00.031	&	G012.0260$-$00.0317	&	G012.025$-$00.031	&	YSO	&	0.36	\\
AGAL012.198$-$00.034	&	G012.1993$-$00.0342B	&	G012.199$-$00.033	&	YSO	&	0.32	\\
AGAL012.888+00.489	&	G012.8909+00.4938C	&	G012.889+00.489	&	YSO	&	0.79	\\
AGAL015.029$-$00.669	&	G015.0357$-$00.6795	&	G015.034$-$00.677	&	\hii\ region	&	0.99	\\
AGAL017.637+00.154	&	G017.6380+00.1566	&	G017.638+00.157	&	YSO	&	1.16	\\

\hline\\
\end{tabular}\\

Notes: Only a small portion of the data is provided here, the full table is available in electronic form at the CDS via anonymous ftp to cdsarc.u-strasbg.fr (130.79.125.5) or via http://cdsweb.u-strasbg.fr/cgi-bin/qcat?J/MNRAS/.
\
\end{minipage}

\end{center}
\end{table*}
\setlength{\tabcolsep}{6pt}

The methanol masers, MYSOs and \hii\ regions in this sample have astrometry accurate to an arcsec or better and so we are able to distinguish between genuine coincidences, in which more than one of the three tracers is found in the same core, from cases where two tracers are associated with separate cores in the same clump. Using a matching radius of 2\arcsec, we find a total 113 cases in the whole sample of 246 clumps where the methanol maser is considered coincident with the MYSO or \hii\ region present in the same clump. There are 39 of these positional coincidences in the distance limited sample, which corresponds to approximately half of all of the methanol masers associated with the multi-phase clumps (76 in total). As a point of reference, 2\arcsec\ is the typical extent of the spread of groups of methanol-maser spots \citep{caswell2009} and corresponds to a physical size of 0.1\,pc at 10\,kpc. This is also the radius of a typical core that is likely to form either a single star or small group of no more than a few stars. These positional associations are therefore likely to be genuine and may indicate that an evolutionary transition of some sort is taking place. Table\,\ref{tbl:rms_mmb_matches} lists these positional coincidences and the angular offset between them. 

In total, 141 methanol masers are included in the distance-limited sample. Of these, we find 29 that are positionally coincident with an MYSO and 10 are positionally associated with an \hii\ region; these correspond to $\sim$20 and 6\, per\,cent of the associated methanol masers. \citet{urquhart_radio_south} found that the radio continuum emission at 5 and 8\,GHz towards six of these \hii\ regions is unresolved with a mean spectral index of $\sim$1 (values range between 0.3 and 2, with a standard error on the mean of 0.3 and standard deviation of 0.7). The \hii\ regions are therefore optically thick and making these sources good candidate hypercompact (HC) \hii\ regions, which is the stage prior to the formation of the UC region. The radio continuum emission seen towards the other four \hii\ regions is also found to be unresolved (\citealt{purcell2013,walsh1998}), however, these have only been observed at one frequency and their spectral indices cannot be estimated. The association of these compact radio sources with methanol masers may indicate that accretion continues after the formation of the \hii\ regions through ionized infall (\citealt{keto2003,keto2007}) or through the partial shielding of the ionizing radiation by the accretion flow (\citealt{peters2010}). In the full sample we have found a total of 34 sources where the position the methanal is coincident with the \hii\ region and these may provide basis for a more statistical study of HC \hii\ regions.

The proportions of methanol masers associated with MYSOs and \hii\ regions are similar to those reported by \citet{sridharan2002} and \citet{walsh1998}, however, the analysis presented here on a distance limited sample with accurate astrometry is more robust. Methanol masers are therefore present over a large range of the early evolutionary stages, from the beginning of accretion to after the formation of the \hii\ region. However, $\sim$74\,per\,cent of this sample of methanol masers are not within 2\arcsec\ of either an MYSO or \hii\ region and therefore they are likely to be tracing a stage that precedes MYSO's and \hii\ region stages. For completeness we estimate the corresponding number of MYSOs and \hii\ regions that are coincident with a methanol maser to be approximately 28 and 10\,per\,cent, respectively.

Since a methanol maser coincident with a MYSO or \hii\ region is not considered to be tracing a separate stage (i.e., angular offsets $<2$\arcsec), it is reasonable to combine these clumps into the MYSO and \hii\ region subsamples. In the distance-limited sample (i.e., between 3 and 5\,kpc; see Sect.\,4.1.2 for details) there are 65 methanol maser only clumps, 60 MYSO only clumps (41 MYSOs + 19 MYSOs coincident with a methanol maser) and 63 \hii\ region clumps (54 \hii\ regions + 9 \hii\ regions coincident with a methanol maser), 17 clumps that host a MYSO and \hii\ region, 18 clumps that host a methanol maser and \hii\ region, 9 clumps that are associated with a methanol maser and MYSO, and 9 clumps that host all three tracers. The methanol maser, MYSO, \hii\ region and multi-phase subsamples each include $\sim$25\,per\,cent of the clumps, with approximately a third of all tracers being associated with one of the multi-phase clumps. This highlights the need for careful analysis of high-resolution multi-wavelength data in order to properly distinguish between clumps containing multiple evolutionary stages and those genuinely associated with a single, fairly well-defined stage. 

We have now identified three relatively clean subsamples of methanol-maser, MYSO and \hii-region only associated clumps that will be the primary focus of this paper. In the analysis that follows, we will not refer to specific survey matches but to the subsamples they have been used to define.

\section{Derived clump properties}
\label{sect:results}

In this section, we describe the procedures used to derive parameters for all of the associated clumps and present the analysis of these parameters. As well as discussing the statistical properties of the whole sample we will also make comparisons of the clump properties of the methanol maser, MYSO and \hii\ region subsamples. We will restrict our analysis to comparisons primarily of these three subsamples as these provide the best opportunity to identify differences that may be linked to the evolutionary stage of the embedded objects. However, we include multi-phase clumps in all of the plots for completeness and to facilitate comparison with the other subsamples. 

\subsection{Distances}

\subsubsection{RMS-MMB distance comparison}

 \begin{figure}
\begin{center}
\includegraphics[width=0.49\textwidth, trim= 0 0 0 0]{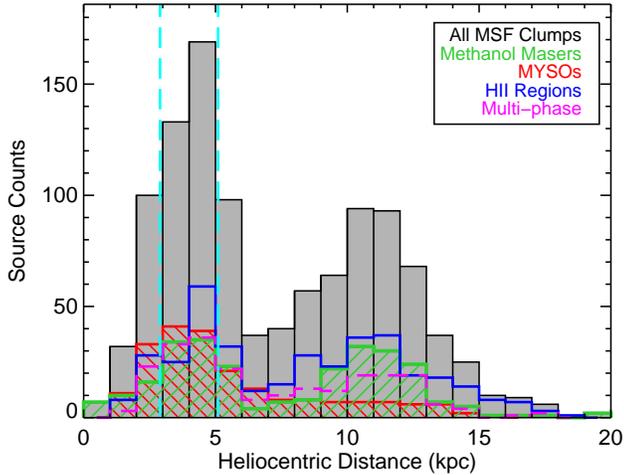}

\caption{\label{fig:atlas_distance_hist} Heliocentric distance distribution for all  MSF-tracer associated ATLASGAL clumps and subsamples. The vertical long-dashed cyan lines indicate the range (3 -- 5\,kpc) from which our distance-limited sample is drawn. The bin size is 1\,kpc.} 

\end{center}
\end{figure}

Distances have already been obtained for the vast majority of the ATLASGAL-RMS matched sources ($\sim$90\,per\,cent; see Urquhart et al. 2014 for details). Although parallax and spectrophotometric distances have been adopted where available, a large number are kinematically derived distances estimated using a source's radial velocity (as measured from NH$_3$ or $^{13}$CO observations) and a model of the Galactic rotation curve. The near/far kinematic distance ambiguities (KDAs) have been resolved in an analysis of archival \hi\ spectra (e.g., \citealt{urquhart2012}).  The MMB survey has conducted a similar analysis using the same \hi\ data and the velocity of the peak maser component as a proxy for the velocity of the natal cloud \citep{green2011b}.  

In total there are 120 ATLASGAL clumps that are associated with both an RMS and MMB source that have an assigned kinematic distance. For 95 of these matches, the kinematic distances are in reasonable agreement. This corresponds to $\sim$80\,per\,cent of this sample, which is similar to the reliability estimated from other studies (e.g., \citealt{busfield2006,anderson2009a}). In Table\,\ref{tbl:rms_mmb_dvlsr} we provide a list of the 25 sources and their associated parameters for which the distance discrepancy is larger than 1.5\,kpc. 

Differences of only a few \kms\ in the velocity of the local standard of rest (\vlsr) can be enough to change the kinematic distance solution from near to far or vice versa, moving the source velocity so that it aligns with an absorption feature in the \hi\ spectrum or breaking the alignment between the source velocity and an absorption feature, leading to an incorrect distance solution. Methanol-maser emission often consists of a number of strong peaks spread over several \kms\ in velocity ($7.2\pm5.3$\,\kms, where the uncertainty is the standard deviation, estimated from the combined MMB catalogue) with the strongest peak changing over time and, although normally close to the systemic velocity of the source, can differ sufficiently to produce errors in the distance solution. Molecular-line observations of thermally excited transitions provide a more reliable measurement of the clump systemic velocity and we have therefore adopted the RMS distances in cases where the MMB and RMS kinematic distances disagree. 

\begin{figure*}
\begin{center}
\includegraphics[width=0.98\textwidth, trim= 0 0 0 0]{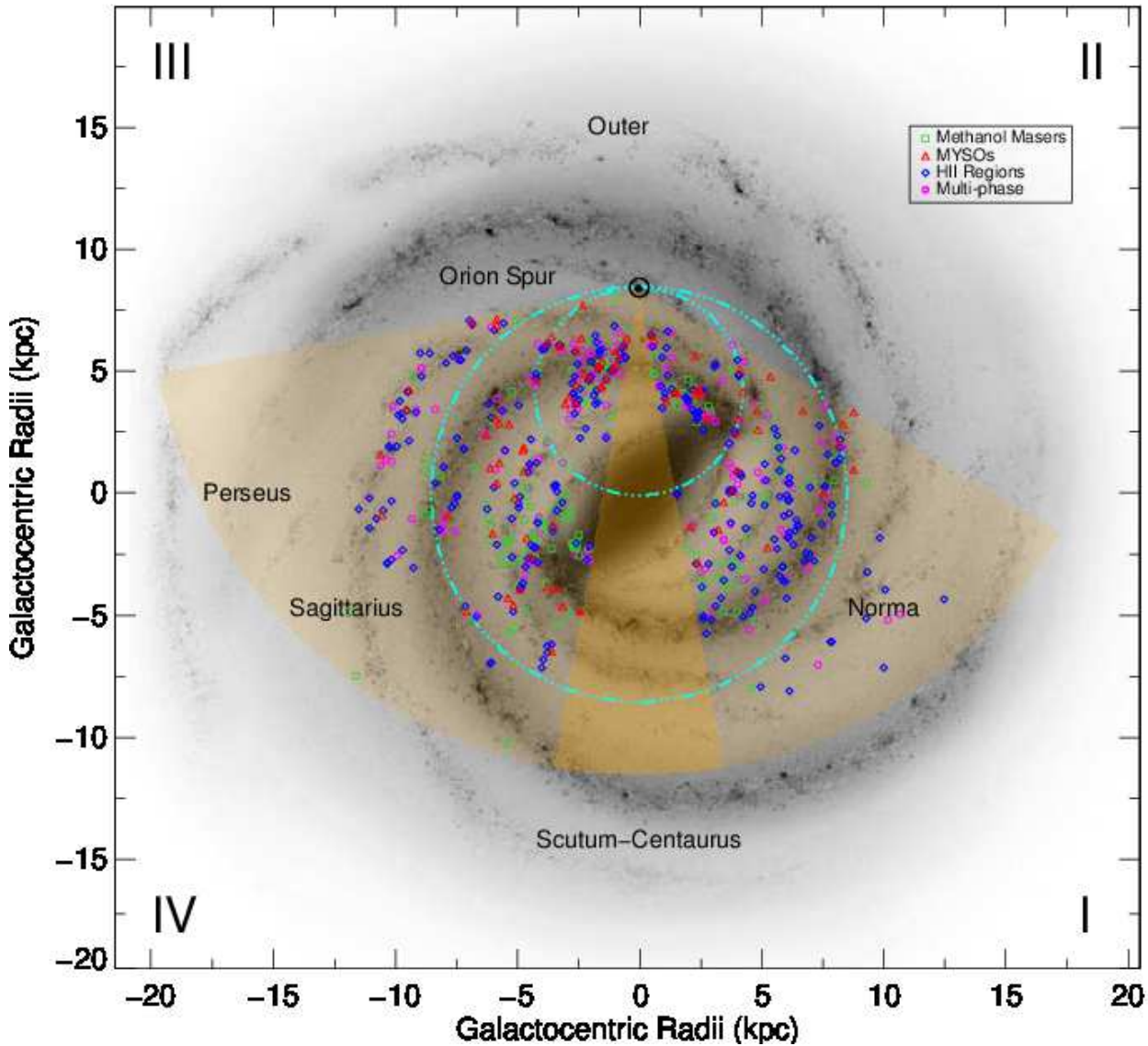}

\caption{\label{fig:gal_distribution} Galactic distribution of all MSF clumps with masses larger than 1000\,\msun. Different symbols and colours are used to indicate the different subsamples (see legend for details). The background image is the same as presented in Fig.\,1. The orange wedges show the ATLASGAL longitude coverage our to a distance of 20\,kpc where is has a clump mass sensitivity of $\sim$1000\,\msun. The smaller of the two cyan dot-dashed circles represent the locus of tangent points, while the larger circle traces the Solar circle.} 

\end{center}
\end{figure*}

\setlength{\tabcolsep}{6pt}

\begin{table*}

\begin{center}\caption{Summary of RMS and MMB source parameters where the kinematic distance ambiguity solutions disagree.}
\label{tbl:rms_mmb_dvlsr}
\begin{minipage}{\linewidth}
\small
\begin{tabular}{ll.....cc}
\hline \hline
  \multicolumn{1}{c}{RMS}&  \multicolumn{1}{c}{MMB}&	\multicolumn{1}{c}{RMS \vlsr}  &\multicolumn{1}{c}{MMB \vlsr}  &\multicolumn{1}{c}{$\Delta$\vlsr}  &\multicolumn{1}{c}{RMS distance}  &\multicolumn{1}{c}{MMB distance}  &\multicolumn{1}{c}{RMS flag}   &	\multicolumn{1}{c}{MMB flag} \\
  
    \multicolumn{1}{c}{name }&  \multicolumn{1}{c}{name }&  \multicolumn{1}{c}{(\kms)} & \multicolumn{1}{c}{(\kms)}  &\multicolumn{1}{c}{(\kms)}  &\multicolumn{1}{c}{(kpc)}  &\multicolumn{1}{c}{(kpc)}  &	\multicolumn{1}{c}{}&\multicolumn{1}{c}{}\\
\hline
G011.0340+00.0629	&	G011.034+00.062	&	15.4	&	20.5	&	5.1	&	14.2	&	2.9	&	F	&	N	\\
G013.6562$-$00.5997	&	G013.657$-$00.599	&	47.4	&	51.2	&	3.8	&	4.1	&	12.0	&	N	&	F	\\
G014.3909$-$00.0216	&	G014.390$-$00.020	&	23.1	&	26.9	&	3.8	&	2.5	&	13.5	&	N	&	F	\\
G018.4608$-$00.0034	&	G018.460$-$00.004	&	52.2	&	49.4	&	-2.8	&	12.1	&	3.9	&	F	&	N	\\
G281.7059$-$01.0986	&	G281.710$-$01.104	&	-1.0	&	0.9	&	1.9	&	1.7	&	4.2	&	N	&	F	\\
G294.5117$-$01.6205	&	G294.511$-$01.621	&	-15.3	&	-11.9	&	3.4	&	2.6	&	1.0	&	F	&	N	\\
G311.2292$-$00.0315	&	G311.230$-$00.032	&	29.2	&	24.8	&	-4.4	&	5.5	&	13.6	&	N	&	F	\\
G314.3197+00.1125	&	G314.320+00.112	&	-48.9	&	-43.5	&	5.4	&	3.6	&	8.8	&	N	&	F	\\
G318.9480$-$00.1969	&	G318.948$-$00.196	&	-33.8	&	-34.6	&	-0.8	&	2.4	&	10.5	&	N	&	F	\\
G319.8366$-$00.1963	&	G319.836$-$00.197	&	-13.2	&	-9.2	&	4.0	&	11.7	&	0.6	&	F	&	N	\\
\hline\\
\end{tabular}\\
Notes: Only a small portion of the data is provided here, the full table is available in electronic form at the CDS via anonymous ftp to cdsarc.u-strasbg.fr (130.79.125.5) or via http://cdsweb.u-strasbg.fr/cgi-bin/qcat?J/MNRAS/.
\end{minipage}

\end{center}
\end{table*}
\setlength{\tabcolsep}{6pt}

\subsubsection{Distance and Galactic distribution}
\label{sect:distances}

The kinematic distances presented in Papers\,I and II were calculated using the \citet{brand1993} Galactic rotation curve, while the RMS distances are calculated using the \citet{reid2009} rotation curve. To maintain consistency with the results presented in the previous two papers we have recalculated the RMS kinematic distances using the \citet{brand1993} model. However, for the vast majority of sources, the differences in distance given by the two rotation curves is smaller than their associated uncertainty and so the statistical results are robust against the choice of model. 

We have collected together distances for 1082 MSF clumps and show the distribution of these distances for all the associated clumps and the four associated subsamples in Fig.\,\ref{fig:atlas_distance_hist} while in Fig.\,\ref{fig:gal_distribution} we show the Galactic distribution as viewed from above the northern Galactic pole. The overall distribution seen in Fig.\,\ref{fig:atlas_distance_hist} consists of two prominent peaks centred at approximately 4.5 and 11.5\,kpc. Looking at the Galactic distribution shown in Fig.\,\ref{fig:gal_distribution} it is clear the 4.5\,kpc peak coincides with the inner part of the Scutum-Centaurus arm and the northern end of the Galactic bar at $\ell \sim 30\degr$, while the 11.5\,kpc peak is associated with the far segment of the Perseus arm between $\ell= 10-60$\degr\ and the location where the latter coincides with the southern end of the Galactic bar also at $\ell \sim 330$\degr. The presence of the dip between these two peaks is due to a combination of several factors: 1) the low number of clumps located within 3-4\,kpc of the Galactic centre;  2) the rotation curves tend to locate sources away from the tangent point; and 3) there is a general lack of any prominent features of Galactic structure in these bins. A detailed discussion of the Galactic distribution of the methanol masers, MYSOs, \hii\ regions has been presented in a number of recent papers (i.e., \citealt{green2011b}, Papers\,I and II and \citealt{urquhart2014}) and so will not be discussed here. 

Comparing the distance distributions of the four subsamples we find that the \hii-region, methanol-maser and multi-phase associated clumps have a similar distribution to each other. KS tests of combinations of these do not reveal any significant differences, however, the MYSO distribution is shown to be significantly different from the other three subsamples, with a much higher fraction being located within 6\,kpc. In Sect.\,4.6 we find that the MYSOs are systematically less luminous than \hii\ regions and so the lower number of more distant MYSOs found is likely to be due to the sensitivity of the 21\,\mum\ MSX band (\citealt{mottram2011a}). As mentioned in Sect.\ref{sect:sensitivity},  
we define a distance-limited sample to be all sources located between 3 and 5\,kpc. This is indicated by the vertical cyan lines shown in Fig.\,\ref{fig:atlas_distance_hist}; this distance range has been selected as it includes the peaks in the distributions of all three subsamples ensuring the largest numbers of sources in each evolutionary stage and the largest spread in clump mass range while minimising the potential bias due to distance.

\subsection{Clump radius}
\label{sect:size}

\begin{figure}
\begin{center}
\includegraphics[width=0.49\textwidth, trim= 0 0 0 0]{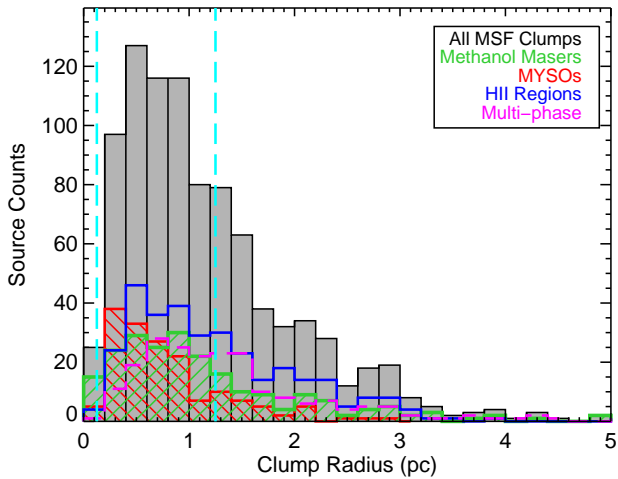}

\caption{\label{fig:physical_size} Distribution of the effective radii of all ATLASGAL massive star forming clumps and the four associated subsamples. The vertical long-dashed cyan coloured lines indicate the notional radii separating cores from clumps (0.125\,pc), and clumps from clouds (1.25\,pc), respectively. The bin size is 0.2\,pc.} 
\end{center}
\end{figure}

In Fig.\,\ref{fig:physical_size} we present the distribution of physical radii for all ATLASGAL associated clumps (grey filled histogram) and the four subsamples. The radius of each clump has been estimated using its heliocentric distance and the effective radius given in the ATLASGAL CSC. To help put these sizes into context we have added two vertical lines at radii of  0.15 and 1.25\,pc. These have been suggested as the nominal boundaries between cores and clumps, and clumps and clouds, respectively (e.g., \citealt{bergin2007}). Our sample includes sources that are large enough to be classified as clouds but, as there are no obvious breaks in the radius distributions, these distinctions appear somewhat arbitrary. For this reason, and to avoid introducing unnecessary confusion, we will simply refer to all of the ATLASGAL sources as clumps, but with the caveat that the sample covers a larger range of sizes than might ordinarily be expected. 

The mean and median radius for the whole sample of massive star forming clumps is $1.18\pm0.03$\,pc and 0.95\,pc, respectively. Although there appears to be a difference between some of the radius distributions of the subsamples, KS tests of the distance-limited subsamples reveal no significant differences between any of them. For the distance-limited sample the mean and median values are 0.8$\pm$0.03\,pc and 0.75\,pc, respectively.

\subsection{Physical separation}
\label{sect:offset_pc}

\begin{figure}
\begin{center}
\includegraphics[width=0.49\textwidth, trim= 0 0 0 0]{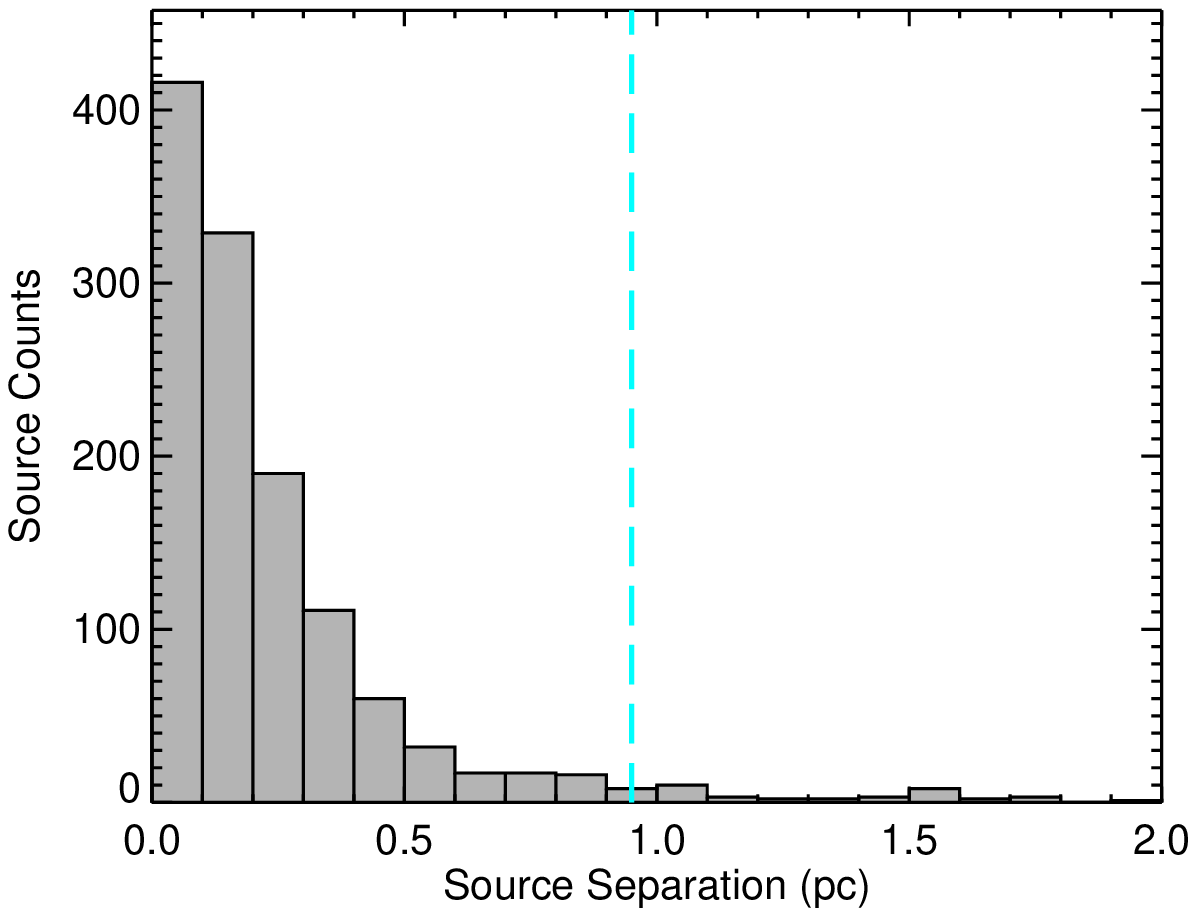}
\includegraphics[width=0.49\textwidth, trim= 0 0 0 0]{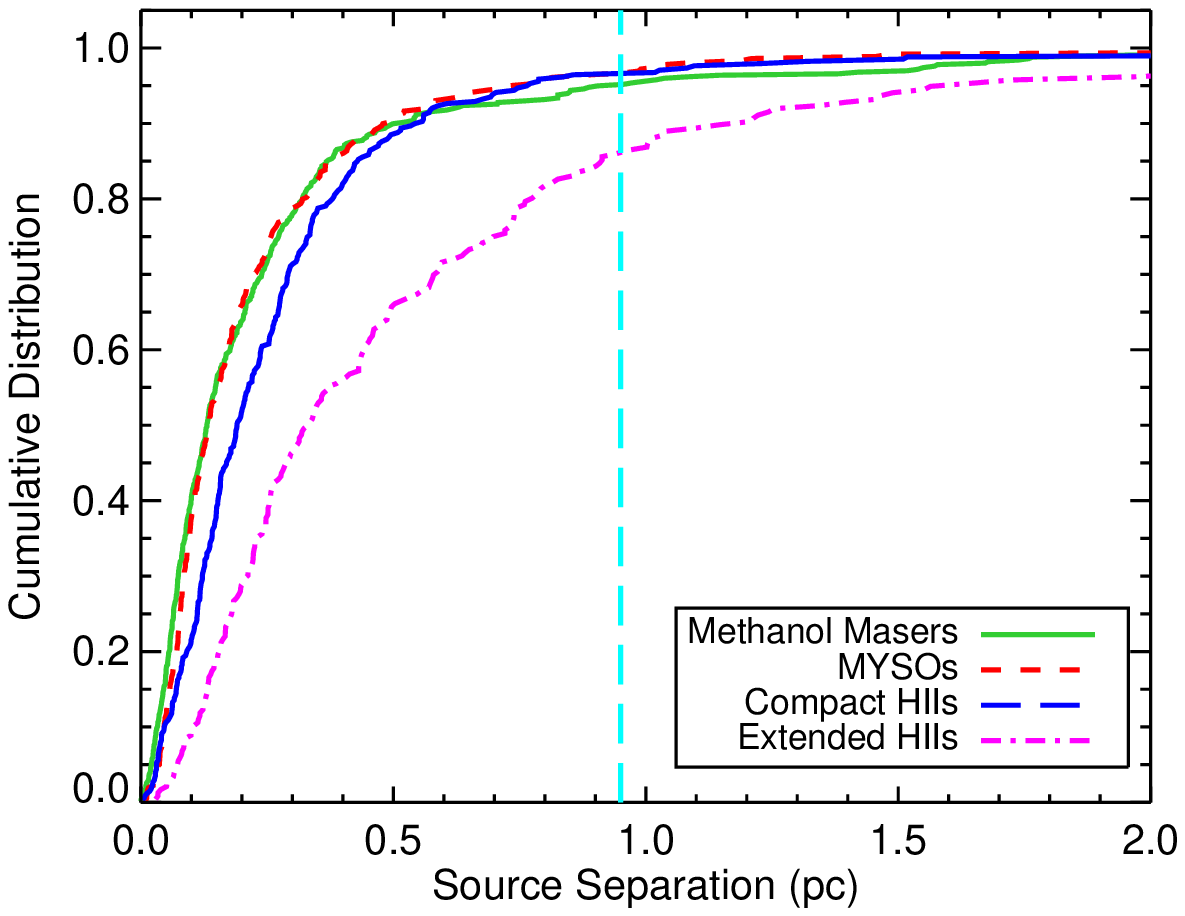}

\caption{\label{fig:offset_pc} Distributions of the projected physical separation between the peak of the dust emission and position of embedded object. The upper panel shows the distribution of all massive star-forming clumps with bin size of 0.1\,pc (filled grey histogram). In the lower panel we present the cumulative distribution function for the three tracers, splitting the \hii\ regions into compact and extended subsamples (see Sect.\,\ref{sect:atlas_rms_assoc}).  The vertical dashed cyan line shown in both plots indicates the median clump radius for the whole sample (i.e., 0.95\,pc).} 

\end{center}
\end{figure}

In Fig.\,\ref{fig:offset_pc} we show the variation in projected physical separation between the position of the peak submillimetre emission and that of the embedded massive stars. In the upper panel we present a histogram of the distribution of the whole sample; this clearly shows a strong correlation between the emission peak within a clump and the position of the embedded objects, with $\sim$90\,per\,cent within 0.5\,pc of each other and $\sim$60\,per\,cent within a couple of tenths of a parsec. Given that the median radius of the clumps is $\sim$1\,pc it confirms that the star formation is concentrated in the innermost part of the host clump where the densities are highest and where the  gravitational potential is deepest. This also suggest that there is little star formation taking place towards the edges of these clumps where we might expect star formation to be triggered by shocks and expanding \hii\ regions.

In the lower panel of Fig.\,\ref{fig:offset_pc} we show the cumulative distributions of the separations in the associated methanol masers, MYSOs and extended and compact \hii\ regions. As discussed in Sect.\,2.3 the distribution of the extended \hii\ regions is very different from that of the other three samples and the feedback from the central star is likely to be having a significant impact on its natal clump. The other three samples look similar and a KS test comparing the methanol masers and MYSOs finds no significant difference between them. However, the KS test comparing the methanol masers and MYSOs with the sample of compact \hii\ regions is able to reject the null hypothesis that they are drawn from the same parent population ($r \ll 0.01$ for both comparisons). The compact \hii\ regions are located at slightly larger distances from the peak of the submillimetre emission ($\sim$0.14\,pc for the methanol masers and MYSOs, and 0.19\,pc for the compact \hii\ regions). This difference is small but statistically significant and probably reflects the fact that even the compact \hii\ region sample includes a range of evolutionary stages with the most evolved starting to affect the structure of their host clumps. 

\subsection{NH$_3$ line widths and kinetic temperatures}
\label{sect:ammonia_obs}

\begin{figure}
\begin{center}
\includegraphics[width=0.49\textwidth, trim= 0 0 0 0]{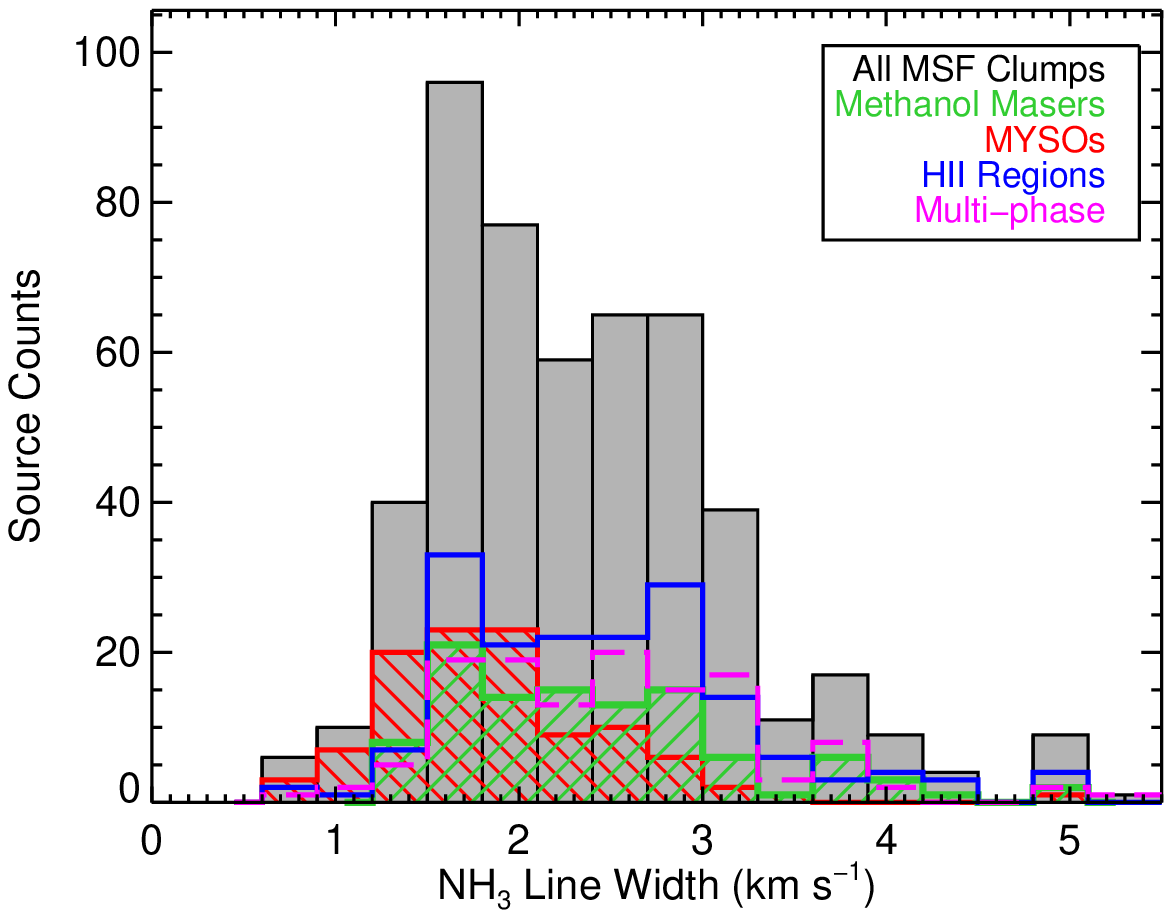}
\includegraphics[width=0.49\textwidth, trim= 0 0 0 0]{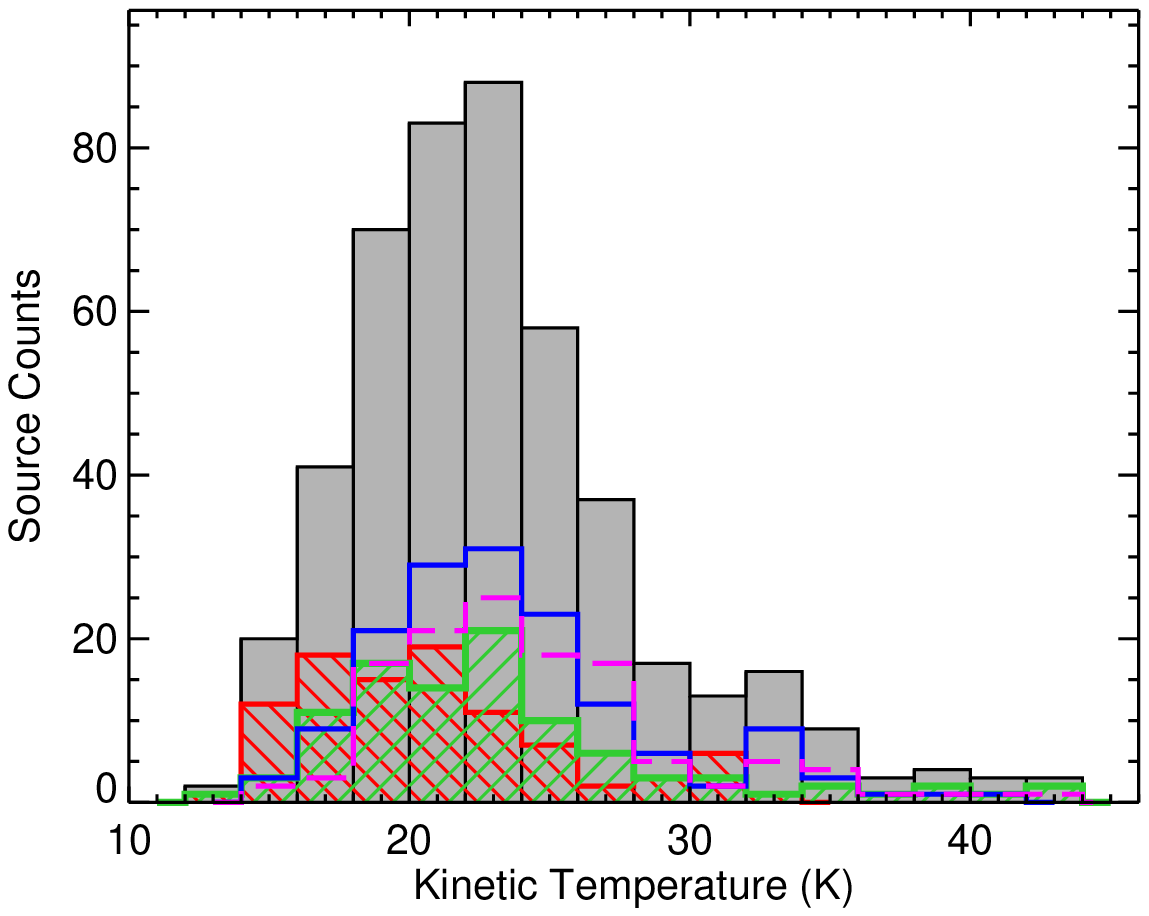}

\caption{\label{fig:line_width_temp} NH$_3$ (1,1) line width and kinetic gas temperature distributions (upper and lower panels, respectively) taken from the observations reported by \citet{urquhart2011b}, \citet{dunham2011} and \citet{wienen2012}. In cases where two or more observations are available for a particular clump we have adopted the mean value. The bin sizes used in the upper and lower plots are 0.3\,\kms\ and 2\,K, respectively.} 

\end{center}
\end{figure}

In Paper\,II we used the line widths of the ammonia (1,1) transition to estimate virial masses for approximately 100 clumps and, by comparing these with the isothermal clump masses, assessed their gravitational stability. Extending this work we have compared the positions of our sample of matched clumps with three recently published ammonia surveys made with the Effelsberg 100-m Telescope (\citealt{wienen2012}) and the Green Bank Telescope (\citealt{urquhart2011b} and \citealt{dunham2011b}) and extracted the line widths and kinetic gas temperatures, derived from the ratio of the NH$_3$ (1,1) and (2,2) lines, for 512 and 467 clumps, respectively. In the upper and lower panels of Fig.\,\ref{fig:line_width_temp} we present the FWHM velocity dispersion and kinetic temperature of the gas, respectively, for the whole matched sample of clumps and the four associated subsamples. We have excluded sources with kinetic temperatures larger than 45\,K as these are not accurately probed by the NH$_3$ (1,1) and (2,2) transitions and are therefore not reliable.

The mean and median values for the line widths of these massive star forming clumps are 2.37$\pm$0.04\,\kms\ and 2.2\,\kms, respectively, and we find no significant difference in the line widths for the four subsamples. These line widths are fairly typical for massive star forming regions and significantly wider than generally found towards quiescent, infrared-dark clouds ($\sim$1.7-1.9\,\kms; \citealt{chira2013,wienen2012}). There is a weak correlation between the line width and distance ($r=0.17$ with a significance $\ll 0.01$) but this is expected given that the size of clumps increases with distance and our data are consistent with the \citet{larson1981} size-line width relation. Since we find no significant difference between the line widths of the various subsamples these large line widths are unlikely to be driven by feedback from the embedded source (e.g., outflows or stellar winds) and likely to be a property of the clump prior to the onset of star formation. 

The mean and median values for the kinetic temperature of the gas towards the centre of a clump is 23.2$\pm$0.25\,K and 22.3\,K, respectively. We find no correlation between the kinetic temperature of the gas with distance ($r=0.02$ with a significance of 0.65); although the beam dilution increases with distance the beam filling factor is relatively constant presumably due to the clumpy substructure of a source. Inspecting the plot presented in the lower panel of Fig.\,\ref{fig:line_width_temp} we see that the distribution of the MYSO subsample is clearly different from the other subsamples, which are broadly similar to each other. 

The median kinetic temperature of the MYSOs is $\sim$3\,K lower than found for the other subsamples. A KS test on the kinetic temperature using a distance-limited sample confirms that the MYSO subsample is significantly different from the \hii\ region subsample but not from the methanol masers subsample ($r$  values are 0.003 and 0.016, respectively). The clumps associated with MYSOs are therefore colder than those associated \hii\ regions. \citet{urquhart2011b} found a correlation between the kinetic temperatures of the gas to the bolometric luminosities and since the \hii\ regions are significantly more luminous than the MYSOs (see Sect.\,\ref{sect:bol_lum}) this would explain the difference in temperature between them. Given the proposed evolutionary sequence we might have expected the methanol masers to be associated with colder clumps and it is therefore unclear why the methanol masers have a similar temperature to the \hii\ regions. 

\subsection{Isothermal clump masses and column densities}
\label{sect:mass}

\begin{figure}
\begin{center}

\includegraphics[width=0.49\textwidth, trim= 0 0 0 0]{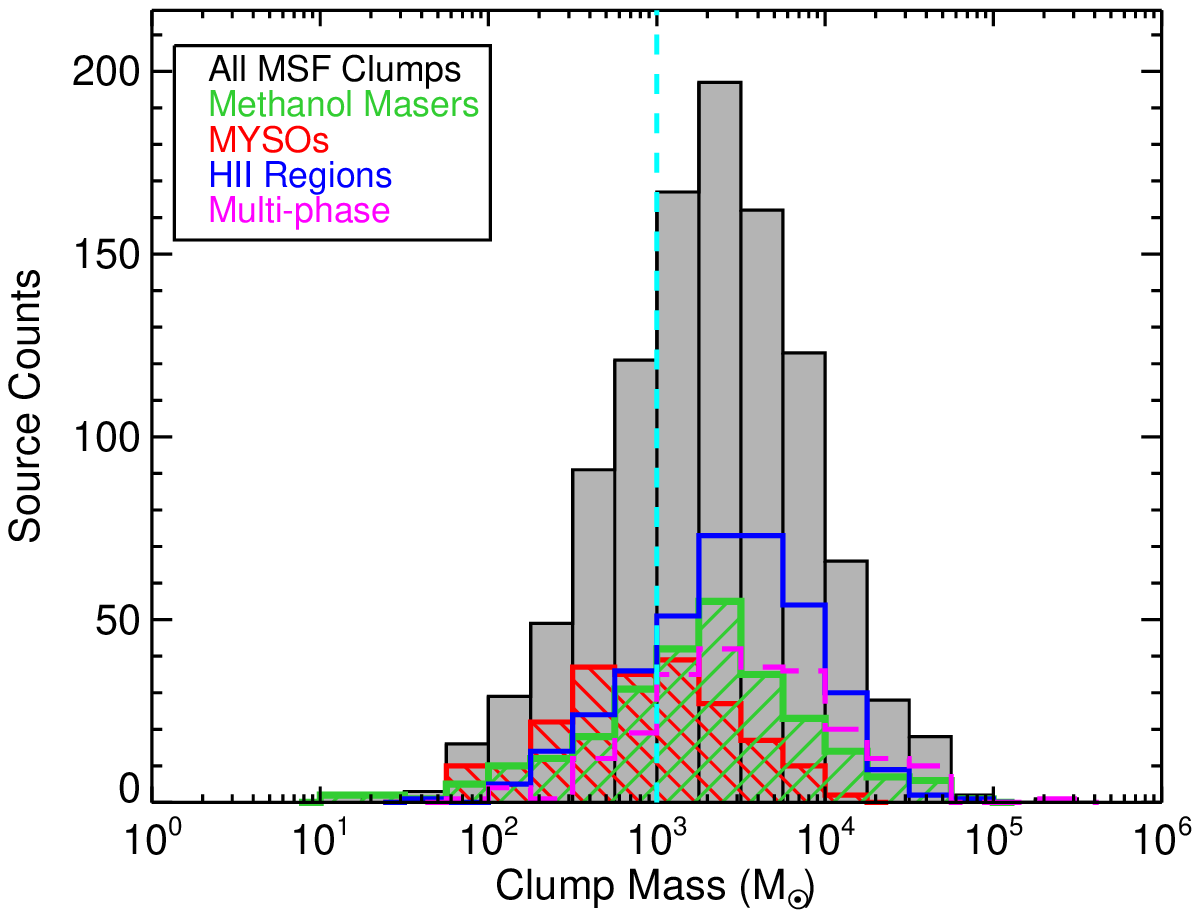}
\includegraphics[width=0.49\textwidth, trim= 0 0 0 0]{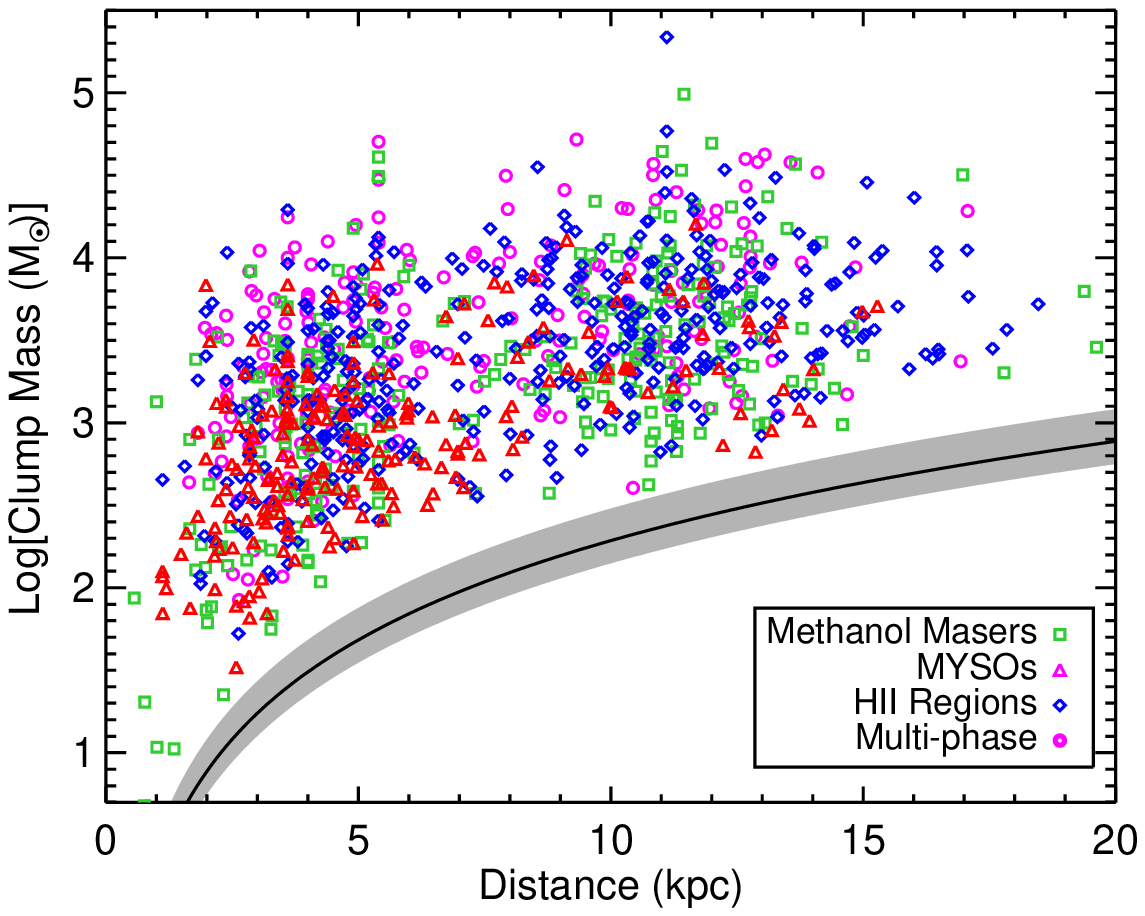}

\caption{\label{fig:clump_mass} Upper panel: The isothermal clump-mass distribution of all matched clumps with a distance and the four subsamples. The distribution of the whole sample is shown by the filled grey histogram while the distributions of the methanol-maser, MYSO, \hii-region and multi-phase subsamples are shown in green, red, blue and magenta, respectively. A dust temperature of 20\,K has been assumed. The vertical dashed cyan line indicates the ATLASGAL sensitivity limit at 20\,kpc. The bin size is 0.5 dex.  Lower panel: Mass-distance distribution of all associated clumps. See legend for colours and shapes of individual source types. The solid black line and the grey filled region indicated the ATLASGAL surveys clump mass sensitivity limit and its associated uncertainty assuming a dust temperature of 20$\pm5$\,K and for an unresolved source (i.e., 19.2\arcsec\ radius). Nearly all of the clumps are extended with respect to the beam, which produces a systematic offset between the sensitivity curve and the data points.} 

\end{center}
\end{figure}

Detailed descriptions of the determination of clump mass and column density are presented in Paper\,I and so we will only provide a brief summary of the procedure here. The isothermal clump masses are estimated using the \citet{hildebrand1983} method assuming that the total clump mass is proportional to the integrated flux density measured over the source:

\begin{equation}
\label{eqn:mass}
M_{\rm{clump}} \, = \, \frac{D^2 \, S_\nu \, R}{B_\nu(T_{\rm{dust}}) \, \kappa_\nu},
\end{equation}

\noindent where $S_\nu$ is the integrated 870-\mum\ flux density, $D$ is the distance to the source, $R$ is the gas-to-dust mass ratio (assumed to be 100), $B_\nu$ is the Planck function for a dust temperature $T_{\rm{dust}}$, and $\kappa_\nu$ is the dust absorption coefficient taken as 1.85\,cm$^2$\,g$^{-1}$ (\citealt{schuller2009} and references therein).

To be consistent with the analysis presented in Papers\,I and II, and the masses derived from similar studies (e.g., \citealt{motte2007, hill2005}), we assume that the gas and dust are in local thermodynamic equilibrium and adopt a kinetic temperature of 20\,K to estimate the total clump mass and peak column density.

The peak column densities are estimated from the peak flux density of the clumps using the following equation:

\begin{equation}
N_{\rm{H_2}} \, = \, \frac{S_\nu \, R}{B_\nu(T_{\rm{dust}}) \, \Omega \, \kappa_\nu \, \umu\, m_{\rm{H}}},
\end{equation}

\noindent where $\Omega$ is the beam solid angle, $\umu$ is the mean molecular weight of the gas, which we assume to be equal to 2.8, $m_{\rm{H}}$ is the mass of the hydrogen atom, and $\kappa_\nu$ and $R$ are as previously defined. 

As in Papers\,I and II we estimate the uncertainty in the derived clump mass and column density to be $\sim$50\,per\,cent, assuming an error of $\pm$5\,K for the gas temperature and an absolute flux uncertainty of 15\,per\,cent (\citealt{schuller2009}),  added in quadrature. However, as noted in the previous papers, these uncertainties are unlikely to have a significant impact on the overall distributions or the statistical analysis.
 
We have determined distances and masses for 1077 clumps associated with one of the three tracers of interest and these distributions are shown in Fig.\,\ref{fig:clump_mass}. The upper panel of this figure shows the clump-mass distribution while in the lower panel we present the mass distribution as a function of heliocentric distance. The clump masses range from as little as a few $\times 10$\,\msun\ to 10$^5$\,\msun, with a peak in the distribution between 2-3000\,\msun, which probably reflects the completeness level of the survey. Adopting a standard IMF (e.g., \citealt{kroupa2003}) and an upper limit for the star-formation efficiency of 30\,per\,cent (e.g., \citealt{lada2003}) we use a Monte-Carlo simulation to estimate the minimum clump mass required to form a cluster containing at least one massive star (see Paper\,II for details). We find that clump masses of  $\gtrsim$1000\,\msun\ are needed to form a cluster with a $\gtrsim$20\,\msun\ star (cf. \citealt{tackenberg2012}). Approximately 800 of the 1100 MSF clumps for which we are able to estimate a mass for satisfy this criterion, corresponding to $\sim$65\,per\,cent of the sample. We also note that it is still possible for clumps below this mass threshold to form a massive star, but that the fraction that actually does will be lower. 

Fig.\,\ref{fig:clump_mass} shows a significant difference between the masses of clumps associated with MYSOs and the other three subsamples. The MYSO associated clump distribution has a median value of $\sim$1000\,\msun, while the methanol maser, \hii\ region and multi-phase clump associations have values of 1900, 2800 and 3300\,\msun, respectively. A KS test on the distance-limited sample reveals that the distribution of MYSO associated clumps is significantly different from that of the \hii\ region associated clumps ($r= 0.00013$) but not significantly different from the methanol maser associated clumps ($r=0.04$). This suggests that there is a real difference between the masses of clumps associated with MYSOs and those with \hii\ regions. 

\subsubsection{Column densities}
\label{sect:col_den}

\begin{figure}
\begin{center}
\includegraphics[width=0.49\textwidth, trim= 0 0 0 0]{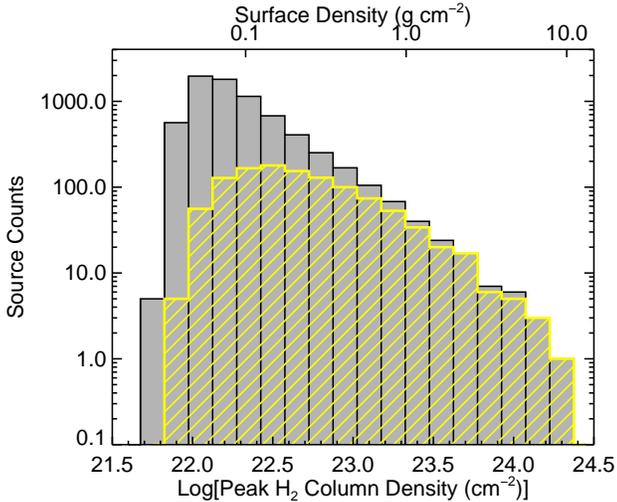}

\caption{\label{fig:col_den} $N_{\rm{H_2}}$ distribution for the whole ATLASGAL CSC is shown by the grey filled histogram and the massive star forming clump subsample is shown as the hatched yellow histogram. The bin size is 0.15\,dex. For reference the upper axis gives the peak surface densities of the clumps.} 

\end{center}
\end{figure}

Because there are relatively few unresolved ATLASGAL sources, distance is therefore not a factor in the column-density detection limits and, by assuming a constant dust temperature, we are able to estimate the column densities for all ATLASGAL sources and identify \emph{all} of the most compact ($R_{\rm{eff}}\lesssim100\arcsec$) and highest column-density (Log($N$(H$_2$)) $\gtrsim 22$\,cm$^{-2}$), pre-stellar and protostellar clumps located across the inner Galactic disc. Fig.\,\ref{fig:col_den} shows the distribution of estimated column densities for the whole CSC and for the MSF-associated subsample. The quiescent clumps have significantly lower column densities compared to those associated with MSF; median values are $\sim1.5\times10^{22}$\,cm$^{-2}$ and $\sim$$4.1\times10^{22}$\,cm$^{-2}$, respectively. Furthermore, the distribution for the MSF and quiescent clumps are very different and a KS test is able to reject the null hypothesis that these two samples are drawn from the same parent population ($r \ll 0.01$). 

Similar differences in column and surface densities have been reported by \citet{butler2012} from a comparison of starless cores to the sample of more evolved luminous MYSOs and \hii\ regions studied by \citet{mueller2002}, and by \citet{giannetti2013} from a comparison of pre-stellar and protostellar clumps. This would suggest that the central density increases before star formation begins or that star formation can only occur in the highest density clumps. This is consistent with our finding that no significant difference exists between the MSF subsamples with respect to their morphology and structure (see Sect.\,\ref{sect:size}), which also suggests that the increase in density would need to precede the onset of star formation. Another possibility is that the clumps do evolve after the star formation begins, but do so in a self-similar way, i.e., the density and \textit{radial distribution} is maintained. Alternatively, it could also mean that the MSF requires higher density clumps, which would imply that the non-MSF clumps might be sterile rather than simply being younger and less evolved.

\begin{figure}
\begin{center}
\includegraphics[width=0.49\textwidth, trim= 0 0 0 0]{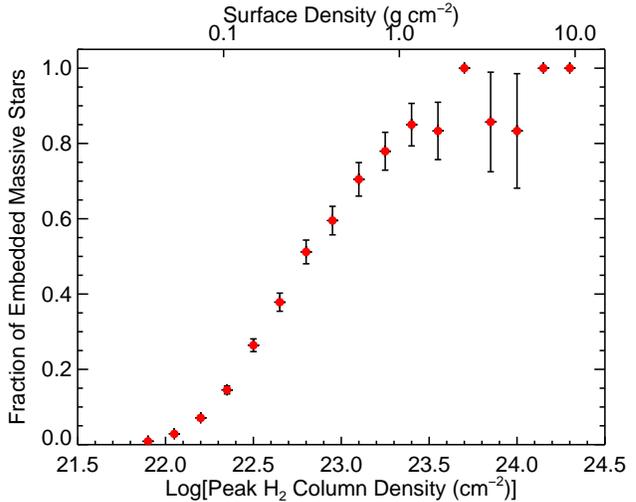}

\caption{\label{fig:col_den_ratio} Fraction of massive star forming clumps as a function of peak column density. The bin size is 0.15\,dex and the uncertainties are estimated using binomial statistics.} 

\end{center}
\end{figure}

The upper axis of Fig.\,\ref{fig:col_den} gives the peak mass surface densities of the clumps, $\Sigma_{\rm{clump}}$, which range from $\sim$0.1 to 10\,g\,cm$^{-2}$. These values are similar to those reported by \citet{battersby2011} from a $Herschel$ study of IRDCs (0.2-5\,g\,cm$^{-2}$). \citet{krumholz2008} calculated that, for the monolithic accretion model, radiative feedback is needed to prevent massive cores from fragmenting. In their model this feedback comes from the high accretion luminosities from a surrounding low-mass protostellar population and requires a minimum clump surface density of $\Sigma_{\rm{clump}} \gtrsim 1$\,g\,cm$^{-2}$. This value is significantly larger than the peak surface density values determined for much of our sample and the clump-averaged surface densities are significantly lower still. This might present a problem for the monolithic accretion model, however, strong magnetic-field support ($\sim$mG) would reduce the need for radiative feedback and for a minimum surface-density threshold (\citealt{tan2014}).

Fig.\,\ref{fig:col_den_ratio} shows the fraction of massive star forming clumps in the CSC as a function of column density in Fig.\,\ref{fig:col_den}. This plot demonstrates that the massive stars are primarily forming in the highest column-density clumps of the Galactic disc and that the fraction of MSF clumps increases rapidly with increasing column density, saturating above $3 \times 10^{23}$\,cm$^{-2}$ ($\sim$1\,g\,cm$^{-2}$). This is consistent with the finding of \citet{csengeri2014} from an analysis of the mid-infrared properties of ATLASGAL sources. It implies that the free-fall time decreases as a function of increasing column density (i.e., $n \propto N$) and may provide an explanation why no very massive pre-stellar clumps have been identified outside of the Galactic centre (e.g., \citealt{tackenberg2012,ginsburg2012}). This again suggests a fast star-formation process with the speed being related to the peak column density of the clump.

Looking into this in a little more detail, we find that there are only 70 clumps located in the Galactic disc with column densities larger than $1\times10^{23}$\,cm$^{-2}$ that are not associated with one of the massive star-formation tracers we have considered so far, and only 12 of these have column densities above $2\times10^{23}$\,cm$^{-2}$. However, this is not to say that they are not already forming stars, since it is possible that they have simply not been picked up by one of the three surveys investigated in this series of papers. \citet{csengeri2014} compared the ATLASGAL GaussClump catalogue with the \textit{WISE} and \textit{MSX} point-source catalogues and determined a conservative lower limit to the fraction of star forming clumps of $\sim$33\,per\,cent. This is a factor of two larger than we have determined here for MSF clumps and suggests that many of the high column density clumps that have no associations may still be forming stars. 

Comparing the sample of the 70 highest-density unassociated clumps with the \textit{WISE} catalogue, we are able to match 57 with mid-infrared point sources indicating that most are currently forming stars. Further investigation of the three highest column-density clumps reveals that  AGAL287.596$-$00.629 is associated with the Eta Carina cluster, AGAL337.916$-$00.477 is associated with the extended green object G337.91$-$0.48 (EGO; \citealt{cyganowski2008}) and AGAL327.301$-$00.552 is associated with an extended \hii\ region that has a complicated structure and so is not in the RMS sample. A full analysis of these high column density sources is beyond the scope of this work but will be reported in a future paper, however, from a preliminary examination of this sample we can already conclude that there are likely to be few very massive pre-stellar clumps in the Galaxy. This is consistent with the finding of \citet{ginsburg2012} from a search of the BGPS for massive proto-clusters located in the northern Galactic plane. 

One important caveat is that we have assumed a temperature of 20\,K for the whole population of clumps and, although this is reasonable for the massive star-forming clumps  it is likely to underestimate the column densities of the quiescent clumps where the temperatures are likely to be lower ($\sim$15\,K; e.g., \citealt{chira2013,wienen2012}). However, as mentioned above, a reduction of 5\,K will result in an increase of $\sim$50\,per\,cent in the derived column densities. Even allowing for a difference of  10\,K between the star-forming and quiescent clumps would result in the column densities of the latter being underestimated by at most a factor of two, corresponding to 0.3\,dex. This is equivalent to the width of two of the bins in Fig.\,\ref{fig:col_den}, and will have little impact on the dominance of the high column-density part of the distribution by the massive star-forming clumps. Furthermore, these column-density estimates are averaged over the beam, which samples a larger physical scale at larger distances, resulting in the column densities being inversely proportional to distance. The fact that many of the MSF clumps are also some of the most distant is likely to flatten the overall distribution and lower average column density of the sample.

\begin{figure}
\begin{center}
\includegraphics[width=0.49\textwidth, trim= 0 0 0 0]{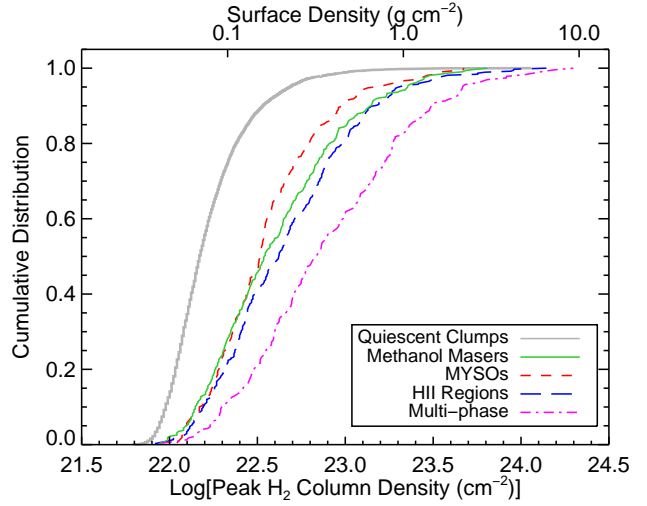}

\caption{\label{fig:col_den_CDF} Cumulative distribution of column densities for the whole quiescent and MSF clump subsamples.} 

\end{center}
\end{figure}

In Fig.\,\ref{fig:col_den_CDF} we show the cumulative distribution functions all MSF-associated clumps and the unassociated quiescent clump population for the whole ATLASGAL catalogue. KS tests comparing the $N$(H$_2$) distributions of the methanol maser, MYSO and \hii\ regions associated clumps in the distance-limited subsamples we find that the methanol-maser and MYSO distributions are not significantly different ($r=0.016$) and neither are those of the methanol masers and \hii\ regions ($r=0.58$). However, the MYSO and \hii\ region distributions are significantly different ($r=0.0004$) and are therefore unlikely to be drawn from the same population. 

We have found several significant differences between the MYSO and \hii\ region associated clumps --- the former are significantly colder, less massive and have lower column densities --- and these do not appear to be due to any distance bias. However, the methanol-masers associated clumps have not been found to be significantly different from the MYSO and \hii\ region associated clumps, which would suggest that their properties are somewhere in between these other two subsamples. 

The multi-phase clumps are clearly associated with significantly more massive and higher column density clumps than the other three subsamples, consistent with these objects being more active star forming regions.

\begin{figure}
\begin{center}

\includegraphics[width=0.49\textwidth, trim= 0 0 0 0]{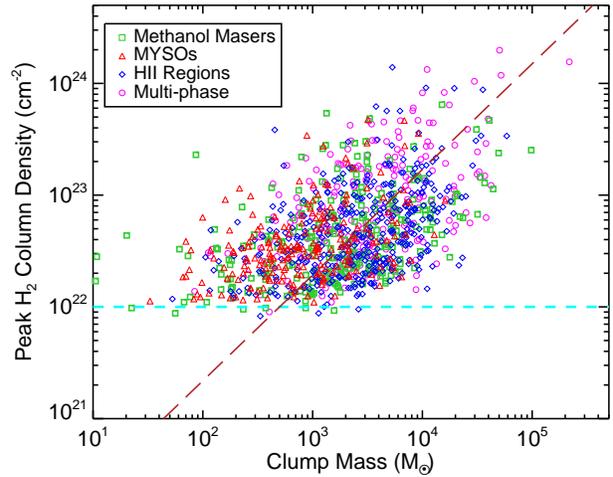}

\caption{\label{fig:mass_col_den} Comparison of clump mass and column density for the four subsamples of massive star-forming clumps. The dashed horizontal cyan line indicates the nominal column density sensitivity of the ATLASGAL survey ($N$(H$_2$) $\sim$10$^{22}$\,cm$^{-2}$) while the long-dashed red line shows the result of a linear fit to the data. The slope has a value of $0.95\pm1.29$.} 

\end{center}
\end{figure}

In Fig.\,\ref{fig:mass_col_den} we compare the clump masses and column densities of the four MSF subsamples. We evaluate the observed correlation between these two parameters using a partial Spearman correlation test to account for the mutual distance dependence in both parameters (see \citealt{yates1986}, \citealt{collins1998} and Paper\,I for more details). This gives a correlation coefficient of 0.81 with an associated significance of  $t \ll 0.01$, where $t$ is the student-t statistic,  which confirms a strong correlation between these two parameters. Our sample is therefore likely to include not only all of the highest column-density clumps but also all of the most massive clumps in the Galactic disc. It also means that we are unlikely to have missed any embedded massive protostars that might have remained hidden due to high levels of extinction as there are no very massive clumps that are not already included in our sample.

\subsubsection{Clump mass function}

\begin{figure}
\begin{center}
\includegraphics[width=0.49\textwidth, trim= 0 0 0 0]{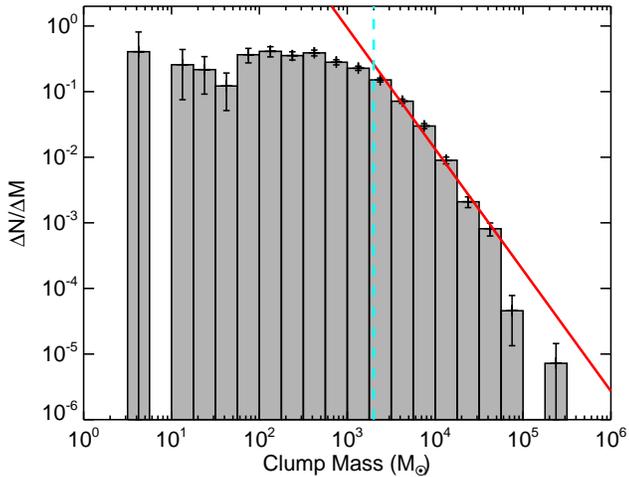}

\caption{\label{fig:mass_radius_dndm_histogram} Differential clump-mass distribution for all MSF clumps. The red line shows a linear fit to the data and has a slope of $-$1.85$\pm$0.15 which is similar to that found in a number of other studies. The vertical dashed cyan line indicates the mass completeness of the sample; the latter approximates the turnover of the distribution and bins below this value are not included in the fit. The uncertainties are estimated using \poi\ statistics and the bin size used is 0.25\,dex.} 

\end{center}
\end{figure}

In Fig.\,\ref{fig:mass_radius_dndm_histogram} we present the differential mass distribution for the ATLASGAL clumps associated with at least one of the three tracers of interest. On this plot, the vertical cyan dashed line indicates the nominal sensitivity limit and the solid red line shows the results of a linear least-squares fit to the bins above the peak in the mass distribution shown in the left panel of Fig.\,\ref{fig:clump_mass} ($\sim$2000\,\msun). This line provides a reasonable fit to all of the bins above the peak of the mass distribution except for the two highest mass bins, which is suggestive of a break in the power-law or perhaps that the distribution is log normal, however, there are too few bins populated investigate this further. The derived exponent ($\alpha$ where ${\rm{d}}N/{\rm{d}}M \propto M^{\alpha}$) of the fit to the high-mass tail for all associations is $-1.85\pm0.15$. 

We have repeated this process for the methanol-maser and \hii-region associations, fitting the slope to the mass bins above the mass completeness limit. Since the majority of the MYSO distribution falls below the nominal mass completeness of the survey we do not include them in this analysis.  We obtain values for the slope of $-1.91\pm0.09$ and $-1.90\pm0.22$ for the methanol-maser and HII-region subsamples, respectively.  Comparing the slopes and their associated uncertainties, we find no significant difference in these two subsamples. These values are in good agreement with those derived for samples covering similar evolutionary stages as presented here (e.g., \citealt{williams2004}, \citealt{reid2005}, \citealt{beltran2006}) and starless ATLASGAL clumps (\citealt{tackenberg2012}); these have reported power-law exponents with values between $-2.0$\ to $-2.3$. From this we conclude that the clump-mass function does not change significantly as the embedded massive star evolves.

The slope of the clump mass function is consistent with the upper end of the range found for Galactic samples of giant molecular clouds (GMCs), which are generally found to be between $-$1.5 and $-$1.8 (e.g., \citealt{solomon1987, heyer2001, roman-duval2010}). Recent measurements have found the clump formation efficiency, the ratio of mass in dense clumps to the  total cloud mass, is relatively invariant to galactic location, cloud mass and environment ($\sim$7-15\,per\,cent; \citealt{eden2012,eden2013,heyer2014}). The similarity of the ratio of dense clump mass to cloud mass for all clouds, and difference in the slopes of the cloud and clump mass functions, suggests that lower mass dense clumps are preferentially forming and very mass dense clumps are rare, even in very massive GMCs. 

\subsubsection{Virial masses}
\label{sect:virial_mass}

A detailed description of the virial mass calculation is given in Paper\,II and so will only be briefly covered here. We first make a correction to the measured ammonia (1,1) line width so that it better reflects the average velocity dispersion of the total column of gas (i.e., \citealt{fuller1992}):

\begin{equation}
\Delta v^{2}_{\rm{avg}} \; = \; \Delta v^{2}_{\rm{corr}}+8ln2\times \frac{k_{\rm{b}}T_{\rm{kin}}}{m_{\rm{H}}}\left(\frac{1}{\mu_{\rm{p}}}-\frac{1}{\mu_{\rm{NH_3}}}\right)
\end{equation}

\noindent where $\Delta v_{\rm{corr}}$ is the observed NH$_3$ line width corrected for the resolution of the spectrometer, $k_{\rm{b}}$ is the Boltzmann constant, $T_{\rm{kin}}$ is the kinetic temperature of the gas (again taken to be 20\,K) and $\mu_{\rm{p}}$ and $\mu_{\rm{NH_3}}$  are the mean molecular mass of molecular hydrogen and ammonia taken as 2.33 and 17, respectively. We then use the following equation to estimate the virial mass for each clump:

\begin{equation}
\left (\frac{M_{\rm{vir}}}{\rm{M}_\odot} \right) \; = \; \frac{5}{8ln2} \frac{1}{a_1 a_2 G} \left(\frac{R_{\rm{eff}}}{\rm{pc}}\right) \left(\frac{\Delta v_{\rm{avg}}}{\rm{km\,s}^{-1}}\right)^{2}
\end{equation}

\noindent where $R_{\rm{eff}}$ is the effective radius of the clump, $G$ is the gravitational constant and $a_1$ and $a_2$ are corrections for the assumptions of uniform density and spherical geometry, respectively (\citealt{bertoldi1992}). For aspect ratios less than 2 the value $a_2 \sim 1$, which is suitable for this sample of clumps, and the value of $a_1 \sim 1.3$ (e.g., see \citet{beuther2002} and Paper\,II for more details). We estimate the uncertainty in the virial mass to be of order 20\,per\,cent allowing for a $\sim$10\,per\,cent uncertainty each in the distance, the fitted line width, and spectrometer channel width, and the uncertainty in the measured source size.

\begin{figure}
\begin{center}

\includegraphics[width=0.49\textwidth, trim= 0 0 0 0]{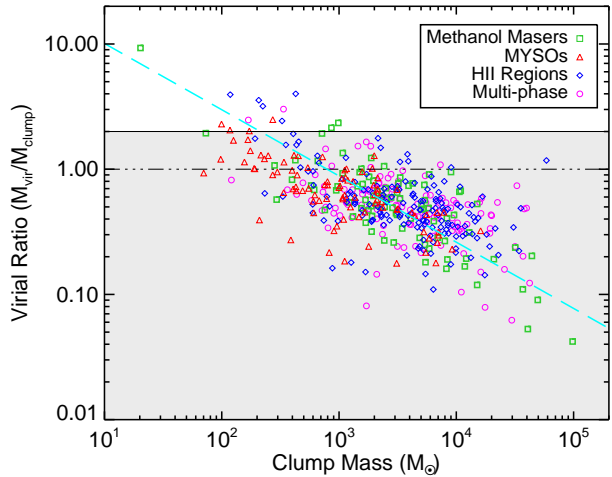}

\caption{\label{fig:virial_mass} Virial ratio ($\alpha$) as a function of clump mass ($M_{\rm{clump}}$) is shown for the four subsamples. The solid and dash-dotted lines indicate the critical values of $\alpha$ for an isothermal sphere in hydrostatic equilibrium with and without magnetic support, respectively. The light grey shading indicates the region where clumps are unstable and likely to be collapsing without additional support from a strong magnetic field. The long-dashed cyan line shows the result of a linear fit to these data.} 

\end{center}
\end{figure}

In Fig.\,\ref{fig:virial_mass}, we plot the virial ratio ($\alpha = M_{\rm{vir}}/M_{\rm{clump}}$) versus the clump mass for a sample of 466 MSF clumps for which we have a radius and that have been detected in one of the three ammonia surveys discussed in Sect.\,\ref{sect:ammonia_obs}. In a recent paper, \citet{kauffmann2013} have shown that the critical virial parameter ($\alpha_{\rm{cr}}$) for an isothermal sphere that is in hydrostatic equilibrium (i.e., a Bonnor-Ebert sphere; \citealt{ebert1955,bonnor1956}) that is not supported by a magnetic field is $\alpha_{\rm{cr}}$ = 2. Clumps with $\alpha > 2$ are subcritical and unless pressure confined by their local environment will expand. Clumps with $\alpha < 2$ are supercritical and unstable to collapse unless supported by a strong magnetic field. The solid black horizontal line shown in Fig.\,\ref{fig:virial_mass} indicates the $\alpha = 2$ locus and the grey shaded area shows the region of parameter space where clumps are gravitationally unstable. These virial ratios are a useful measure of a clump's stability and although not definitive for any particular clump they do provide some measure of the stability of the population as a whole.  

Even allowing for an equal amount of magnetic and kinetic support (i.e., $\alpha_{\rm{cr}}$ = 1; \citealt{bertoldi1992}) we find that the vast majority of the clumps are likely to be unstable and in a state of gravitational collapse. Since all of these clumps are in the process of forming massive stars we should expect smaller regions to be collapsing locally. However, the fact that parsec scale clumps are unstable implies that they are also undergoing global collapse.  

The distribution of the whole sample in Fig.\,\ref{fig:virial_mass} reveals a strong trend for decreasing values of $\alpha$ with increasing clump masses, which implies that the most massive clumps are also the least gravitationally stable. A log-log linear regression fit to the distribution gives a slope of $-$0.53$\pm$0.16, which is similar to values reported by \citet{kauffmann2013} from their analysis of massive star formation studies in the literature (e.g., \citealt{sridharan2002,wienen2012}). As noted by \citet{kauffmann2013} the trend for decreasing virial ratios with increasing clump mass provides a relatively simple explanation to the relative sparsity of very massive pre-stellar clumps, of which very few have been found outside the Galactic centre (\citealt{ginsburg2012,tackenberg2012}). The larger values of $\alpha$ found for lower mass clumps may indicate that feedback from the embedded stars is able to play a bigger role in stabilising their natal environment on smaller scales through the injection of turbulence and heating of the circumstellar envelope.

Comparing these observed values of $\alpha$ (i.e., $\alpha \ll 2$) \citet{kauffmann2013} found them to be inconsistent with the monolithic accretion model, which requires $\alpha \geq 2$. They concluded that magnetic fields would need to be included but that this would represent a major modification to the original model. Conversely, the observed values of $\alpha$ do satisfy the conditions required for competitive accretion to be viable for a large proportion of the MSF clumps (where $\alpha < 1$; \citealt{krumholz2005}). Parsec-sized, globally collapsing clumps provide the flow of material that is funneled into the centre of the proto-cluster and onto the competing protostars (e.g., \citealt{schneider2010}).  However, as \citet{kauffmann2013} point out, this does not rule out the monolithic accretion model, and there are a number of lower-mass clumps (i.e., $<$1000\,\msun; \citealt{lada2003}) that are unlikely to form a stellar cluster and therefore competitive accretion is not viable for all clumps. 

The distributions of the four subsamples presented in Fig.\,\ref{fig:virial_mass} reveal a clear difference between the MYSOs and \hii\ regions with the former generally having lower clump masses and higher values of $\alpha$ compared to the latter. However, these are largely due to the significant difference in clump mass of the two subsamples that has already been discussed and so we will not dwell further on this difference here.

\subsection{Luminosities}

\subsubsection{RMS Bolometric luminosities}
\label{sect:bol_lum}

We calculate the luminosities of RMS sources using the model spectral energy distribution (SED) fitter developed by \citet{robitaille2007} and applying a similar method to that discussed in \citet{mottram2011a}. For this process, we start with near-infrared photometry from the 2MASS \citep{Skrutskie2006}, UKIDSS \citep{lucas2008} or Vista-VVV \citep{Minniti2010} surveys and mid-infrared photometry taken from MSX \citep{egan2003} and WISE \citep{Wright2010} --- see \citet{lumsden2013} for a discussion on the preferred choice of values when two or more are available for a particular wavelength. We complement these near- and mid-infrared measurements with \textit{Herschel} far-infrared photometry provided by the Hi-GAL survey \citep{molinari2010}, and submillimetre data from the ATLASGAL CSC \citep{contreras2013} and distances as discussed in Sect.\,\ref{sect:distances}.

\begin{figure}
\begin{center}

\includegraphics[width=0.49\textwidth, trim= 0 0 0 0]{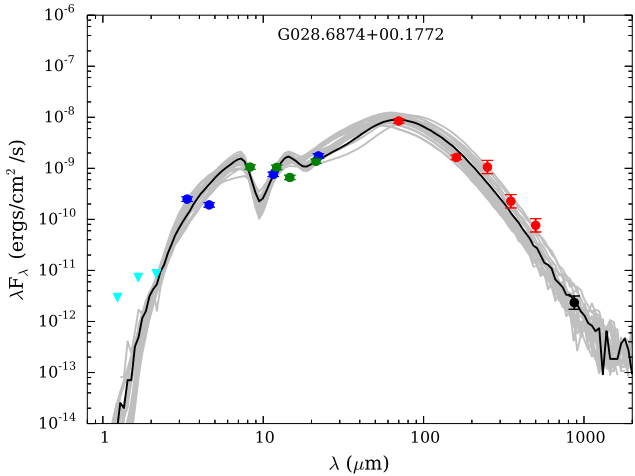}

\caption{\label{fig:sed_example} Spectral energy distribution for the RMS source G028.6874+00.1772. The near-infrared fluxes (2MASS, VVV, or UKIDSS) are shown in cyan, WISE and MSX mid-infrared fluxes are shown in blue and green, respectively, \textit{Herschel} Hi-GAL PACS/SPIRE fluxes are shown in red and the integrated ATLASGAL flux is shown in black. The best-fit model obtained from the SED fitter is plotted as a solid black curve while the 9 next best fits are shown as solid grey curves.} 

\end{center}
\end{figure}

The fluxes are input into the SED fitter along with the allowed range of distances which we set to the RMS distance assuming an uncertainty of $\pm$1\,kpc. The fitter searches through 20000 models which probe a range of properties of both the central (proto)star and surrounding material \citep{robitaille2006}, each of which can be viewed from 10 different viewing angles, resulting in possible 200000 SEDs. The fitter also takes into account foreground extinction and the different beam sizes of the various observations, returning the best fit and all models within a given tolerance through chi-squared minimisation. We calculate the luminosity and uncertainty using chi-squared weighting as discussed in \citet{mottram2011a}. The mean chi-squared value for the best fit is $\sim$12 with a mean delta-chi-squared value for the best 10 fits of $\sim$3. Fig.\,\ref{fig:sed_example} shows the measured flux values and the model fit obtained from the SED fitter for G028.6874+00.1772, as a typical example. We are primarily concerned with obtaining a good fit to these data and an accurate estimate of the bolometric luminosity and so the precise details of the models are somewhat irrelevant and may be inappropriate anyway.

In a number of cases where there are multiple MYSOs or \hii\,regions associated with a particular MSX source the individual embedded sources are not resolved in the longer wavelength surveys (i.e., \textit{MSX}, \textit{Herschel} 250\,\mum, 350\,\mum, 500\,\mum\ and ATLASGAL). For these sources we only use the longer wavelength photometric measurements to estimate the total bolometric luminosity of all the embedded sources. The luminosity is then split between the sources present using the ratio of their relative mid/far-infrared fluxes measured from the images.\footnote{Note this is an important difference to the method previously used (\citealt{mottram2011a}) where the mid/far-infrared fluxes were apportioned before the SEDs were fitted.} The \textit{Herschel} fluxes have been used to fit SEDs to 845 RMS sources, which corresponds to $\sim$88\,per\,cent of the RMS sample considered here. The full results of this process, along with details of flux determination and analysis of the infrared colours will be presented in Mottram et al. (2014). Bolometric luminosities for the other 119 RMS sources are taken from \citet{mottram2011a}.

\begin{figure*}
\begin{center}

\includegraphics[width=0.49\textwidth, trim= 0 0 0 0]{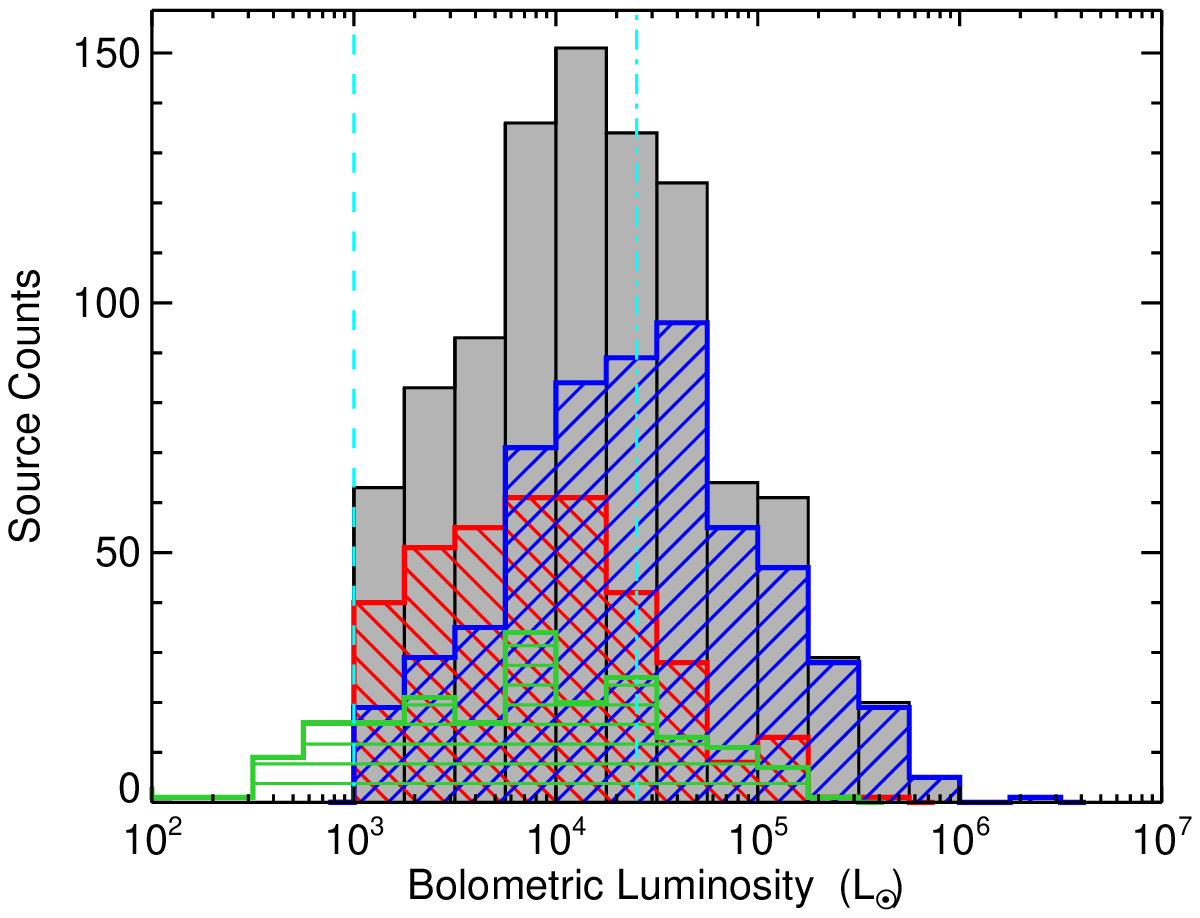}
\includegraphics[width=0.49\textwidth, trim= 0 0 0 0]{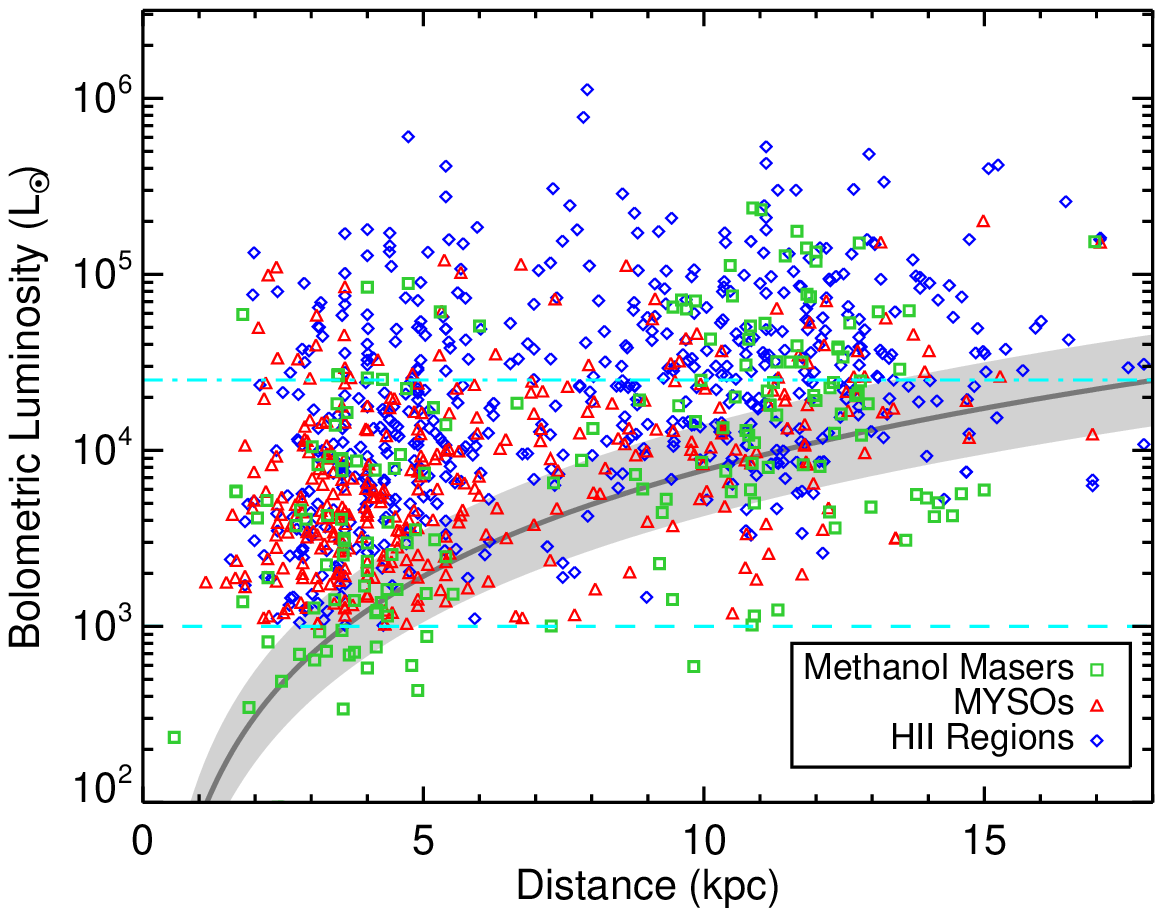}

\caption{\label{fig:bol_lum} Left panel: The bolometric luminosity distribution of all matched RMS sources is shown in grey while that of the methanol-maser, MYSO and \hii-region samples are shown in green, red and blue, respectively.  The dashed and dash-dot vertical cyan lines indicate the luminosity cut applied to the RMS sample to insure only massive star-forming clumps are selected (i.e., \lbol\ $>$ 1000\,\lsun) and the RMS completeness limit at $\sim$20\,kpc, respectively. The bin size is 0.25\,dex. Right panel: The luminosity distribution as a function of heliocentric distance. The dark line and light grey shaded region indicate the limiting sensitivity of the MSX 21-\mum\ band and its associated uncertainty. The horizontal cyan lines are the same as described for the left panel. Where multiple sources have been identified within the MSX beam, the luminosity has been apportioned between them resulting in some sources being located below the sensitivity limit (see \citealt{mottram2011a} for details).} 

\end{center}
\end{figure*}

Fig.\,\ref{fig:bol_lum} shows the luminosity distribution of all matched RMS sources as well the MYSO and \hii-region subsamples. The distribution peaks at 1-$2\times10^4$\,\lsun.  However, the distributions of the two subsamples are significantly different, with the MYSO and \hii-region distributions peaking at $\sim1\times10^4$\,\lsun\ and $\sim4\times10^4$\,\lsun, respectively. This difference in the luminosity function of these two samples and the lack of very luminous MYSOs has been previously noted and discussed by \citet{mottram2011b}.  The latter show that MYSOs with masses larger than  $\sim$30\,\msun\ have a Kelvin-Helmholtz contraction time that is shorter than their accretion timescale, resulting in the rapid formation of an \hii\ region. It is interesting to note that there are very few MYSOs with luminosities above 10$^5$\,\lsun and none that have luminosities larger than $2\times 10^5$\,\lsun, which corresponds to a zero age main sequence (ZAMS) star with a mass of $\sim$30\,\msun\ \citep{mottram2011b}. This is consistent with the models of \citet{hosokawa2009} and \citet{hosokawa2010} described in the introduction of this paper and, indeed, it is clear from the plots shown in Fig.\,\ref{fig:bol_lum} that the MYSOs are approximately an order of magnitude less luminous than the more evolved compact \hii\ regions.  This also means that we are less sensitive to MYSOs with increasing distance. A KS test of the distance-limited sample reveals that the luminosity distributions of these two subsamples are significantly different ($r \ll 0.01$), but this is also nicely explained by the Hosokawa et al. models and to be expected as the MYSOs transition into \hii\ regions.

\subsubsection{Methanol maser luminosities}
\label{sect:mmb_lum}

\begin{figure}
\begin{center}

\includegraphics[width=0.49\textwidth, trim= 0 0 0 0]{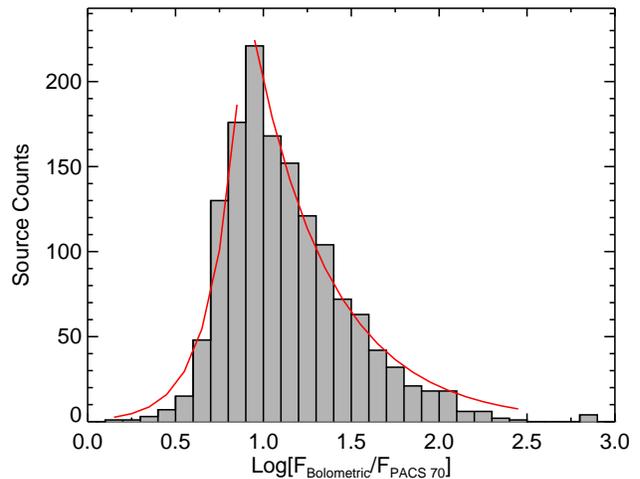}

\caption{\label{fig:pacs70_ratio} Comparison of the RMS bolometric fluxes obtained from SED fits to the infrared and submillimetre photometry and their PACS 70\,\mum\ flux. The red curve show the results of exponential fits to either side of the peak of the distribution from which we estimate the uncertainties. The bin size is 0.1\,dex. } 

\end{center}
\end{figure}

Bolometric luminosities are not currently available for the majority of the methanol-maser associated embedded sources. However, following the work of \citet{mottram2011a} we are able to obtain a rough estimate by scaling up their infrared luminosity. Since it is likely that the methanol masers are more embedded than the MYSOs and \hii\ regions, we use far-infrared wavelengths where the extinction is less likely to affect our analysis. We have therefore compared the bolometric fluxes of a large sample of MYSOs and \hii\ regions identified by the RMS survey with their \textit{Herschel} 70-\mum\ fluxes. In Fig.\,\ref{fig:pacs70_ratio} we show the results of this comparison in the form of a histogram showing the distribution of ratio values obtained. The distribution is skewed and so we take the peak of the distribution as the conversion factor and fit an exponential function to each side of the peak to estimate the asymmetric uncertainties; this gives a $F_{\rm{bol}}/F_{\rm{PACS\,70\,\umu m}}$ conversion factor $9.1_{-3}^{+16}$. Note that in applying this conversion we are implicitly assuming that the  $F_{\rm{bol}}/F_{\rm{PACS\,70\,\umu m}}$ is approximately the same for both methanol maser sources and MYSOs/\hii\,regions.

We have used the publicly available level 2.5 70\,\mum\ tiles of the Hi-GAL open-time Key Project (\citealt{molinari2010}) obtained from the \textit{Herschel} Science Archive.\footnote{http://herschel.esac.esa.int/Science\_Archive.shtml.} We have used the high pass filtered maps which have been calibrated and reduced with SPG v10.3.0 and combine both scan directions (parallel and orthogonal). From these tiles we have extracted postage-stamp images for all the methanol masers and performed aperture photometry to estimate their PACS 70\,\mum\ emission.The main objective here is not to provide a complete set of reliable bolometric luminosities for the methanol masers (this is being done by the Hi-Gal team and will be published separately) but to provide a reasonable indication of their range and distribution.

We have used the 70-\mum\ fluxes to estimate the bolometric luminosities for $\sim$200 methanol-maser associated clumps and the distribution of values is shown in Fig.\,\ref{fig:bol_lum}. Although the luminosity of any particular methanol maser is less reliable than those of RMS sources, the overall luminosity distribution should be robust (i.e., the sample size is large enough that the mean should be well determined). The luminosities cover the same range as those of the MYSO subsample but the overall distribution is significantly flatter. A KS test on the distance-limited subsamples is unable to reject the null hypothesis that the methanol masers and MYSOs are drawn from the same population at the required 3$\sigma$ confidence level ($r= 0.003$). We do find a significant difference between the luminosity distribution of the \hii\ regions and that of the methanol masers ($r\ll 0.001$). We note that many of the methanol masers are located above the MSX 21-\mum\ sensitivity limit (or would be if they had an MYSO like SED) but do not have an MSX counterpart, which is approximately 70\,per\,cent of the sample. This is consistent with methanol masers being associated with very luminous objects that are more embedded than the MYSOs and \hii\ region subsamples, and therefore likely to precede these two more evolved and less embedded stages.

\section{Empirical Relationships}

As previously noted here and in Papers\,I and II, all of the associated clumps appear to be rather spherical in structure, are centrally condensed and the vast majority are associated with a single massive star-formation tracer. Even in cases where multiple tracers are found, these tend to be clustered towards the centre of their host clump (e.g., see lower panels of Fig.\,2). These clumps typically have masses of a few 10$^3$\,\msun\ and radii of $\sim$1\,pc, which are similar to the physical sizes and masses expected to form stellar clusters \citep{lada2003}. It is therefore likely that the majority of clumps are in the process of forming a stellar cluster. It follows that the methanol masers, mid-infrared point sources and radio continuum emission are associated with the most massive members of a young proto-cluster. The derived clump properties (e.g., mass, density, radius etc) and the bolometric luminosities of the embedded sources are therefore much more likely to be related to the embedded cluster rather than to a single star.

Before we begin investigating these relationships we caution that the mere existence of any correlations between the observed properties of clumps does not imply causation (i.e., post hoc ergo propter hoc).  Most are simply the result of the trivial fact that most massive clumps tend to have more extremes within them. So more massive clumps tend to be larger and contain more luminous stars and this alone may not be meaningful and care must therefore be taken not to over-interpret these correlations.  There is, however, potentially useful information about physics in the gradients of any correlations and in the positions of sources in the relevant parameter space.

\subsection{Mass-Luminosity relation}

\begin{figure*}
\begin{center}

\includegraphics[width=0.99\textwidth, trim= 0 0 0 0]{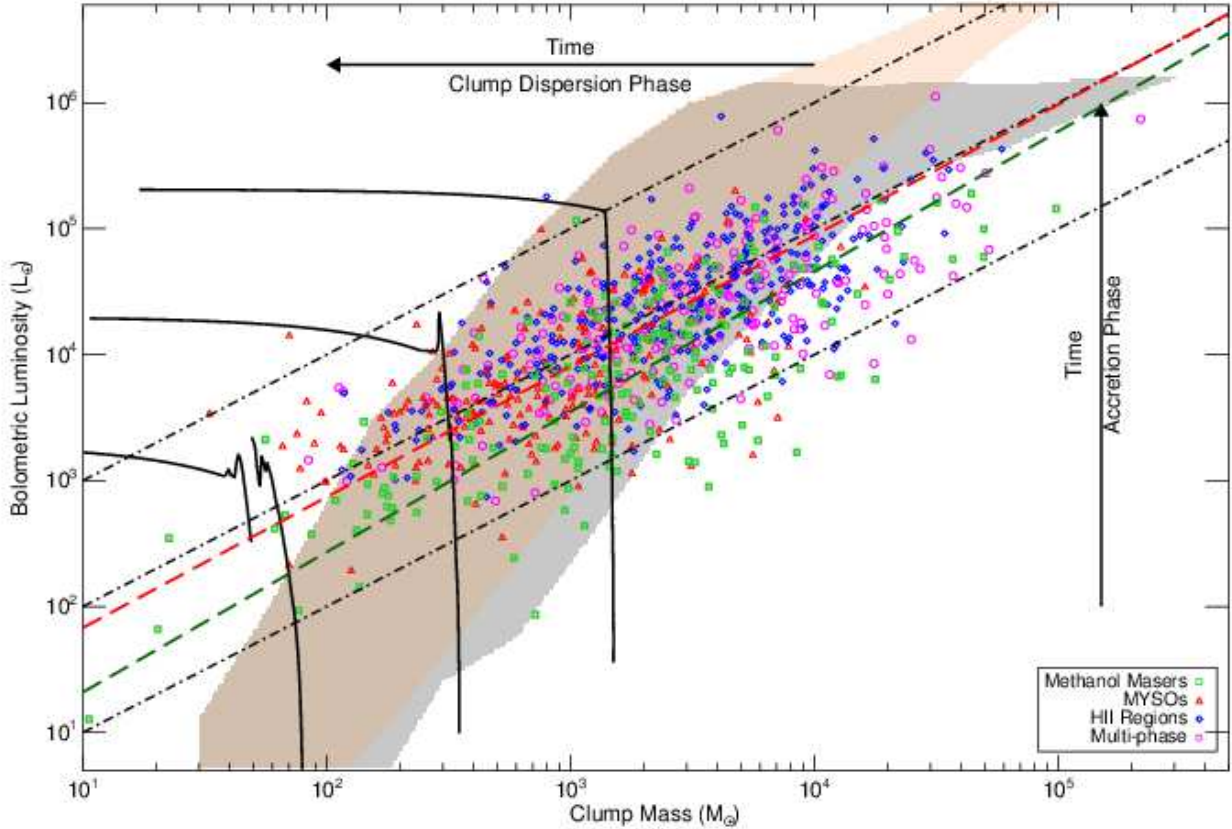}

\caption{\label{fig:mass_relations} Clump mass-bolometric luminosity relationship for all MSF clumps and their associated subsamples. The orange shaded region indicates the area in which 90\,per\,cent of the Monte-carlo simulations of ZAMS stellar clusters are located, while the region highlighted in grey indicates the area where the most massive star of each {simulated} cluster is located. The lower, middle and upper diagonal dash-dot lines indicate the \lbol/\mclump\,=\,1, 10 and 100\,\lsun/\msun, respectively. The solid black curves, running respectively left to right,  show the model evolutionary tracks calculated by \citet{molinari2008} for stars with final masses of 6.5, 13.5 and 35\,\msun, respectively. The long-dashed diagonal red and dark green lines show the results of a log-log, outlier-resistant, linear fit to the MYSO and \hii-region associated clumps and methanol-maser associated clumps, respectively.}

\end{center}
\end{figure*}

\begin{figure}
\begin{center}

\includegraphics[width=0.49\textwidth, trim= 0 0 0 0]{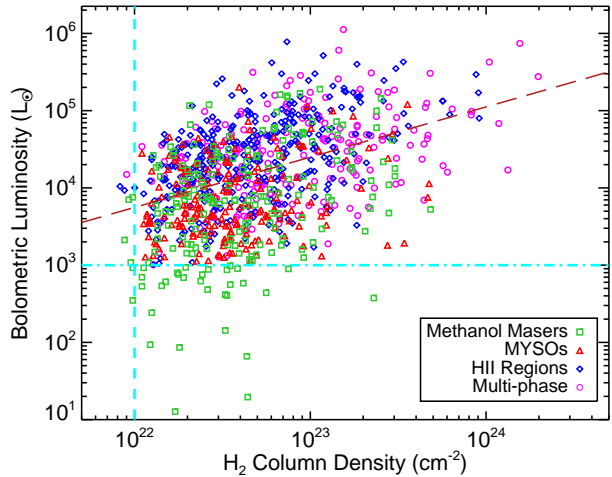}

\caption{\label{fig:lum_col_den} Comparison of peak column density and bolometric luminosities for the four subsamples of massive star-forming clumps. The dashed vertical and dash-dotted horizontal cyan lines indicate the nominal column density sensitivity of the ATLASGAL survey ($\sim$10$^{22}$\,cm$^2$) and the luminosity section cut applied to the RMS sources ($>$1000\,\lsun). The long-dashed red line shows the result of a linear fit to the data.} 

\end{center}
\end{figure}

In this subsection we compare the clump masses and the bolometric luminosities of the embedded proto-clusters and examine the relationship between the physical properties of the clumps and their associated star formation. Mass-luminosity (\mclump-\lbol) diagrams have previously been used in studies of low-mass (\citealt{saraceno1996}) and high-mass star-forming regimes (\citealt{molinari2008, giannetti2013}) and are taken to be a useful diagnostic tool with which to separate different evolutionary stages. Fig.\,\ref{fig:mass_relations} shows the mass-luminosity distribution of all associated clumps for which a distance and luminosity are available. We use different colours and symbols to show the distributions of the methanol-maser, MYSO and \hii-region associated clouds and those associated with multiple evolutionary stages (green, red, blue and magenta, respectively). In cases where a clump is associated with multiple RMS sources we have integrated the luminosity of each embedded source to obtain a total bolometric luminosity for each clump. In Fig.\,\ref{fig:lum_col_den} we present a similar plot showing the correlation between luminosity and column density. It is clear from this plot and the \mclump-\lbol\ diagram in Fig.\,\ref{fig:mass_relations} that the most massive stars are forming in the most massive and dense clumps.

\citet{molinari2008} developed a simple model for the formation of massive stars based on the observational evidence that their formation is a scaled-up version of the inside-out collapse model that has been successfully applied to low-mass star formation. Using the turbulent core-collapse model with accelerating accretion rates proposed by \citet{mckee2003} as their starting point, they calculated how the clump mass and source bolometric luminosity should change as the protostar evolves, producing evolutionary tracks for stars with final masses between $\sim$6 and 35\,\msun. These tracks consist of a vertical and horizontal component that \citet{molinari2008} refer to as the \emph{main accretion} and the \emph{envelope clean-up} phases, which we will also adopt here. Note that the efficiency with which the clump mass is converted into stellar mass determines the position of the apex in mass-luminosity space, but is not predicted by the model and are included for illustrative purposes only and no attempt is made to constrain the models.

Assuming that the accretion rate increases with the mass of the protostar (as favoured by the population synthesis model; \citealt{davies2011}), and that this accelerates towards the end of its vertical track it will reach the end of its main accretion phase with increasing rapidity. However, once the main accretion phase has ended it takes a long time before the embedded star begins to disrupt its natal clump sufficiently to make much progress along its horizontal track. A consequence of this is that a source will spend a relatively large proportion of its embedded lifetime at the apex where these two tracks intersect and we therefore expect to find that a significant fraction of embedded sources will collect along the locus of apices. 

The MYSO, \hii\ region and the multi-phase subsamples all combine to form a continuous distribution that is concentrated towards the \lbol/\mclump\ = 10\,\lsun/\msun\ locus that extends over approximately 3 orders of magnitude in both axes. There is a broad spread in the \lbol/\mclump\  values but the upper and lower envelopes of this range are well fitted by value of 100 and 1\,\lsun/\msun, respectively. These three lines of constant \lbol/\mclump\  (i.e., 1, 10 and 100\,\lsun/\msun) are overlaid on Fig.\,\ref{fig:mass_relations}. We evaluate the correlation between the luminosity and clump masses of these three subsamples using a partial Spearman correlation test and obtain a correlation coefficient of $r=0.64$ with a significance of $t \ll 0.001$. The strong correlation observed is therefore statistically significant and not the result of distance-related biases. A log-log linear regression fit to these data has a slope of $1.03\pm0.05$. To check that the slope is not affected by a distance bias we repeated the fit for the distance-limited sample and obtained a slope of $1.09\pm0.13$. These values are slightly lower but consistent with the slope reported by \citet{molinari2008} ($\sim$1.27) from a similar fit to a sample of 27 infrared-bright sources and consistent with the results of a similar analysis presented by \citet{giannetti2013}. Our analysis builds on the results of these previous studies confirming their main findings and showing this correlation holds for masses and luminosities an order of magnitude larger than previously considered.
 
The methanol-maser associated sample covers a similar range of masses as the MYSO, \hii-region and multi-phase subsamples, however, there is a clear offset to lower bolometric luminosities for the same clump mass than for the other three subsamples. On average, the methanol-maser associated clumps are approximately a factor of two less luminous than the other associated clumps. The distribution is also significantly more scattered. However, the mass and luminosity of these clumps are still correlated with each other although not as strongly ($r=0.58$ with a significance of $t\ll 0.001$). From the log-log linear fit to these parameters we determine the slope of $1.06\pm0.07$, which is consistent with the slope found for the MYSOs and \hii\ regions.

\subsubsection{Evidence for an evolutionary sequence?}

\setlength{\tabcolsep}{6pt}
\begin{table*}


%

\begin{center}\caption{Summary of the MSF clump parameters for sources located between $280\degr < \ell < 350\degr$ and $10\degr < \ell < 60\degr$.}
\label{tbl:suumary_of_derived_clump_para}
\begin{minipage}{\linewidth}
\small
\begin{tabular}{lrcrcrcrcr}
\hline \hline
  \multicolumn{1}{c}{Association}&  \multicolumn{1}{c}{Clump}&	\multicolumn{2}{c}{Distance}  &\multicolumn{2}{c}{Log[$L_{\rm{bol}}$]}&\multicolumn{2}{c}{Log[$M_{\rm{clump}}$]}  &\multicolumn{2}{c}{$L_{\rm{bol}}$/$M_{\rm{clump}}$}  \\
  
    \multicolumn{1}{c}{Type }&  \multicolumn{1}{c}{\#}&  \multicolumn{1}{c}{(kpc)} & \multicolumn{1}{c}{\#}  &\multicolumn{1}{c}{(\lsun)}  &\multicolumn{1}{c}{\#}  &\multicolumn{1}{c}{(\msun) }  &	\multicolumn{1}{c}{\#}&\multicolumn{1}{c}{Mean(\lsun/\msun)}& \multicolumn{1}{c}{\#} \\
\hline
Methanol masers 	&	314	&	$6.8\pm0.2$	&	298	&	$4.3\pm0.3$	&	197	&	$3.7\pm0.2$	&	269	&	$6.8\pm0.7$	&	195	\\
MYSO				&	210	&	$5.6\pm0.2$	&	210	&	$4.2\pm0.3$	&	210	&	$3.2\pm0.2$	&	210	&	$15.7\pm1.3$	&	210	\\
\hii\ regions		&	375	&	$8.2\pm0.2$	&	373	&	$4.8\pm0.3$	&	343	&	$3.7\pm0.2$	&	373	&	$16.4\pm1.1$	&	343	\\
Multi-phase & 231	&	$7.2\pm0.3$	&	230	&	$5.0\pm0.4$	&	219	&	$3.9\pm0.3$	&	230	&	$15.4\pm1.1$	&	219	\\
\hline
All MSF clumps &1130	&	$7.1\pm0.1$	&	1111	&	$4.8\pm0.2$	&	772	&	$3.7\pm0.1$	&	1082	&	$16.0\pm0.7$	&	772	\\
\hline\\
\end{tabular}\\
\end{minipage}

\end{center}
\end{table*}
\setlength{\tabcolsep}{6pt}

In this section we explore the hypothesis that the methanol masers, MYSOs and \hii\ regions represent distinct evolutionary stages of massive star formation (as outlined in Sect.\,1.1). We will put to one side the multi-phase clumps as these contain multiple evolutionary signatures and therefore cannot provide any insight into changes in the clump properties as the formation procedes. As previously mentioned, the luminosity-to-mass ratio has been proposed as a diagnostic tool for distinguishing between different evolutionary states and therefore, if the methanol masers, MYSOs and \hii\ regions are distinct stages we should expect to find a trend of increasing \lbol/\mclump\ ratios for these subsamples. 

\begin{figure}
\begin{center}

\includegraphics[width=0.49\textwidth, trim= 0 0 0 0]{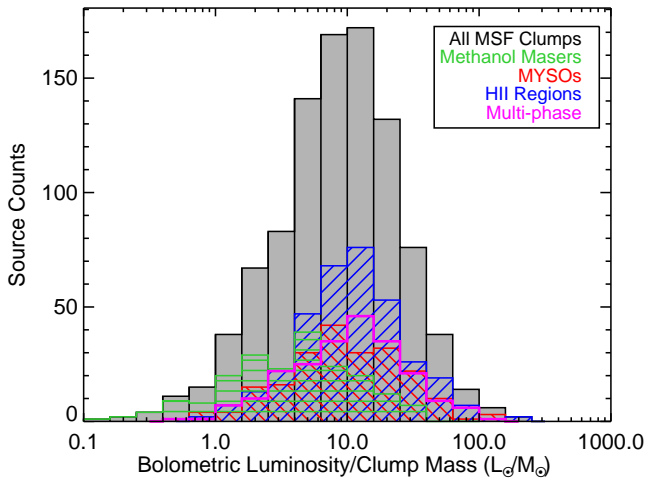}

\caption{\label{fig:lum_mass_histo} Ratio of total bolometric luminosity (integrated luminosity of all objects embedded in the clumps)  to clump mass for all ATLASGAL clumps associated a methanol maser, MYSO or \hii\ region (filled grey histogram). Green, red and blue hatched histograms show the \lbol/\mclump\ ratio distributions of the methanol-maser, MYSO and \hii-region associated clumps, respectively. The bin size is 0.2\,dex.} 

\end{center}
\end{figure}

Fig.\,\ref{fig:lum_mass_histo} shows the distribution of luminosity-mass ratios for the whole sample of MSF clumps and for the clumps associated with the methanol maser, MYSO and \hii\ region subsamples and present a summary of their properties in Table\,\ref{tbl:summary_of_derived_clump_para}. There is a clear difference between the methanol-maser associated subsample and the MYSO and \hii-region associated clumps. The latter have almost the same distribution in \lbol/\mclump\ with approximately the same mean value ($\sim$16\,\lsun/\msun). The \lbol/\mclump\ distribution for the methanol masers is shifted to lower values with a significantly lower mean of $\sim$7\,\lsun/\msun.  Methanol masers are widely thought to trace an earlier stage in the massive star-formation process than the mid-infrared-bright MYSOs and \hii\ regions and their significantly lower \lbol/\mclump\ ratio is consistent with this hypothesis. 

Differences in the \lbol/\mclump\ ratio for \hii\ regions and IR-selected HMPOs have been previously reported in the literature (e.g., \citealt{sridharan2002}).  However, no clear difference between the MYSO and \hii-region associated subsamples is seen in Fig.\,\ref{fig:lum_mass_histo} and the KS test is unable to reject the null hypothesis that they are drawn from the same parent population ($r=0.026$). The similarities in the \lbol/\mclump\ distributions suggest that these two subsamples cover a similar range of evolutionary stages. As previously noted, the MYSOs and \hii\ regions are concentrated around the \lbol/\mclump\,=\, 10\,\lsun/\msun\ locus, which, according to the scenario outlined by \citet{molinari2008}, is likely to correspond to the apex between the main accretion and envelope clean-up phases. This would suggest that the majority of the MYSOs and \hii\ regions are nearing the end of their main accretion phase, or that the accretion has already been halted. \citet{davies2011} arrived at a similar conclusion independently by modeling the Galactic star-formation rate from the RMS source counts. They found that an intermediate radio-quiet phase was required between the end of the main accretion phase and the formation of the \hii\ region to obtain a satisfactory qualitative fit to the lower-luminosity MYSO distribution. This may also provide an explanation for the relatively low association rate of MYSOs and \hii\ regions with methanol masers that are thought to be radiatively pumped in the accretion disc ($\sim$30 and 10\,per\,cent for MYSOs and \hii\ regions, respectively; see Sect.\,3.3). 

Although the MYSOs and \hii\ regions have similar \lbol/\mclump\ ratios and form a continuous distribution in the M-L diagram (Fig.\,\ref{fig:mass_relations}), the MYSOs dominate the lower clump-mass and luminosity end of the parameter space while the \hii\ regions occupy the higher end of both parameter ranges. However, there is no obvious break in the combined distribution of the two subsamples that might be expected if the luminosities of the two groups originated in different physical processes.

In previous sections we have found that the MYSO population covers a lower luminosity range than the \hii\ regions and, since the most massive clumps tend to be located at larger distances, the lower number of MYSOs found to be associated with the more massive clumps may simply reflect this. In this case we might expect there to be a population of massive clumps associated with MYSOs that have yet to be found, however, there are a few points that limit this possibility. First, comparing distance-limited subsamples we still find that the MYSOs are associated with clumps that have significantly lower mass and column density clumps than the \hii\ regions, which suggests that the difference is not a selection effect. Second, in Sect.\,\ref{sect:col_den} we found that there are practically no high column-density clumps in the Galactic disc that have not already been associated with massive star formation. We also found a strong positive correlation between clump mass and peak column density with more massive clumps being associated with high column densities (see Fig.\,\ref{fig:mass_col_den}). We are therefore also very likely to already have identified the vast majority if not all of the most massive (i.e., $M_{\rm{clump}} \gtrsim 1000$\,\msun\ out to 20\,kpc), compact clumps  located within the inner Galactic disc, which means there is limited scope for the existence of large population of MYSOs embedded in more massive clumps in this part of the Galaxy. Finally, we find a similar correlation and slope  to that in Fig.\ \ref{fig:mass_relations} if we consider the distance-limited samples of MYSOs and \hii\ regions (the slope is 1.08$\pm$0.13 and correlation coefficient is 0.63) and so this is unlikely to be due to limited flux sensitivity.

There are no MYSOs associated with very massive clumps, however, it is likely that a significant number of the methanol masers found to be associated with massive clumps are actually MYSOs given that $\sim$20\,per\,cent of methanol masers are coincident with an MYSO ($<$2\arcsec; as discussed in Sect.\,\ref{sect:multiplicity}). Therefore a number of the more distant methanol maser sources are likely to be tracing luminous MYSOs that fall below the MSX 21\,\mum\ detection threshold. However, even taking this into account it does not change the differences in the overall distribution of the MYSO and \hii\ region subsamples seen in Fig.\,\ref{fig:mass_relations}, which are still present in the distance-limited analysis. Furthermore, we have found significant differences in the clump mass, column density and kinetic temperature of clumps associated with MYSOs and \hii\ regions with the former having lower values for all parameters. There are clearly some differences between the clump properties of these two subsamples.

The simplest explanation is that the massive stars forming in the most massive clumps evolve much faster than those embedded in lower mass clumps, which is consistent with the decreasing timescale with increasing MYSO luminosity found by \citet{mottram2011b}. In this case, the lack of very luminous MYSOs associated with the most massive clumps would be because such sources evolve into an \hii\ region so quickly that we are unlikely to observe the MYSO stage. This would also explain why the methanol maser associated clumps appear to have similar intermediate properties to both the \hii\ region and MYSO associated clumps, if they are tracing an earlier evolutionary stage of both the MYSOs and \hii\ regions. 

It is clear from the strong correlations between bolometric luminosity and both clump mass and column density that the most massive stars are forming in the most massive and dense clumps. \citet{mckee2003} derived a formula for the accretion rate as a function of time for a free-falling envelope:

\begin{equation}
\dot{m}_\star \; = \; 4.6 \times 10^{-4} \left(\frac{m_{\star\rm{f}}}{30\,{\rm{M_\odot}}}\right)^{3/4} \Sigma_{\rm{clump}}^{3/4} \left(\frac{m_\star}{m_{\star\rm{f}}}\right)^{0.5} \; \left(\frac{{\rm{M_\odot}}}{{\rm{yr}}}\right)
\end{equation}

\noindent where $\dot{m}_\star$ is the time-dependent accretion rate, $m_\star$ and $m_{\star\rm{f}}$ are the current and final mass of the star in \msun, respectively, and $\Sigma_{\rm{clump}}$\, is the clump surface density in g\,cm$^{-2}$. For a particular $m_{\star\rm{f}}$ the accretion rate is proportional to $\Sigma_{\rm{clump}}^{3/4}$ and therefore stars forming in the most massive and dense clumps will accrete mass more rapidly than stars forming in a lower mass clump and will arrive on the main sequence much sooner. For example, if we consider two stars with the same final mass but forming in clumps where the surface density is different by an order of magnitude. In this case the star in the higher surface density clump will arrive on the main sequence ($\sim$30\,\msun) a factor of 3--5 faster than the one embedded in the lower surface density clump. 

Differences of {bf two orders} of magnitude in the peak surface densities are seen in our data (i.e., Fig.\,\ref{fig:col_den}) and therefore such large variations in timescale seem plausible. Furthermore, the lower mass clumps appear to be forming less massive stars and since the accretion rate is also dependent on the final mass of the star, the difference in the time taken to arrive on the main sequence is likely to be significantly longer than the factor of 3-5 estimated for stars evolving in the lower mass clumps. 

The faster evolution times for stars forming in the most massive and dense clumps means they arrive on the main sequence and form an \hii\ region in a fraction of the time taken for less massive stars forming in lower-mass clumps. The lifetime of the MYSO stage is therefore anti-correlated with clump mass. This nicely explains all of the differences seen in the properties of the MYSOs and  \hii-region associated clumps and the similarities of both of these subsamples with those of the methanol-maser associated clumps.

An important implication is that the MYSO stage for the most massive stars may be so short due to fast accretion rates and very high extinction that we are unlikely to observe it. All of the MYSOs identified will go on to form \hii\ regions, however, and if the Molinari model is correct then they are already near the end of their main accretion phase and their final luminosity is unlikely to increase significantly from its current value. This sample of MYSOs are therefore  unlikely to be the progenitors of the most massive early O-type stars. However, there are still a significant number of MYSOs associated with the multi-phase clumps that have not been considered in the comparison presented here. Some of these multi-phase clumps are also among some of the most luminous and massive in the Galactic disc and may host MYSOs that are possible precursors to early O-type stars. There are also a number of very massive clumps that are associated with methanol masers that might also contain the earliest stages of the most massive stars in the Galaxy. Higher-resolution interferometric observations of these MYSOs and methanol masers are required to investigate these potential early O-type star precursors in more detail. 

\subsubsection{Comparison with expected cluster luminosity}
 
The orange and grey shaded regions shown in Fig.\,\ref{fig:mass_relations} indicate the possible range of bolometric luminosities of an embedded stellar cluster and of its most massive star, respectively, forming in a clump of a given mass. These have been generated using a Monte-Carlo simulation to randomly sample a standard IMF (\citealt{kroupa2001}) with a stellar-mass range of 0.1 to 120\,\msun\ and assuming ZAMS luminosities and a star-formation efficiency (SFE) of 10\,per\,cent (see Paper\,II for a more detailed description of the model). 

Comparing the measured source luminosities to the model values, we find good agreement for MYSOs and \hii\ regions associated with clumps with masses between $\sim$100 and a few thousand \msun. However, the match is significantly poorer for clumps above and below this range, with the lower-mass clumps observed to have excess luminosity and the higher-mass clumps being under-luminous compared to the model predictions. A similar disparity was seen by \citet{sridharan2002} from their analysis of HMPOs. This would suggest that the instantaneous SFE decreases as a function of increasing clump mass and is much lower for the more massive clumps than the 10\,per\,cent SFE assumed in our model. In previous studies, low SFE has been associated with the least evolved clusters and is thought to increase as it evolves with an upper limit of roughly 30\,per\,cent (\citealt{lada2003} and references therein). 

If we assume that the low- and high-star formation in a given cluster is coeval, then we would expect the most massive stars to evolve into \hii\ regions on short timescales but that the rest of the associated cluster members, particularly the lower-mass stars, will take much longer before they begin to make a significant contribution to the total cluster luminosity. The higher accretion rates for the more massive stars results in a similar formation time for the whole cluster regardless of the large range of masses of the individual stellar members; approximately $3\pm1 \times 10^5$\,yr (\citealt{offner2011}).  However, as previously mentioned the most massive stars in the cluster reach the main sequence and form an \hii\ region while still deeply embedded and continue to accrete material for a significant period before attaining their final mass. The MYSO lifetime for the most massive stars ($>$30\,\msun) is estimated to be of order $\sim1 \times 10^5$\,yr after which they join the main sequence and begin to form a \hii\ region (\citealt{mottram2011a,davies2011}). 

So although the overall formation time is similar for all cluster members the most massive star is likely to reach the main sequence significantly ahead of the lower mass stars and dominate the cluster luminosity. For a typical cluster consisting of 1500\,\msun\ of stars we would expect to find 1$\times$60\,\msun, 1-2$\times$30\,\msun, 5-10$\times$15\,\msun, 10-20\,\msun\ with the rest of the mass residing in low- and intermediate mass stars (cf. Table\,2 of \citealt{zinnecker2007}). When fully formed approximately 50\,per\,cent of the cluster luminosity will be generated by the most massive 60\,\msun\ star, with majority of the other high-mass stars contributing the majority of the rest of the cluster's luminosity. If we now consider their contribution to the cluster's luminosity at the time the most massive star joins the main sequence (i.e., when it has a stellar mass of $\sim$30\,\msun). At this stage, due to the flattening off of the $L/M$ ratio with increasing stellar mass, the most massive star is producing approximately a third of its final mass luminosity, however, since luminosity is a strong function of stellar mass the rest of the clusters massive star population are only producing approximately 10\,per\,cent of their final mass luminosity. As a consequence, at the time the most massive star joins the main sequence it is contributing up to 75\,per\,cent of the cluster's total luminosity and result in a lower SFE than would otherwise be expected.

We also find that the more massive stars are forming in the most massive clumps and are therefore likely to join the main sequence faster than less massive stars forming in lower mass clumps. The difference between the time taken for the most massive stars to reach the main sequence and cluster formation will be larger for the most massive stars (i.e., clusters are less evolved). Since the most massive stars are forming in the most massive clumps we would expect the SFE to decrease with increasing clump mass, because the contribution from the lower mass cluster members is likely to be much smaller (as discussed in the previous paragraph). Furthermore, the cluster formation times are estimated assuming that the surface density is constant for a give cluster, however, this is unlikely to be true as there is likely to be strong density gradients. We have already found that the most massive stars are forming in the highest column density regions towards the centre of the clumps where the gravitational potential well is deepest. The majority of the lower mass stars are therefore likely to be forming in lower density regions, which would lead to longer formation times and further lower the measured SFE. 

The difference in cluster formation timescale and the time taken for the most massive stars to join the main sequence would explain the low SFE seen in our data. Support for this hypothesis comes from the fact that, although the luminosities measured for the most massive clumps fall well short of the expected ZAMS cluster luminosities, however,  there is a better match between the expected luminosities of the most massive stars in a given cluster and the measured bolometric luminosity. Furthermore, in Paper\,II we found that there is a good correlation between the bolometric luminosities and the luminosities derived independently from the measured radio-continuum emission from the compact \hii\ regions. Since the ionizing flux traced by the radio emission in a cluster is totally dominated by the most massive star, it would strongly suggest that the bolometric luminosity measured for the most massive clumps is also dominated by the most massive stars and that the clusters are less evolved and the rest of the lower mass members of the cluster are yet to make a significant contribution. Conversely, the lower mass clumps have the highest SFE and are much less strongly bound, which may indicate that these are much more evolved. 

These results therefore suggest a scenario where the most massive stars form towards the centre of very massive gravitationally unstable clumps with the lower mass stars forming either coevally or at later times as the proto-cluster is fed from the massive globally collapsing clump. However, more detailed modeling of the early stages of cluster evolution are need to investigate this further; this work is ongoing and will be reported in a subsequent publication.   

\subsection{Mass-Radius relation}

\begin{figure*}
\begin{center}
\includegraphics[width=0.99\textwidth, trim= 0 0 0 0]{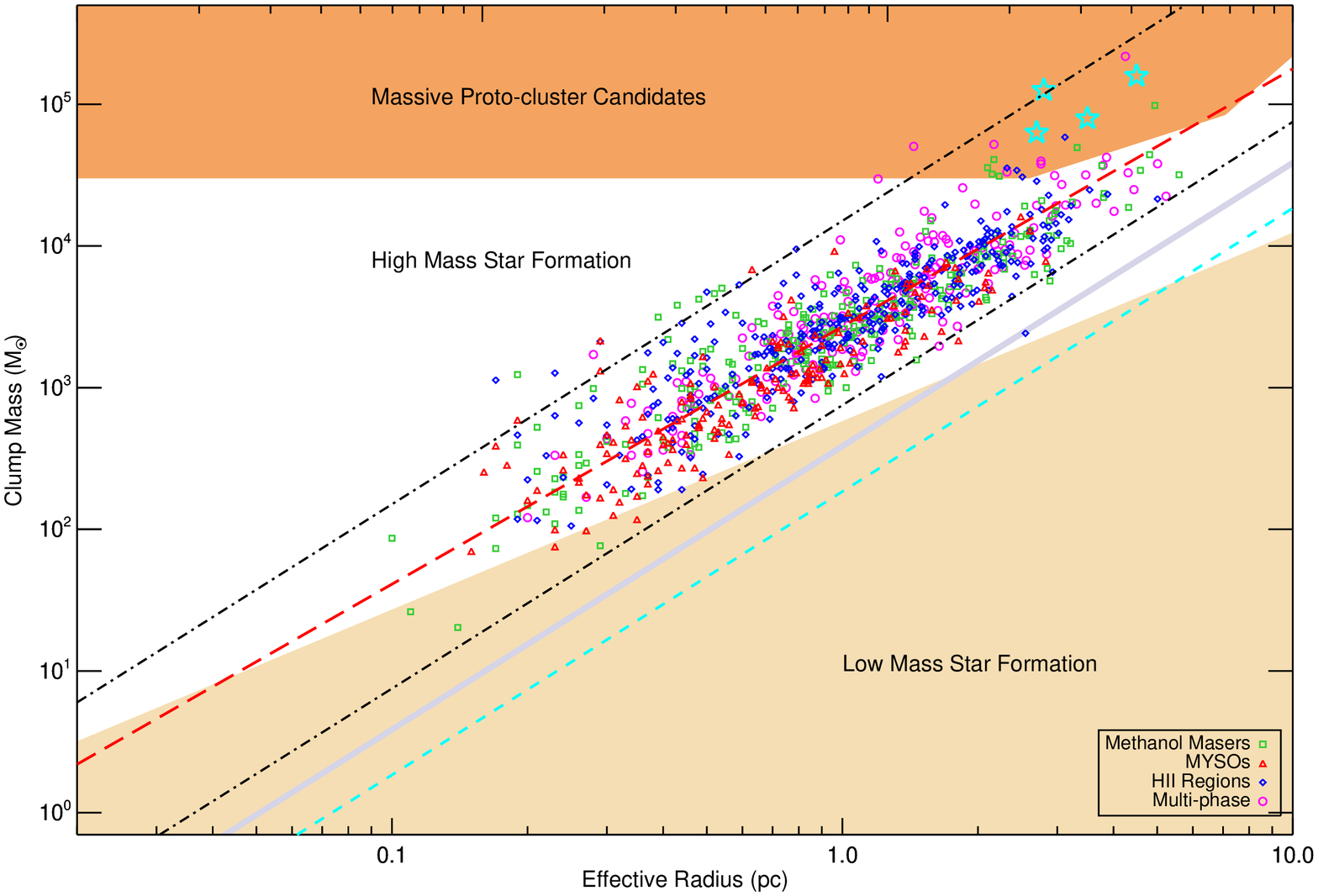}

\caption{\label{fig:mass_radius_distribution} The mass-size relationship of ATLASGAL clumps associated with methanol masers, MYSOs and \hii\ regions. The colours and symbols are explained in the legend, with the exception of the cyan stars which indicate the distribution of the MPC candidates found towards the Galactic centre (\citealt{longmore2012a,immer2012}). The beige shaded region shows the part of the parameter space found to be devoid of massive star formation that satisfies the relationship $m(r) \le 580$\,\msun\, $(R_{\rm{eff}}/{\rm{pc}})^{1.33}$ (cf. \citealt{kauffmann2010c}). The orange shaded area towards the top of the diagram indicates the region of parameter space where young massive cluster progenitors are expected to be found (i.e., \citealt{bressert2012}). The long-dashed red line shows the result of a linear power-law fit to the whole sample of associated clumps. The dashed cyan line shows the sensitivity of the ATLASGAL survey ($N_{\rm{H_2}} \sim 10^{22}$\,cm$^{-2}$) and the upper and lower dot-dashed lines mark surface densities of 1\,g\,cm$^{-2}$ and 0.05\,g\,cm$^{-2}$, respectively. The diagonal light blue band fills the gas surface density ($\Sigma(\rm{gas})$) parameter space between 116-129\,\msun\,pc$^{-2}$ suggested by \citet{lada2010} and \citet{heiderman2010}, respectively, to be the threshold for ``efficient'' star formation.}

\end{center}
\end{figure*}

In Fig.\,\ref{fig:mass_radius_distribution} we present a mass-radius (\mclump-\reff) diagram of the whole sample of ATLASGAL sources for which a distance has been determined. We presented similar diagrams in Papers\,I and II and found that the methanol-maser and \hii-region associated samples were strongly correlated with each other. Here we build on that work with increased sample sizes for the latter and inclusion of the MYSO-associated clumps (the combined sample consists of $\sim$1000 massive star-forming clumps). As seen in the two previous papers, there is a strong correlation between these parameters. Again using a partial Spearman correlation test to remove any dependence of the correlation on distance yields a coefficient value of 0.85 with a $t$-value $\ll 0.001$. We fit these parameters using a power-law which yields log($M_{\rm{clump}}) = 3.42\pm0.01 + (1.67\pm0.025)\times {\rm{Log}}(R_{\rm{eff}}$); the long-dashed red line shows the fit to these data. The fit agrees within the uncertainty with those determined in Papers\,I and II  and we find no significant differences between the different subsamples, which all combine to form a continuous distribution over two orders of magnitude in radius and almost 4 orders of magnitude in clump mass. Interestingly, the slope is similar to that found for cluster mass as a function of radius as determined by \citet{pfalzner2011} (i.e., $1.71\pm0.07$) suggesting a fairly constant SFE regardless of Galactic location and environment, or clump mass ($\sim$15\,per\,cent).

In our earlier work we found nearly all of our associated sources were located within a relatively narrow range of surface densities and this is also the case for the larger and more complete data set presented here. The diagonal dashed-dotted lines overlaid on Fig.\,\ref{fig:mass_radius_distribution} trace the lines of constant surface density, $\Sigma({\rm{gas}})$, of 1\,g\,cm$^{-2}$ and 0.05\,g\,cm$^{-2}$, respectively; these provide a reasonable empirical fit to the upper and lower range of our sample of MSF clumps. We also show the minimum threshold of $\sim$116-129\,\msun\,pc$^{-2}$ proposed by \citet{lada2010} and \citet{heiderman2010} (hereafter LH threshold) for ``efficient'' star formation in nearby molecular clouds (d $\le 500$\,pc); this range is indicated by the thick light blue band on the \mclump-\reff\ diagram, however, the 0.05\,g\,cm$^{-2}$ that effectively traces the lower envelope of the ATLASGAL clumps is approximately twice the LH threshold. This suggests that a surface density of 0.05\,g\,cm$^{-2}$ might better represent a minimum threshold of ``efficient'' massive star formation. 

The shaded regions shown in the lower-right part of  Fig.\,\ref{fig:mass_radius_distribution} correspond to the regions of parameter space where massive stars are not typically found (empirically determined by \citealt{kauffmann2010b}). Comparing the distribution of our sources with the \citet{kauffmann2010b} threshold (i.e., $m(r) \ge 580$\,\msun\ $(R_{\rm{eff}}/{\rm{pc}})^{1.33}$)\footnote{Note we have reduced the value of the mass coefficient given by \citet{kauffmann2010b} by 1.5 to match the dust opacity value used in this paper to estimate clump masses} we find good agreement with all but a handful of sources located below this level, however, we note that the empirical fit to the lower envelope of our distribution (i.e., 0.05\,g\,cm$^{-2}$) provides a tighter constraint for clumps larger than $\sim$0.5\,pc and more massive than 500\,\msun.

\subsubsection{Massive proto-cluster candidates}
\label{sect:mpc_candidates}

\setlength{\tabcolsep}{3pt}

\begin{table*}

\begin{center}\caption{Derived parameters for massive proto-cluster (MPC) candidates. Masses have been estimated assuming a dust temperature of 20\,K for all sources except G002.53+00.016 for which a range of temperatures between 19 and 27\,K estimated from SED fits to individual pixels across the clump (see \citealt{longmore2012a} for details).}
\label{tbl:MPC_derived_para}
\begin{minipage}{\linewidth}
\begin{tabular}{lcc......c}
\hline \hline
  \multicolumn{1}{c}{ATLASGAL}&  \multicolumn{1}{c}{Complex}&\multicolumn{1}{c}{Association}&	\multicolumn{1}{c}{Distance}&\multicolumn{1}{c}{\reff}  &	\multicolumn{1}{c}{Log[\mclump]} & \multicolumn{1}{c}{Log[N(H$_2$)]}& \multicolumn{1}{c}{Log[\lbol]} & \multicolumn{1}{c}{\lbol/\mclump} & \multicolumn{1}{c}{Reference}\\

  \multicolumn{1}{c}{name}& \multicolumn{1}{c}{name}&\multicolumn{1}{c}{type}& \multicolumn{1}{c}{(kpc)}&	\multicolumn{1}{c}{(pc)}  &	\multicolumn{1}{c}{(\msun)}&	\multicolumn{1}{c}{(cm$^{-2}$)}&\multicolumn{1}{c}{(\lsun)}&\multicolumn{1}{c}{(\lsun/\msun)}&\\
 \hline
 \multicolumn{10}{c}{Galactic Disc}\\
  \hline
AGAL010.472+00.027	&	GAL 010.46+00.02	&	MM/HII	&	8.6	&	2.3	&	4.55	&	23.94	&	5.54	&	9.73	&	1,2,4	\\
AGAL012.208$-$00.102	&	\multicolumn{1}{c}{$\cdots$}	&	MM/HII	&	13.6	&	2.8	&	4.58	&	23.46	&	\multicolumn{1}{c}{$\cdots$}	&	\multicolumn{1}{c}{$\cdots$}	&	1,4$^a$	\\
AGAL019.609$-$00.234	&	\multicolumn{1}{c}{$\cdots$}	&	MM/HII	&	12.7	&	2.8	&	4.60	&	23.68	&	5.66	&	11.54	&	1,3,4$^a$	\\
AGAL032.797+00.191	&	G32.80	&	HII	&	13.3	&	2.5	&	4.49	&	23.49	&	5.55	&	11.65	&	1,4$^a$	\\
AGAL043.148+00.014	&	W49A	&	MM/HII	&	11.1	&	2.3	&	4.52	&	23.16	&	5.62	&	12.63	&	1,3	\\
AGAL043.164$-$00.029	&	W49A	&	HII	&	11.1	&	3.1	&	4.77	&	23.53	&	5.68	&	8.22	&	1,3,4	\\
AGAL043.166+00.011	&	W49A	&	MM/HII	&	11.1	&	4.2	&	5.34	&	24.19	&	6.04	&	5.04	&	1,3,4	\\
AGAL043.178$-$00.011	&	W49A	&	MM	&	11.1	&	2.2	&	4.51	&	23.16	&	\multicolumn{1}{c}{$\cdots$}	&	\multicolumn{1}{c}{$\cdots$}	&	1	\\
AGAL049.472$-$00.367	&	W51 (A\&B)	&	MM	&	5.4	&	2.2	&	4.61	&	23.67	&	\multicolumn{1}{c}{$\cdots$}	&	\multicolumn{1}{c}{$\cdots$}	&	1,3,4	\\
AGAL049.482$-$00.402	&	W51 (A\&B)	&	MM	&	5.4	&	2.2	&	4.49	&	23.59	&	\multicolumn{1}{c}{$\cdots$}	&	\multicolumn{1}{c}{$\cdots$}	&	1	\\
AGAL049.489$-$00.389	&	W51 (A\&B)	&	MM/HII	&	5.4	&	1.4	&	4.70	&	24.30	&	5.57	&	7.33	&	1,3,4	\\
AGAL328.236$-$00.547	&	\multicolumn{1}{c}{$\cdots$}	&	MM	&	11.4	&	4.9	&	4.99	&	23.40	&	5.16	&	1.47	&	1,2	\\
AGAL329.029$-$00.206$^\star$	&	\multicolumn{1}{c}{$\cdots$}	&	MM	&	12.0	&	3.3	&	4.69	&	23.38	&	4.78	&	1.21	&	1,2	\\
AGAL330.954$-$00.182	&	 GAL 331.03-00.15	&	MM/HII	&	9.3	&	2.2	&	4.72	&	24.07	&	5.75	&	10.84	&	1	\\
AGAL337.704$-$00.054	&	\multicolumn{1}{c}{$\cdots$}	&	MM/HII	&	12.3	&	2.4	&	4.53	&	23.53	&	5.16	&	4.21	&	1	\\
AGAL350.111+00.089	&	\multicolumn{1}{c}{$\cdots$}	&	MM	&	11.4	&	2.1	&	4.55	&	23.27	&	\multicolumn{1}{c}{$\cdots$}	&	\multicolumn{1}{c}{$\cdots$}	&	1,2	\\
\hline
\multicolumn{10}{c}{Galactic Centre}\\
  \hline
G000.253+00.016 & ``The Brick'' & Starless? & 8.4 & 2.8 & 5.1  &23.39&\multicolumn{1}{c}{$\cdots$}&\multicolumn{1}{c}{$\cdots$}& 5,6 \\
AGAL000.411+00.051 (d)$^b$ & \multicolumn{1}{c}{$\cdots$} & Starless? & 8.4 & 3.5 & 4.9  &23.23&\multicolumn{1}{c}{$\cdots$}&\multicolumn{1}{c}{$\cdots$}& 6 \\
AGAL000.476$-$00.007 (e)$^b$ & \multicolumn{1}{c}{$\cdots$} & Starless? & 8.4 & 4.5 &5.2 &23.53&\multicolumn{1}{c}{$\cdots$}&\multicolumn{1}{c}{$\cdots$}& 6 \\
AGAL000.494+00.019 (f)$^b$ & \multicolumn{1}{c}{$\cdots$} & MM & 8.4 & 2.7& 4.8   &23.26&\multicolumn{1}{c}{$\cdots$}&\multicolumn{1}{c}{$\cdots$}& 6 \\
\hline\\
\end{tabular}\\
References: (1) this work, (2) Paper\,I, (3) Paper\,II, (4) \citet{ginsburg2012}, (5) \citet{longmore2012a}, (6) \citet{immer2012} \\
Notes: $^a$ Identified by \citet{ginsburg2012} as a massive clump but the derived mass was slightly less than the $3\times 10^4$\,\msun\ required for inclusion in their sample of MPC candidates. $^b$ The source names have been taken from the ATLASGAL CSC (\citealt{contreras2013}) while the letters in parentheses are the nomenclature used by \citet{lis1999}. $\star$ The distance to this source is uncertain and should be used with caution. 

\end{minipage}

\end{center}
\end{table*}
 
\setlength{\tabcolsep}{6pt}

The shaded area towards the top of Fig.\,\ref{fig:mass_radius_distribution} indicates the region of the parameter space where massive proto-cluster (MPC) candidates are thought to be found  (see \citealt{bressert2012} for details). Comparing the upper end of the observed mass--radius distribution, we find 16 clumps satisfying the MPC criteria that are potential precursors to the next generation of young massive clusters (YMCs; i.e., \citealt{bressert2012}); these are broadly defined as having stellar masses $\gtrsim$10000\,\msun\ and younger than 100\,Myr (\citealt{longmore2012a} and references therein). The ATLASGAL region covers approximately 70\,per\,cent of the Galactic disc (assuming the star formation is distributed between 1-15\,kpc; \citealt{ginsburg2012}) and therefore this sample should include a large fraction of the total Galactic population of MPC candidates. We find similar numbers of MPC candidates in the northern and southern Galactic plane, which is to be expected since the division is arbitrary. Table\,\ref{tbl:MPC_derived_para} lists the derived properties for these ATLASGAL clumps and also includes four additional MPC candidates associated with the Galactic centre region identified by \citet{longmore2012a} and \citet{immer2012}. These objects are also plotted in Fig.\,\ref{fig:mass_radius_distribution} as cyan stars.

In Paper\,I, two additional clumps were identified as potential MPC candidates (i.e., AGAL351.774$-$00.537 and AGAL352.622$-$01.077).  However, the distances to these sources are no longer considered reliable and they have therefore been excluded from this updated version of the table. A maser velocity of 1.3\,\kms\ was used to determine the distance to AGAL351.774$-$00.537, placing it in the outer Galaxy, but the velocity from an ammonia measurement of $-3$\,\kms\ and correlation with \hi\ data suggest that a near distance is more likely; this source has been studied in detail by \citealt{leurini2007} and is thought to be located at a distance of 1.8\,kpc (\citealt{snell1990}). For AGAL352.622$-$01.077, a positive radial velocity would place this source in the outer Galaxy at a distance of $\sim$19\,kpc, however, this would also locate it too far from the Galactic mid-plane, and so a near distance is more likely in this case as well; this was also noted by \citet{green2011b}. 

The MPC candidates located in the northern Galactic plane have been previously identified by \citet{ginsburg2012} and so can be considered reasonably robust. We have checked the distances assigned for the southern Galactic-plane candidates and find them to be reliable, with the exception of AGAL329.029$-$00.206, where the distance looks a little less certain. Inspection of the mid-infrared images towards this clump reveals that it is associated with some infrared extinction, which may indicates an association with an infrared-dark cloud and therefore possibly located at the near kinematic distance.  \citet{green2011b} place this clump at the far distance with a high level of confidence and so we include it in Table\,\ref{tbl:MPC_derived_para}, with a footnote to indicate that its distance is less robust than the others.

In the previous section we used the \lbol/\mclump\ ratio as a means to evaluate the current evolutionary state of the embedded star formation, with higher values being associated with more evolved objects. However, this ratio can also be used as a proxy for the SFE of a given clump. Given that some of the clumps identified as MPC are associated with complexes previously described as ``mini-starbursts" (e.g., W49 and W51) it is somewhat surprising to find that their SFE values are rather less than spectacular. We might have expected these mini-starbursts to stand out from the general population of MSF clumps and have much higher \lbol/\mclump\ ratios. \citealt{moore2012} compared \lbol/\mco\ ratios for RMS sources matched with molecular clouds identified by the Galactic Ring Survey (GRS; \citealt{jackson2006}) and found a ratio for W51 similar to that estimated in this paper from the sub-millimetre continuum ($7.6\pm2.3$\,\lsun/\msun).  In W49, however, they found a value of $32\pm6$\,\lsun/\msun, approximately 4 times larger than the ratio derived from the dust emission. Taken at face value this implies that the W49 molecular clouds have a significantly larger fraction of mass in high-density clumps but that the SFE and YSO luminosity distribution within those clumps is not much different from that in other regions. 

The distribution of the Galactic-centre and disc samples occupy approximately the same region of parameter space in Fig.\,\ref{fig:mass_radius_distribution} and, comparing the clump properties in Table\,\ref{tbl:MPC_derived_para}, we find them to be very similar. The column densities and sizes are very closely matched but the Galactic-centre MPC candidates are marginally more massive.  However, a KS test comparing these three parameters reveals no significant differences ($r=0.02$, 0.23 and 0.01, respectively). In contrast, there is a big difference in the level of star-formation activity associated with the two samples, with very little evidence for \emph{any} star formation taking place within the Galactic centre MPC candidates (\citealt{longmore2012a,immer2012}), while all of the candidates in the disc are extremely active star-forming regions. 

We have identified our sample of massive star-forming clumps by matching the ATLASGAL survey results with catalogues of star-formation tracers and so it is not surprising that all of the MPC candidates that we have identified in the disc are star-forming. Two recent detailed studies by \citet{ginsburg2012} and \citet{tackenberg2012} searched for massive starless clumps in the BGPS and in a subsample of the ATLASGAL survey (i.e., $10\degr < \ell < 20\degr$) and, although \citet{ginsburg2012} identified a handful of MPC candidates (e.g., W43 and W49; see Table\,\ref{tbl:MPC_derived_para} for full list), they found none that are starless. Since a detailed study of all potential MPC candidates has not yet been made of the southern Galactic plane, similar to that performed by \citet{ginsburg2012} in the northern Galactic plane, it is possible that there are some starless MPC candidates located in the disc that have yet to be discovered, the number is likely to be very small and perhaps even zero. 

It is currently unclear why the observed level of star formation is so very different for the Galactic centre and disc populations, particularly since their clump properties are so similar. This would not be expected if the formation processes were so very different. The most significant difference is that the velocity dispersion is significantly larger and gas temperatures are higher for the Galactic-centre population (a factor or a few) compared to those in the disc. It seems more plausible that the formation mechanism is the same but that the much more extreme Galactic-centre environment is somehow impeding the star formation in its local MPC candidate population. The difference in star-formation activity has led to suggestions that the mode of star formation is fundamentally different in the Galactic centre compared to the disc (see \citealt{longmore2014} for a detailed discussion).

\section{Summary and conclusions}

This is the third in a series of papers that use the ATLASGAL survey to conduct a detailed and comprehensive investigation of massive star-formation environments. The main aim of these papers is to use the unbiased coverage and uniform sensitivity of the ATLASGAL survey to map the dust emission over the inner Galactic Plane to connect the results derived from samples selected using different high-mass star formation tracers. We have combined the ATLASGAL and RMS surveys to identify a large sample of massive, submillimetre-continuum-traced clumps associated with the formation of massive stars. These MYSO and \hii-region associated clumps are complemented with similar clumps identified through their association with either methanol masers or compact radio-continuum emission (for details see Papers\,I and II). We have matched the ATLASGAL sources with additional methanol masers identified outside the MMB region to obtain a more complete sample covering the whole inner Galactic plane. In total we have identified $\sim$1700 embedded massive stars with $\sim$1300 clumps, which is the largest sample of massive star-forming clumps yet compiled and includes a large fraction of all embedded massive stars in the Galaxy. This represents $\sim$15\,per\,cent of the whole catalogue of clumps identified by ATLASGAL in the regions in which the surveys are complete ($280\degr < \ell < 350\degr$  and $10\degr < \ell < 60\degr$).

Using distances drawn from the literature, we derive clump masses and column densities, bolometric luminosities, physical sizes and virial masses for the majority of these associated clumps. These clumps have typical sizes and masses of $\sim$1\,pc and a few thousand \msun, respectively. The embedded sources have luminosities of $\sim$10$^4$\,\lsun\ and appear to be warming their local environments with typical gas temperatures of $\sim$20\,K.

We find that approximately a third of all  clumps are associated with two or more massive-star formation tracers and are therefore likely to host multiple evolutionary stages. These clumps present a problem when trying to attribute derived physical properties to a specific evolutionary stage and so have been excluded from the detailed analysis. We have identified $\sim$100 methanol masers that are positionally coincident ($<$2\arcsec) with a MYSO or \hii\ region; this angular size corresponds to less than 0.1\,pc at 10\,kpc and therefore these sources are located in the same core within the larger clump and very likely to be tracing the same embedded object. These ``multi-phase'' sources may represent interesting evolutionary transition objects but also reveal that methanol masers are associated with a wide range of evolutionary stages. Approximately 20\,per\,cent of the methanol masers are positionally associated with a MYSO and a further $\sim$10\,per\,cent are associated with an \hii\ region and therefore care needs to be taken to identify those associated with the pre-MYSO and \hii\ region stages. For the purposes of our analysis, if a methanol maser is found to be coincident with a MYSO or \hii\ region it has been classified as a MYSO or \hii\ region, respectively.

We separate the sample of the clumps into four observationally distinct subsamples; these are methanol-maser, MYSO, \hii-region and multi-phase associated clumps. We compare statistical properties of these subsamples to look for evidence to support the commonly assumed evolutionary sequence: methanol maser $\rightarrow$\ MYSO $\rightarrow$\ \hii\ region. Our main findings are as follows:

\begin{enumerate}

\item Clumps associated with a massive-star formation tracer are more spherical in morphology and centrally condensed than the rest of the quiescent population of ATLASGAL clumps. We find that the young massive stars identified are tightly correlated with the position of the peak of the submillimetre emission found towards the centres of their host clumps, with typical projected separations of $\sim$0.14\,pc. The most massive stars are therefore predominately found towards the centres of spherical, centrally condensed clumps.

\item The massive star-forming clumps identified are strongly associated with the highest column-density clumps in the Galaxy. Essentially {\emph all} clumps with column densities in excess of 2--3 $\times 10^{23}$ cm$^{-2}$ are host to current massive star formation. There is also a significant correlation between column density and clump mass with the most massive clumps also having the highest column densities. Our samples of MSF clumps therefore include all of the most massive clumps located in the Galactic disc.

\item No significant difference is found between the $Y$-factors, aspect ratios or relative location within the clump, with respect to the submm peak, for any of the MSF subsamples, which would suggest that the clump structure changes little as the embedded objects evolve from the protostellar, through the MYSO to the compact \hii-region stages. There are clearly differences between the clumps currently associated with massive star formation and the more quiescent population of clumps, which suggests that the structure of the pre-stellar clumps must evolve significantly before star formation begins but, once underway, it proceeds faster than the clump can evolve. This is consistent with a fast star-formation process (\citealt{csengeri2014}).

\item Almost all of the subsample of massive star forming clumps for which we have ammonia linewidths have a virial parameter $\alpha<2$ and, unless supported by strong magnetic fields, are gravitationally unstable against collapse. These clumps are thus likely to be undergoing global infall.  Furthermore, we find a strong anti-correlation between clump mass and the virial parameter, with the virial parameters decreasing with increasing clump mass (cf. \citealt{kauffmann2013}). The most massive clumps are therefore the least stable against gravity, which  provides a simple explanation as to why no massive pre-stellar clumps have yet been identified in the Galactic disc.

\item We find a very strong correlation between clump mass and the bolometric luminosity of the embedded methanol-maser sources ($r=0.58$) and MYSOs and \hii\ regions ($r=0.64$). The bolometric luminosity for all embedded sources is linearly dependent on clump mass. We therefore conclude that the most massive stars are forming in the most massive and highest column-density clumps.

\item The MYSOs, \hii\ regions and multi-phase subsamples form a continuous distribution over three orders of magnitude in mass and four orders of magnitude in luminosity. All of these clumps are clustered at approximately \lbol/\mclump\ = 10\,\lsun/\msun, which, according to the scenario outlined by \citet{molinari2008}, is the transition point the main accretion and envelope clean-up phases in the formation of massive stars. This suggests that the MYSOs and \hii\ regions occupy a similar range of evolutionary stages with the majority near the end of their main accretion phase.

\item Despite the close correspondence of the \lbol/\mclump\ distributions of clumps containing MYSOs and \hii\ regions, these two subsamples dominate different ends of the mass-luminosity parameter space. The MYSOs dominate the lower-mass and lower-luminosity part of the distribution while the \hii\ regions occupy the more massive and luminous end. However, these two subsamples together form a continuous distribution with no break of slope, which might be expected if different underlying processes were producing the observed luminosity. 

\item  MYSOs are associated with significantly lower mass and  column-density clumps compared with the \hii\ region subsample. This is also true for a distance-limited subset of the data and so is unlikely to stem from any distance-related sensitivity bias. However, the differences between these two subsamples are likely to result from the higher column densities associated with the most massive clumps as these are theoretically predicted to lead to higher accretion rates and faster evolution times. We conclude that the MYSOs and \hii\ regions cover a similar range of evolutionary stages, and the main difference between them is due to a decreasing MYSO lifetime with increasing clump mass as also suggested by \citet{mottram2011b} and \citet{davies2011}. A consequence of the faster evolutionary times may be that the MYSO stage may not be observable for the most massive stars.

\item The properties of the clumps hosting methanol masers and \hii\ regions are much more closely matched. The KS test is unable to distinguish between their clump-mass and column-density distributions.  However, the methanol masers are significantly less luminous than the \hii\ regions. The \lbol/\mclump\ ratio for the methanol masers is $\sim$7\,\lsun/\msun, which is approximately half the value found for the  MYSOs or \hii\ region associated clumps and so these are likely to be in an earlier evolutionary stage. The methanol masers associated with the most massive clumps are therefore a promising sample of early O-type star precursor candidates.

\item The bolometric luminosities determined for the most massive clumps fall far below those expected for an embedded stellar cluster for a fixed SFE at all clump masses, but are more consistent with the luminosity expected from the most massive stars in a cluster. They are also consistent with ZAMS luminosities derived independently from the radio emission from the compact \hii\ regions. This suggests that the observed luminosities are primarily coming from the most massive embedded stars in the forming cluster and that the rest of the lower-mass stars have yet to make a significant contribution to the cluster luminosity. 

\item We compare the clump masses and radii to the empirical criterion derived by \citet{kauffmann2010b} for high-mass star formation and those derived by \citet{lada2010} and \citet{heiderman2010} for ``efficient'' low-mass star formation. We find a strong correlation ($r=0.85$) for all associated clumps that covers 2 orders of magnitude in radius and four orders of magnitude in clump mass. We find no significant difference between the different subsamples. All sources are above the threshold for ``efficient'' low-mass star formation and all but a handful of sources satisfy the \citet{kauffmann2010b} criterion and so these clumps have the potential to form massive stars. 

\item We have identified a sample of 16 MPC candidates located in the Galactic disc. This sample is likely to consist of a large fraction of the whole Galactic population of these objects. We compare the properties with those of a similar sample found towards the Galactic centre and find most of their properties to be similar (mass, size and densities). However, while the MPC candidates found in the disc are very actively forming massive stars, with a few being classified as mini-starbursts, the Galactic-centre candidates are effectively starless. 

\end{enumerate}

Finally ATLASGAL provides a uniform set of measurements that can be used to put the result of these previous studies in a more Galaxy-wide context and in the developing evolutionary sequence for the formation of massive stars. This sample of massive-star-forming clumps we have compiled should provide an ideal starting point for more detailed studies in future, particularly with ALMA.

\section*{Acknowledgments}

We thank Steve Longmore, Adam Ginsburg and Padelis Papadopoulos for discussions concerning the MPC candidate sources, and Mark Heyer and Rosie Chen for some interesting discussions on an earlier draft of the paper. We would also like to thank the referee for their comments and suggestions that have helped to significantly improve this manuscript. The ATLASGAL project is a collaboration between the Max-Planck-Gesellschaft, the European Southern Observatory (ESO) and the Universidad de Chile. This research has made use of the SIMBAD database operated at CDS, Strasbourg, France. This work was partially funded by the ERC Advanced Investigator Grant GLOSTAR (247078) and was partially carried out within the Collaborative Research Council 956, sub-project A6, funded by the Deutsche Forschungsgemeinschaft (DFG). This paper made use of information from the Red MSX Source survey database at http://rms.leeds.ac.uk/cgi-bin/public/RMS\_DATABASE.cgi which was constructed with support from the Science and Technology Facilities Council of the UK. L. B. acknowledges support from CONICYT project Basal PFB-06.

\bibliography{rms}

\bibliographystyle{mn2e_new}

\end{document}